\documentclass[aps,english,nofootinbib,notitlepage,showpacs,longbibliography]{revtex4-1}
\usepackage{amstext,amsmath,amssymb,amsfonts,bbm}
\usepackage[latin1]{inputenc}
\usepackage{graphicx}
\usepackage{epstopdf}
\usepackage{amsthm}
\usepackage{tocvsec2}
\usepackage{enumerate}
\usepackage{subfigure}
\usepackage{color}

\usepackage{multirow} \usepackage{rotating}
\usepackage[colorlinks,citecolor=red,linkcolor=red,urlcolor=red]{hyperref}
\usepackage{ae,aecompl}

\usepackage{babel}
\usepackage{float}

\usepackage{amssymb}
\usepackage{graphicx}
\usepackage{esint}

\def\beq{\begin{equation}}
\def\be{\begin{equation}}
\def\ee{\end{equation}}
\def\bes{\begin{eqnarray}}
\def\ees{\end{eqnarray}}

\makeatletter

\providecommand{\tabularnewline}{\\}

\makeatother

\begin{document}

\title{(2+1) Regge Calculus: \\
Discrete Curvatures, Bianchi Identity, and Gauss-Codazzi Equation}

\author{Seramika Ariwahjoedi$^{1}$, Freddy P. Zen$^{1,2}$\vspace{1mm}
}

\affiliation{$^{1}$Theoretical Physics Laboratory, THEPI Division, Institut Teknologi
Bandung, Jl. Ganesha 10 Bandung 40132, West Java, Indonesia.\\
$^{2}$Indonesia Center for Theoretical and Mathematical Physics (ICTMP),
Indonesia.}
\begin{abstract}
\noindent The first results presented in our article are the clear
definitions of both intrinsic and extrinsic discrete curvatures in
terms of holonomy and plane-angle representation, a clear relation
with their deficit angles, and their clear geometrical interpretations
in the first order discrete geometry. The second results are the discrete
version of Bianchi identity and Gauss-Codazzi equation, together with
their geometrical interpretations. It turns out that the discrete Bianchi
identity and Gauss-Codazzi equation, at least in 3-dimension, could
be derived from the dihedral angle formula of a tetrahedron, while
the dihedral angle relation itself is the spherical law of cosine
in disguise. Furthermore, the continuous infinitesimal curvature 2-form,
the standard Bianchi identity, and Gauss-Codazzi equation could be
recovered in the continuum limit.
\end{abstract}
\maketitle

\section{Introduction}

The research of quantum gravity, as an attempt to consistently quantize
the gravitational field, has been growing fast in many directions.
The first step of the modern work in the field was started with the
phase-space variables and Hamiltonian of general relativity \cite{key-1,key-2}.
This was done canonically through the construction of a 3-dimensional
hypersurface embedded in spacetime, introduced by Arnowitt, Deser,
Missner, in the second order formulation of gravity, where the fundamental
variable is the 3-dimensional spatial metric \cite{key-3,key-4}.
The quantization of the phase space of gravity was carried directly
by Dirac and Bergmann \cite{key-5,key-6,key-7,key-8,key-9,key-10},
resulting in the Wheeler de Witt equation which is difficult to solve
\cite{key-11,key-12}. Other attempt to write gravity in the form
similar to Yang-Mills field fibre bundle seems to give a promising
path, this is known as the first order formulation, where the fundamental
variables are the spatial connection and triads \cite{key-13}. The
dynamical equations arising from the first order formulation are a
set of constraint equations. Attempt to write the constraints first
class leads to the definition of the Ashtekar new variables, based
on the Plebanski approach \cite{key-14,key-15,key-16,key-17,key-18}.
The use of the new variables leads to a set of solution on the kinematical
level; known as the Rovelli-Smolin loop representation \cite{key-19,key-20}.
This in turn gives rise to the field of loop quantum gravity \cite{key-21,key-22,key-23}.

In the fundamental level, loop quantum gravity predicts that space
are discrete and fuzzy \cite{key-24,key-25,key-26}. The discreteness
is due to the compactness of the $SU(2)$ group, as the gauge group
of the 3-dimensional space. The spacetime continuum in the classical
general relativity picture is obtained asymptotically in the continuum
limit of the theory, where the size and number of grains of space
are extremely large \cite{key-27,key-28,key-29}. In between the Plank
scale and classical continuous general relativity scale, the mesoscopic
scale is defined as the scale where the space behave classically but
discrete. This is the scale of the large size and finite numbers of
the grains of space, which also known as the semi-classical limit
\cite{key-30,key-31,key-32}. The behaviour of spacetime in this scale
could be well-approximated by the theory of discrete gravity \cite{key-32}.

Discrete gravity had first been studied by Regge, in the second order
formulation \cite{key-33,key-34}. The powerful approach of Regge
calculus, which is different from other discrete theories, is in the
writing of general relativity formulation without the use of coordinate,
i.e., using scalar variables such as angle and norm of area, instead
of vectorial variables. Furthermore, it had been shown that discrete
gravity will coincide with general relativity in the classical limit,
at the level of the action, when the discrete manifold converges to
Riemannian manifold \cite{key-33,key-34,key-35}. In the other hand,
attempt to write discrete gravity in first order fomulation is done
by Barret \cite{key-36}.

A part of the theory which is not entirely clear and needs more attention
is the ADM splitting in Regge calculus. The ADM formalism is based
by an older theorem of Gauss, widely known by mathematicians as the
Gauss-Codazzi relation, which describe the relation between curvatures
of a manifold with its embeddded submanifold, or, in the language
of general relativity, between spacetime and its hypersurface foliation.
Works on canonical formulation of Regge calculus had been started
by \cite{key-37,key-38,key-39,key-40,key-41}, and specifically, on
the hypersurface foliation and Gauss-Codazzi equation in discrete
geometry by \cite{key-42,key-43}, with the recent works by \cite{key-44,key-44a}.
Our work is an attempt to clarify some parts of these previous results.

A complete understanding in the $\left(3+1\right)$ formulation of
discrete geometry is needed to understand completely the canonical
formulation of quantum gravity, for instance, the evolution of spin-network
in loop quantum gravity, and its relation with spinfoam theory. Some
problems which are not entirely clear include the procedure to define
the hypersurface in the Regge simplicial complex and the relation
between the deficit angle as discrete curvature with the curvature
2-form in the first order formulation. Partial results to clarify
these issues can be found in \cite{key-45,key-46,key-47}. Another
problem which is partially unclear is the definition of extrinsic
curvature, which had been studied in \cite{key-48}, and more recently
in \cite{key-49,key-50}. The curvatures need to satisfy some geometrical
relations, which are, the Bianchi identity and the Gauss-Codazzi relation.
The discrete version of the first has been studied extensively, for
instance \cite{key-51,key-52,key-53}, and the latter in \cite{key-42,key-43,key-44a},
but the discrete geometrical interpretation of these relations remain
unclear.

Our work is an attempt to clarify these problem. The first result
presented in this article are the clear definitions of both intrinsic
and extrinsic discrete curvatures in terms of holonomy and plane-angle
representation, a clear relation with their deficit angles, and their
clear geometrical interpretations in the first order formulation of
discrete geometry. All of these are done with the use of minimal assumptions.
The second result is related to the identities and relation between
these curvatures. The relation between the Bianchi identity and the
law of cosine is already indicated in \cite{key-53A}. We show that
this indication is correct, by obtaining the discrete version of Bianchi
identity and its geometrical interpretation. It turns out that the
discrete Bianchi identity and Gauss-Codazzi equation, at least in
3-dimension, could be derived from the dihedral angle formula of a
tetrahedron, while the dihedral angle relation itself is the spherical
law of cosine in disguise. Moreover, we show that the continuous infinitesimal
curvature 2-form, the standard Bianchi Identity, and Gauss-Codazzi
relation could be recovered in the continuum limit.

The article is structured as follows. In Section II, we reviewed the
definition of curvatures in fibre bundle, this include the ADM procedure
in the first order formulation. Section III is a brief review of discrete
geometry and its formulation in the lattices, which include the definition
of the abstract (combinatorial) dual-lattice. Section IV is the main
result of out works, which consists the definition of intrinsic curvature
2-form, extrinsic curvature, Bianchi Identity, and Gauss-Codazzi relation
in the discrete Regge calculus setting. In Section V, the continuum
limit is recovered, altogether with the discussions relevant to our
results.

\section{Curvatures on Fibre Bundle}

\subsection{The Curvature 2-Form}

Suppose we have a standard vector bundle $M\times\mathbb{R}^{n},$
with $M$ is an $n$-dimensional base manifold equipped with a Riemannian
metric $g,$ and $\mathbb{R}^{n}$ is the $n$-dimensional vector
space. Let $E$ be a fibre bundle locally trivial to $M\times\mathbb{R}^{n}$,
equipped with connection $\boldsymbol{A}.$ The $n$-dimensional intrinsic
curvature of the connection could be described by the \textit{curvature
2-form}, which is a map acting on sections of a bundle:
\begin{eqnarray*}
\boldsymbol{F}\left(\boldsymbol{s}_{1},\boldsymbol{s}_{2}\right):E_{p} & \rightarrow & E_{p}\\
\boldsymbol{V} & \mapsto & \boldsymbol{V}'.
\end{eqnarray*}
$\boldsymbol{F}\left(\boldsymbol{s}_{1},\boldsymbol{s}_{2}\right)$
is defined as the derivative of the connection:
\begin{equation}
\boldsymbol{F}\left(\boldsymbol{s}_{1},\boldsymbol{s}_{2}\right)\boldsymbol{V}=d_{D}\boldsymbol{A}\:\boldsymbol{V}=\left[D_{\boldsymbol{s}_{1}},D_{\boldsymbol{s}_{2}}\right]\boldsymbol{V}-D_{\left[\boldsymbol{\boldsymbol{s}_{1},\boldsymbol{s}_{2}}\right]}\boldsymbol{V},\label{eq:00}
\end{equation}
with $d_{D}$ is the exterior covariant derivative, $\boldsymbol{s}_{1},\boldsymbol{s}_{2}\in T_{p}M,$
are base space vectors with origin $p\in M,$ and $\boldsymbol{V},\boldsymbol{V}'\in E_{p}$
are section of a bundle at $p$. Let $\left\{ \partial_{\mu},dx^{\mu}\right\} $
and $\left\{ \xi_{I},\xi^{I}\right\} $ respectively, be the local
coordinate basis on $M$ and $\mathbb{R}^{n},$ then (\ref{eq:00})
could be written in a local coordinate of $E_{p}$ as follows:
\[
\boldsymbol{F}\left(\partial_{\mu},\partial_{\nu}\right)=\underset{F_{\mu\nu J}^{I}\,\xi_{I}\wedge\xi^{J}}{\underbrace{\partial_{\mu}A_{\nu J}^{I}-\partial_{\nu}A_{\mu J}^{I}+A_{\mu K}^{I}A_{\nu J}^{K}-A_{\nu K}^{I}A_{\mu J}^{K}}}.
\]

The curvature 2-form can be geometrically interpreted as an infinitesimal
rotation of a test vector $\boldsymbol{V}$ by a \textit{rotation
bivector} (the $\xi^{I}\wedge\xi^{J}$ components, also known as the
plane of rotation), if $\boldsymbol{V}$ is parallel-transported around
an infinitesimal square loop $\delta\gamma$ of an infinitesimal plane
(the $dx^{\mu}\wedge dx^{\nu}$ components) \cite{key-54}. It carries
the intrinsic property of the curvature of the connection, through
a parallel transport of tangent vector around a closed curve, see
FIG 1. 
\begin{figure}[H]
\begin{centering}
\includegraphics[scale=0.55]{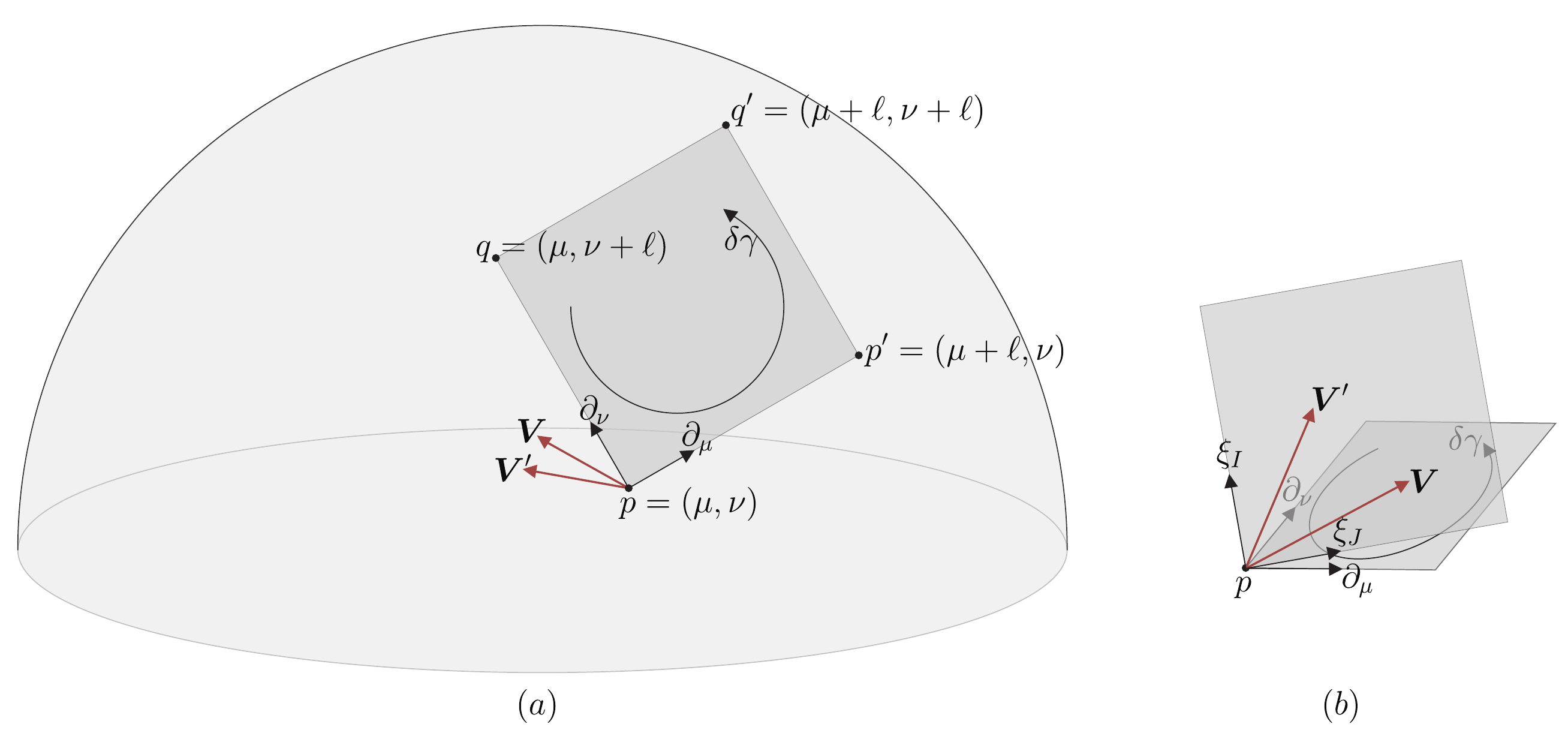}
\par\end{centering}

\centering{}\caption{(a). Geometrical interpretation of curvature 2-form. A test vector
$\boldsymbol{V}$ is carried around an infinitesimal loop $\delta\gamma$
circling plane $dx^{\mu}\wedge dx^{\nu}$. The resulting vector is
$\boldsymbol{V}'$, which does not coincide with $\boldsymbol{V}$
if the suface is curved.  (b). The rotation bivector and loop orientation
in general do not coincide. The test vector $\boldsymbol{V}$ is rotated
according to the rotation bivector, with only its component parallel
to the plane will get rotated.}
\end{figure}

$\boldsymbol{F}$ is antisymmetric by the permutation of the base
space and section indices:
\[
F_{\mu\nu J}^{I}=-F_{\mu\nu I}^{J}=-F_{\nu\mu J}^{I}.
\]
If the torsionless condition is satisfied, it satisfies the \textit{Bianchi
identity}:
\begin{equation}
d_{D}\boldsymbol{F}=0,\label{eq:0}
\end{equation}
which states that the second exterior derivative of any general $p$-form
$\boldsymbol{\omega}$ is zero:
\[
d_{D}^{2}\boldsymbol{\omega}=0.
\]
In terms of components, (\ref{eq:0}) can be written as:
\begin{equation}
d_{D}\boldsymbol{F}=\frac{1}{3!}\left(D_{\mu}F_{\nu\lambda}+D_{\nu}F_{\lambda\mu}+D_{\lambda}F_{\mu\nu}\right)dx^{\mu}\wedge dx^{\nu}\wedge dx^{\lambda}.\label{eq:prev}
\end{equation}

\textbf{Theorem I. }\textit{The Lie algebra $\mathfrak{so(n)}$ of
the rotation group $SO\left(n\right)$ is spanned by antisymmetric
tensor}.

By Theorem I, for each point $p$ of $E_{p}$, the curvature 2-form
carries \textit{two} planes: the rotation bivector and the loop orientation,
where both of them are elements of Lie algebra \textit{$\mathfrak{so(n)}$}.

Let us consider the components of $\boldsymbol{F}$ for some low-dimensional
cases. In dimension two, $\boldsymbol{F}$ has a \textit{single} non-zero
component $F_{xy2}^{1}$ (with its symmetries), written in local coordinate
$\left(x,y\right)$ and $\left(1,2\right)$. Therefore, to describe
completely the intrinsic curvature of a 2-dimensional surface, one
needs a single infinitesimal loop with an \textit{$\mathfrak{so(2)}$}
algebra attached, on each point of the surface. In dimension three,
$\boldsymbol{F}$  has, in general, nine distinct non-zero components
(with its symmetries), thus to describe completely a curvature in
3-dimensional space, one needs \textit{three} infinitesimal loops,
with \textit{$\mathfrak{so(3)}$} algebras attached on each one of
them. Written in coordinates, $F_{xyJ}^{I},$ $F_{yzJ}^{I},$ $F_{zxJ}^{I}$
are $3\times3$ matrices, elements of \textit{$\mathfrak{so(3)}$}.
Any vector $\boldsymbol{V},$ carried along $\delta\gamma$ circling
the $dx\wedge dy$ component of the plane, will be rotated into $\boldsymbol{V}'$
by:
\begin{equation}
V'^{I}=F_{xyJ}^{I}V^{J}.\label{eq:1}
\end{equation}

$\boldsymbol{F}$ could be defined on a general infinitesimal rectangular
loop $\delta\gamma\left(\tau\right)$ circling infinitesimal plane
$\boldsymbol{s}_{1}\wedge\boldsymbol{s}_{2}$, with $\boldsymbol{s}_{1}=s_{1}^{\mu}\partial_{\mu}$,
as:
\begin{equation}
\boldsymbol{F}\left(\boldsymbol{s}_{1},\boldsymbol{s}_{2}\right)=s_{1}^{\mu}s_{2}^{\nu}F_{\mu\nu J}^{I}\xi_{I}\wedge\xi^{J},\label{eq:2}
\end{equation}
 such $\boldsymbol{V}$ carried along $\delta\gamma$, will be rotated
into:
\begin{eqnarray*}
V'^{I} & = & u^{\mu}v^{\nu}F_{\mu\nu J}^{I}V^{J}.
\end{eqnarray*}
The rotation bivector component $\boldsymbol{F}_{IJ}$ defines a \textit{plane
of rotation}, which are labeled as $\delta\hat{\boldsymbol{J}}$.
The 'axis' of the rotation perpendicular to $\delta\hat{\boldsymbol{J}}$
is labeled as $\star\delta\hat{\boldsymbol{J}}$, where the star $\star$
is used to define the combinatorial (topological) dual of a geometrical
quantity; this will be clear in the next section. 

The 'axis' $\star\delta\hat{\boldsymbol{J}}$ is well-defined; we
called this, using Regge terminology, as a (infinitesimal) \textit{hinge}
\cite{key-33,key-34,key-55,key-57}. The hinge depends on the dimension
of the space; if the dimension of space is $n$, then the hinges are
$\left(n-2\right)$ forms. This way of viewing the curvature 2-form
 as pair of planes is important when one consider the curvatures in
Regge calculus.

\subsection{Gauss-Codazzi Equation in Fibre Bundle}

This subsection contains a brief review of Gauss-Codazzi equation,
which describe the relation between the curvatures of a manifold with
its submanifold. The original Gauss-Codazzi equation is defined on
manifold, where, using terminology in general relativity, it is written
in a second order formulation. Nevertheless, the concept can be adopted
to fibre bundles, such that the Gauss-Codazzi equation can be written
in the first order formulation. 

The $(n+1)$ split of the fibre bundle of gravity can be done quite
similarly to the fibre bundle of a Yang-Mills field, say $E\sim M\times\mathcal{F}$.
In Yang-Mills theory, one splits the spacetime $M\sim\mathbb{R}\times\Sigma$
to obtain the electric and magnetic part of the curvature of the fibre,
say, $\boldsymbol{E}\wedge dt$ and $\boldsymbol{B}$, which are the
curvatures projected on $\mathbb{R}$ and $\Sigma$, respectively.
The main difference is, in the Yang-Mills field, one does not split
the fibre, because Yang-Mills theory is background-dependent. In the
other hand, gravity is a background independent theory where the field
and spacetime are indistinguishable entities \cite{key-21,key-22}.
Therefore, spliting the base space will induce the spliting on the
fibre. 

Let $e$ be a local \textit{trivialization}, a diffeomorphism map
between the trivial vector bundle $M\times\mathbb{R}^{n}$ and the
tangent bundle over $M$: $E\sim TM=\cup_{p}\left\{ p\times T_{p}M\right\} $.
Instead of spliting the base space, one starts with spliting the fibre
(which is easier). Let $\xi^{I},$ $I=0,1,2,..,$ be a local coordinate
on $\mathbb{R}^{n}.$ The next step is to construct an embedded $n$-dimensional
hypersurface $\Sigma\subset M$ by selecting $\xi^{0}$ as a normal
to, also, an $n$-dimensional hypersurface $\Omega\subset\mathbb{R}^{n}$.
The diffeomorphism $e$ will sends normals $\xi^{0}\in\mathcal{F}\sim\mathbb{R}^{n}$
to $e^{0}=e\left(\xi^{0}\right)\in T_{p}M.$ Then the $(n+1)$ split
generated by the normal $\xi^{0}$ on vector space $\mathbb{R}^{n}$
will induce split on $T_{p}M$ generated by $e^{0}=e\left(\xi^{0}\right).$ 

The following derivation will be based on our previous work \cite{key-44a};
$e^{0}$ in general will be a linear combination of coordinate basis
vector in $T_{p}M$:
\[
e^{0}=e_{\mu}^{0}dx^{\mu}=\underset{N}{\underbrace{e_{0}^{0}}}dx^{0}+\underset{N_{i}}{\underbrace{e_{i}^{0}}}dx^{i},
\]
with $N$ and $N_{i}$ are, respectively, the lapse and shift functions.
Let us choose a local coordinate in $T_{p}M$ such that:
\begin{equation}
e\left(\xi^{0}\right)=e^{0}=\delta_{\mu}^{0}dx^{\mu}=dx^{0}.\label{eq:gc1}
\end{equation}
This means one use the time gauge where the lapse $N=1,$ and the
shift $N_{i}=0.$ 

The $n$-dimensional intrinsic curvature of the connection is labeled
as: 
\[
\,^{n}\boldsymbol{F}=F_{\mu\nu J}^{I}\xi_{I}\wedge\xi^{J}\otimes dx^{\mu}\wedge dx^{\nu}.
\]
In the time gauge, the $(n+1)$ ADM formulation for the curvature
2-form is carried by the spliting $I=0,a,$ and $\mu=0,i$, which
are compatible with (\ref{eq:gc1}). Therefore, the projection of
$\,^{n}\boldsymbol{F}$ on $\Sigma$ is:
\begin{equation}
\left.\,^{n}\boldsymbol{F}\right|_{\Sigma}=F_{ijb}^{a}\xi_{a}\wedge\xi^{b}\otimes dx^{i}\wedge dx^{j},\label{eq:xox}
\end{equation}
 written in a local coordinate as:
\begin{eqnarray*}
F_{ijb}^{a} & = & \partial_{i}A_{jb}^{a}-\partial_{j}A_{ib}^{a}+A_{ic}^{a}A_{jb}^{c}-A_{jc}^{a}A_{ib}^{c}+A_{i0}^{a}A_{jb}^{0}-A_{j0}^{a}A_{ib}^{0}.
\end{eqnarray*}
The closed part of $\left.\,^{n}\boldsymbol{F}\right|_{\Sigma}$ is
clearly the $\left(n-1\right)$-dimensional intrinsic curvature of
connection in $\Sigma$, and the rest is the extrinsic curvature part:
\begin{eqnarray}
\,^{n-1}\boldsymbol{F} & = & \,^{n-1}F_{ijb}^{a}\xi_{a}\wedge\xi^{b}\otimes dx^{i}\wedge dx^{j}=\left(\partial_{i}A_{jb}^{a}-\partial_{j}A_{ib}^{a}+A_{ic}^{a}A_{jb}^{c}-A_{jc}^{a}A_{ib}^{c}\right)\xi_{a}\wedge\xi^{b}\otimes dx^{i}\wedge dx^{j}\label{eq:yoy}\\
\boldsymbol{K} & = & K_{i}^{a}\xi_{\alpha}\otimes dx^{i}=A_{i0}^{a}\xi_{\alpha}\wedge\xi^{0}\otimes dx^{i}.\label{eq:zoz}
\end{eqnarray}
As a result, one has the Gauss-Codazzi relation for a fibre bundle
of gravity:
\begin{equation}
\left.\,^{n}\boldsymbol{F}\right|_{\Sigma}=\,^{n-1}\boldsymbol{F}+\left[\boldsymbol{K},\boldsymbol{K}\right].\label{eq:20}
\end{equation}
The Gauss-Codazzi equation (\ref{eq:20}) is invariant under coordinate
transformation, but it must be kept in mind that the way of writing
$\left.\,^{n}\boldsymbol{F}\right|_{\Sigma}$, $\,^{n-1}\boldsymbol{F}$,
and $\boldsymbol{K}$ in (\ref{eq:xox}), (\ref{eq:yoy}), and (\ref{eq:zoz})
are written in a special gauge condition (\ref{eq:gc1}). In a general
gauge condition, they do not have such simple forms, for instance,
see \cite{key-58}.

\subsection{Rotations and Holonomies}

Another way to describe the curvatures of a manifold is through holonomy.
The holonomy $\boldsymbol{H}_{\gamma}\left(\boldsymbol{A},\gamma\left(\tau\right)\right)$
of connection $\boldsymbol{A}$ along a curve $\gamma\left(\tau\right)$
with parameter $\tau$ is defined as a solution to the following equation:
\[
D_{\gamma'\left(\tau\right)}\boldsymbol{V}\left(\tau\right)=0,\quad\gamma'\left(\tau\right)=\frac{d\gamma\left(\tau\right)}{d\tau},\boldsymbol{V}\left(\tau\right)\in T_{p}M,
\]
which is: 
\begin{eqnarray*}
\boldsymbol{V}\left(\tau\right) & = & \boldsymbol{H}_{\gamma}\left(\boldsymbol{A},\gamma\left(\tau\right)\right)\boldsymbol{V}\left(0\right),\\
\boldsymbol{H}_{\gamma}\left(\boldsymbol{A},\gamma\left(\tau\right)\right) & = & \hat{P}\exp\int_{\gamma\left(\tau\right)}\boldsymbol{A}\left(\gamma\left(\tau\right)\right).
\end{eqnarray*}
$\boldsymbol{A}$ is the spin connection on the fibre bundle $E$,
and $\hat{P}$ is the path ordered operator \cite{key-21,key-22}.
It is clear that a holonomy is a subset of the rotation group parallel-transporting
vectors while preserving their norms. Its relation with the curvature
2-form  can be obtained by considering the holonomy around a closed
curve or loop with origin $\mathcal{O}$:
\begin{equation}
\boldsymbol{H}_{\gamma}\left(\boldsymbol{A},\gamma_{\mathcal{O}}\right)=\hat{P}\exp\oint_{\gamma\left(\tau\right)=\partial a}\boldsymbol{A}\left(\gamma\left(\tau\right)\right).\label{eq:4}
\end{equation}

\textbf{Theorem II (Stokes-Cartan).}\textit{ If $\boldsymbol{\omega}$
is a smooth $\left(n-1\right)$-form with compact support on smooth
n-dimensional manifold with boundary $\Omega$, and $\partial\Omega$
labels the boundary of $\Omega$ given the induced orientation, then:}

\[
\oint_{\partial\Omega}\boldsymbol{\omega}=\int_{\Omega}d_{D}\boldsymbol{\omega},
\]
\textit{with $d_{D}$ is the exterior covariant derivative}.

By Theorem II, (\ref{eq:4}) could be written as:
\begin{equation}
\boldsymbol{H}_{\gamma}\left(\boldsymbol{A},\gamma_{\mathcal{O}}\right)=\hat{P}\exp\int_{a}d_{D}\boldsymbol{A}=\hat{P}\exp\int_{a}\boldsymbol{F},\label{eq:5}
\end{equation}
using  the definition of $\boldsymbol{F}$ as the derivative of the
connection (\ref{eq:00}). A straightforward calculation gives the
Taylor expansion of (\ref{eq:5}), up to the first order:
\begin{equation}
\boldsymbol{H}_{\gamma}\left(\boldsymbol{A},\gamma_{\mathcal{O},a}\right)=\boldsymbol{1}+a\boldsymbol{F}+O\left(a^{2}\right).\label{eq:6}
\end{equation}

The holonomy representation provides a natural way towards the 'finite'
discrete theory: a \textit{regularization} scheme. One notices that,
in contrast with the infinitesimal formulation, there exists only
a single plane in the holonomy representation, which is the rotation
bivector plane, since the loop orientation is 'summed up' by the integral
in (\ref{eq:4}). This can be understood through the 1-dimensional
analogue: a point, which is an infinitesimal curve, is equipped with
a vector, tangent to the curve; but integrating the tangent vectors
to obtain an integral curve, will result in losing the vector as an
exchange. Returning to our case, as a result of the integration, we
have a holonomy on a surface region $a$, instead of a 2-form plane. 

One could start to apply a regularization scheme. The idea is the
following: Each point $p$ of an arbitrary $n$-dimensional manifold
$\mathcal{F}$, is 'blown' into an (abstract) $n$-simplex, which
is an $n$-dimensional analogue to 'triangle' (and tetrahedron). We
label the collections of $n$-simplices connected to each other as
$\star\mathcal{F}_{\Delta}$. Each $n$-simplex is constructed from
$2$-simplices, which are triangles, unless it is trivial. Let us
labe; the triangle (which is a portion of a plane) as $\star\boldsymbol{l}.$
Moreover, one could attach to the triangle a 2-form $\boldsymbol{J}$.
$\frac{\boldsymbol{J}}{\left|\boldsymbol{J}\right|}$ is the rotation
bivector, or the plane of rotation. As explained earlier, one could
define the dual to the plane of rotation as (now, a \textit{finite})
hinge $\star\hat{\boldsymbol{J}}$, which is attached on segment $\boldsymbol{l}$. 

The next step is to define a holonomy $\boldsymbol{H}_{\Delta}$ along
the boundary of each triangle $\star\boldsymbol{l}$, which circles
the hinge $\star\hat{\boldsymbol{J}}$. Since any $n$-simplex is
flat in the interior, we could write $\boldsymbol{H}_{\Delta}$ as
a special case of (\ref{eq:4}) as follows:
\begin{equation}
\boldsymbol{H}_{\Delta}=\exp\boldsymbol{J}\tau=\exp\underset{\hat{\boldsymbol{J}}}{\underbrace{\frac{\boldsymbol{J}}{\left|\boldsymbol{J}\right|}}}\underset{\phi}{\underbrace{\left|\boldsymbol{J}\right|\tau}},\label{eq:7}
\end{equation}
with $\phi$ is the angle of rotation of $\boldsymbol{H}_{\Delta}.$
Finally, one could define an equivalence class of loops  by the following
statement:\textit{ any closed loop circling a same hinge are equivalent
to one another}. Therefore, one obtain a piecewise-linear manifold
where the curvatures are only concentrated on the hinges. See FIG.
2. 
\begin{figure}[H]
\begin{centering}
\includegraphics[scale=0.55]{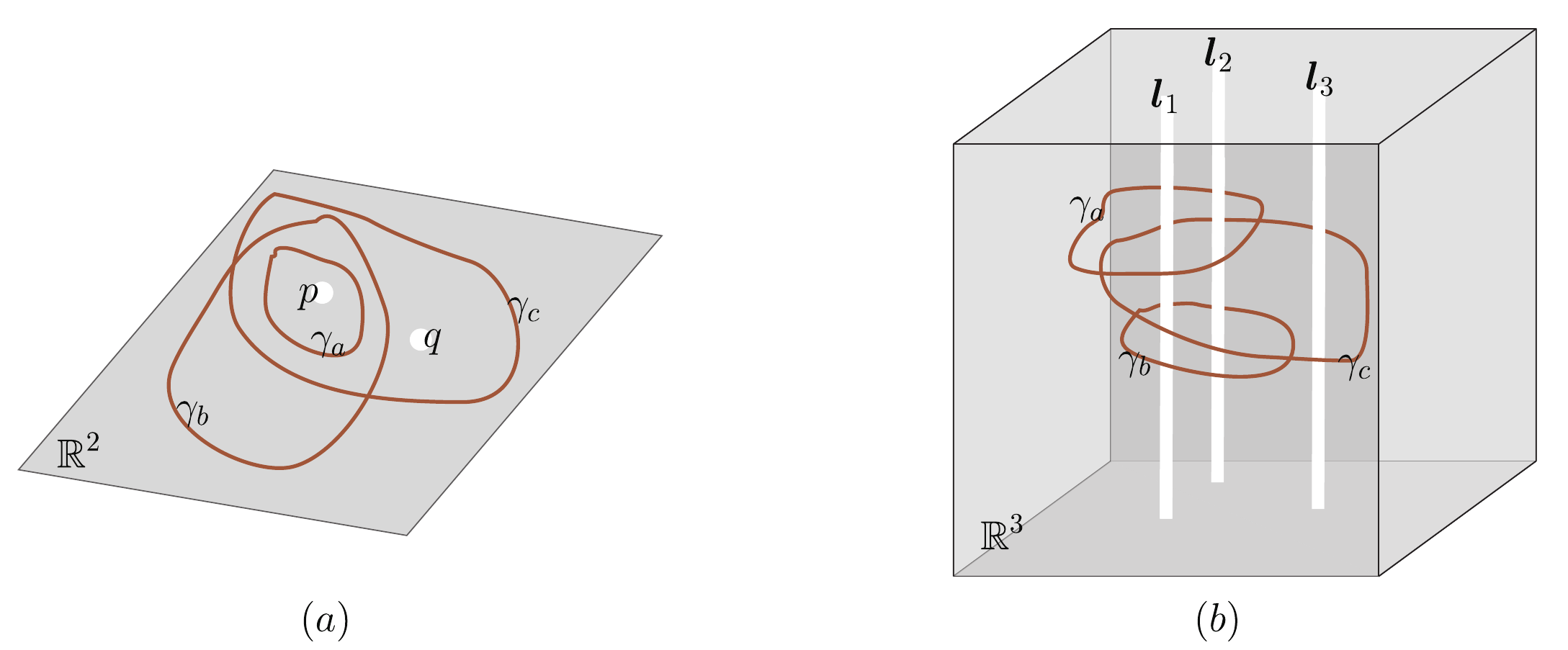}
\par\end{centering}

\caption{Loops circling a same hinge are equivalent; in both (a) and (b), $\gamma_{\alpha}=\gamma_{\beta}\neq\gamma_{c}$.}
\end{figure}
The regularization will be discusses in detail in the next section.

\section{Regge calculus and Discrete Geometry}

In this section, we briefly describe discrete geometry as a discretization
of a differentiable manifold. The simplest discretization is a \textit{simplicial
complex}, where each discrete element is a simplex. A $p$-\textit{simplex}
is the simplest, flat, $p$-dimensional polytope embedded in an $n$-dimensional
space $\mathbb{R}^{n}$, with $n\geq p$. It is constructed from $\left(p+1\right)$
numbers of $\left(p-1\right)$-simplices, such that the lower dimensional
simplices are nested into higher dimensional one. The reason for using
a simplicial complex as a discretization of a continuous manifold
is due to the fact that a simplex is completely determined by their
edges \cite{key-59}.

The discretization of general relativity had first been studied by
Regge, in the second order formulation, this is known as Regge Calculus
\cite{key-33,key-34}. The powerful approach of Regge calculus is
in writing the discrete general relativity formulation without the
use of coordinate, i.e., using scalar variables such as angle, length,
and area, instead of vectorial variables \cite{key-33,key-34}. It
has been shown that discrete gravity will coincide with GR in the
classical limit, at the level of the action, when the discrete manifold
converges to Riemannian manifold \cite{key-33}. However, some aspects
of the theory are not entirely complete, for the research in Regge
gravity is still continued to grow towards many directions.

\subsection{Triangulations: Primal and Dual Lattices.}

The works in discrete geometry and Regge calculus mostly use the Delaunay
lattice for a discretization, such that the vertex of one polytope
is always outside the circumcircles of the others in the lattices
\cite{key-60}. For this reason, the simplicial complex is sometimes
refered as Delaunay \textit{triangulation}. 

A definition of a dual lattice is important for a measurement of the
geometric quantities. This, in return, is important if one needs to
define the action for the dynamical part of the theory \cite{key-55,key-57,key-61}.
Some example of dual lattices commonly used in the literature are
circumcentric or barycentric dual lattices, which are defined by connecting
their circumcenter and barycenters points \cite{key-62}. If the discretization
is a Delaunay triangulation, its circumcentric dual is a Voronoi lattice
\cite{key-62}. Another type of dual lattice, which is important particularly
in loop quantum gravity, it the topological / combinatorial, abstract
dual lattice. The abstract-dual of a complex could be defined as the
circumcentric lattice, but without a fix shape and distance \cite{key-63,key-64}.
In other words, the abstract dual lattice is constructed from graphs,
where only the combinatorial aspects of the graph are important. This
is similar with the framework adopted in the canonical (loop) quantum
gravity \cite{key-21,key-22}. 

In this article, we use the Delaunay triangulation as the primal lattice
and an \textit{abstract-dual }(or combinatorial) lattice as its dual.
The reason for this is explained as follows. Let $\mathcal{F}_{\Delta}$
be the primal lattice of a discretization of an $n$-dimensional manifold
$\mathcal{F}$, and $\star\mathcal{F}_{\Delta}$ be the circumcentric
(or barycentric) dual. Let $\Omega_{\Delta}$ be a discretization
of an $\left(n-1\right)$-dimensional hypersurface $\Omega\subset\mathcal{F}$.
$\Omega_{\Delta}\subset\mathcal{F}_{\Delta},$ such that the $\left(n-1\right)$-simplices
defining $\Omega_{\Delta}$ construct the $n$-simplices of $\mathcal{F}_{\Delta}.$
Moreover, we could define the circumcentric (or barycentric) dual
of $\Omega_{\Delta}$, labeled as $\star\Omega_{\Delta}.$ The reason
of not using both the circumcentric and barycentric dual, is because
it has not been clear if $\star\Omega_{\Delta}\subset\star\mathcal{F}_{\Delta}$,
which is important in our construction of the hypersurface slicing.
Therefore, it is more convenient to use the combinatorial graph, where
the relation $\star\Omega_{\Delta}\subset\star\mathcal{F}_{\Delta}$
could always be defined. 

Let us take a specific example of a primal and dual lattice: Suppose
$\mathcal{F}_{\Delta}$ is a triangulation of a 3-dimensional manifold
$\mathcal{F}.$ $\mathcal{F}_{\Delta}$ is discretized by tetrahedra,
which are described using 3-forms. Embedded in $\mathcal{F}_{\Delta}$,
one could have lower-dimensional simplices: triangles, segments, and
points. With the definition of the abstract-dual lattice, one could
define the following terminologies, adopted from the canonical LQG,
as described in FIG. 3. 
\begin{figure}[H]
\centering{}\includegraphics[scale=0.55]{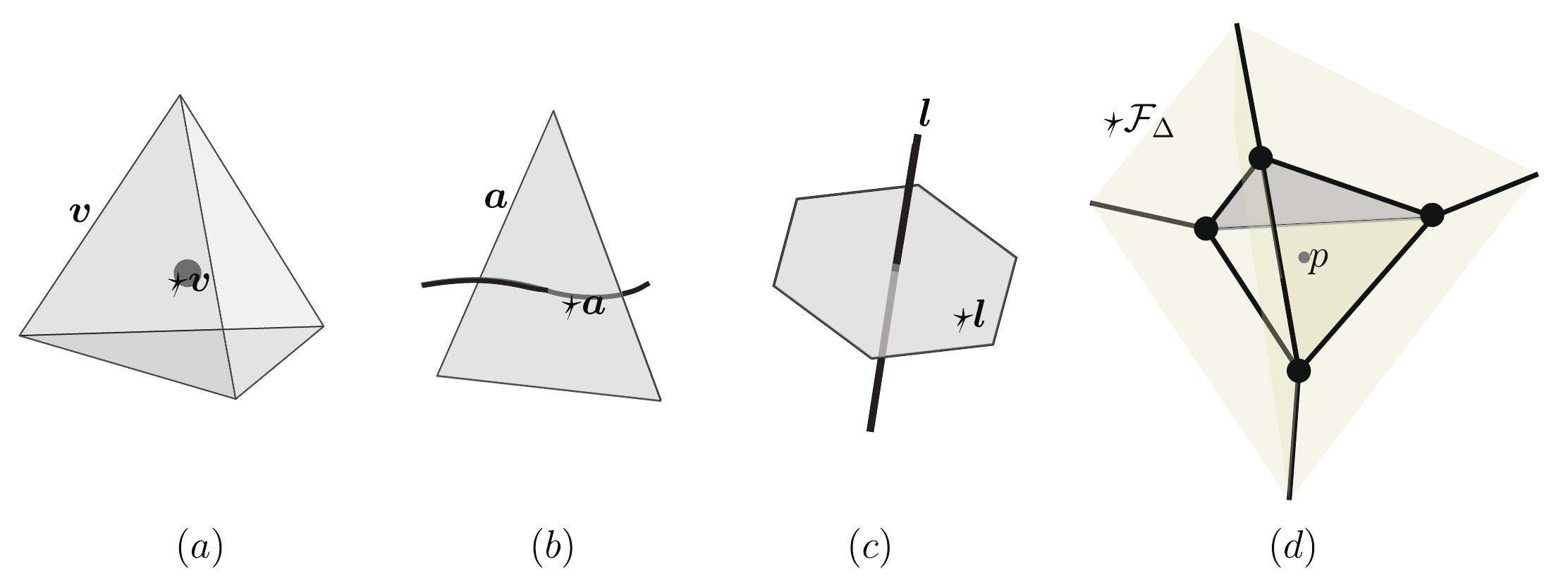}\caption{(a) A tetrahedron in a primal lattice $\mathcal{F}_{\Delta}$, labeled
by its volume $\boldsymbol{v},$ is dual to the \textit{vertex} $\star\boldsymbol{v},$
in $\star\mathcal{F}_{\Delta}$. (b) A triangle in $\mathcal{F}_{\Delta}$,
labeled by its area $\boldsymbol{a}$ is dual to the \textit{edge}
$\star\boldsymbol{a}$ in $\star\mathcal{F}_{\Delta}.$ (c) A segment
$\boldsymbol{l}$ in $\mathcal{F}_{\Delta}$ is dual to the \textit{face}
$\star\boldsymbol{l}$ in $\star\mathcal{F}_{\Delta}.$ (d) A point
$p$ in $\mathcal{F}_{\Delta}$ is dual to the \textit{3D-bubble }in
$\star\mathcal{F}_{\Delta}$.}
\end{figure}

The introduction of the primal and dual cells will be extremely useful
for the rest of this article. In particular, the equivalent class
of loops could be defined using a standard loop, which is naturally
the boundary of the face, dual to the hinge \cite{key-54}. All possible
loops circling hinge $\star\hat{\boldsymbol{J}}$ are represented
by the standard loop.

\subsection{Curvatures}

In the framework of Regge calculus, the length of a geometrical object
has finite minimal size. This is followed by the finiteness of the
size of higher dimensional objects: area, and higher dimensional volumes.

It had been discussed previously that the components of the curvature
2-form are infinitesimal rotations, where the planes of rotation are
dual to the infinitesimal hinges. For discrete geometry, the regularization
is straightforward: \textit{the 'discrete' curvature 2-form is a finite
rotation on a finite hinge}. As for finite rotations, it can be represented
in two standard ways: the plane-angle (or area-angle \cite{key-59})
and the holonomy representation \cite{key-44a}.

\subsubsection{Plane-angle representation.}

In this representation, rotation is describe by a couple $\left(\boldsymbol{J},\delta\phi\right)$,
with $\boldsymbol{J}\in\mathfrak{so(n)}$ is an element of Lie algebra
as the plane of rotation and one (real) parameter group $\tau$ times
the norm of the algebra $\left|\boldsymbol{J}\right|$ for the angle
of rotation $\delta\phi$. In the Regge Calculus picture, the intrinsic
curvature is represented by the angle of rotation, or the \textit{deficit
angle}, located on the hinge:
\[
\delta\phi=2\pi-\sum_{i}\phi_{i}.
\]
Non-trivial value of $\delta\phi$ describe the deviation of a space
from being flat, see FIG. 4 for a 2-dimensional case.

\begin{figure}[H]
\centering{}\includegraphics[scale=0.55]{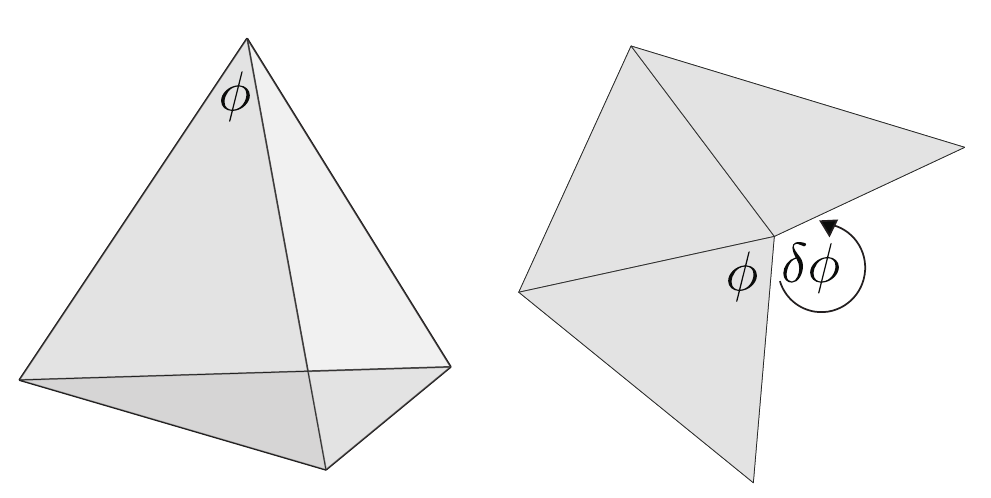} \caption[The hinge and deficit angle.]{Suppose $\Omega_{\Delta}$ is a 2D surface discretized by three triangles.
Point $p$ is the hinge where the deficit angle $\delta\phi$ is located.
$\delta\phi$ is the 2D intrinsic curvature of $\Omega_{\Delta}$
at point $p$.  }
\end{figure}

Any components of a test vector $\boldsymbol{V}$ carried around a
loop $\gamma$ will be rotated by the deficit angle $\delta\phi$
in the direction of the plane of rotation, which in turns, is perpendicular
to the hinge \cite{key-54}. The plane-angle representations give
a natural definition of curvature in the background independence picture
of gravity.

\subsubsection{Holonomy Representation, Exponential and Differential Map.}

Relation (\ref{eq:7}), is a map sending the holonomy, which is an
element of a rotation group, $\boldsymbol{H}\in SO\left(n\right)$,
 to the plane-angle representation, commonly refered as the \textit{exponential
map}. As explained earlier, holonomy representation provides a natural
way to go from the 'infinitesimal' continuous to the 'finite' discrete
theory. For a special case where area inside the loop $a$ is chosen
to be a square $\boldsymbol{s}_{i}\boldsymbol{s}_{j}$, with $\boldsymbol{s}_{i}=\ell\hat{\boldsymbol{s}}_{i}$
and $\hat{\boldsymbol{s}}_{i}$ is a unit vector, (\ref{eq:6}) describe
a direct 'finite' version of the curvature 2-form as follows:
\begin{equation}
\boldsymbol{H}_{\gamma}\left(\boldsymbol{A},\gamma_{\mathcal{O},a}\right)=\boldsymbol{1}+\frac{\ell^{2}}{2}\boldsymbol{F}\left(\hat{\boldsymbol{s}}_{i},\hat{\boldsymbol{s}}_{j}\right)+O\left(\ell^{3}\right),\label{eq:9}
\end{equation}
in other words, the curvature 2-form is the holonomy on an infinitesimal
loops.

As the inverse of the exponential map, one has the \textit{differential
map}, which sends plane-angle representation to the holonomy representation.
This can be obtained from the following procedure: the angle of rotation
$\delta\phi$ can be obtained from the \textit{trace} of the holonomy,
which for a special case $\mathcal{G}\sim SO\left(n\right)$, gives:
\begin{equation}
\textrm{tr}\boldsymbol{H}_{\gamma}=n-2\left(1-\cos\delta\phi\right),\label{eq:trace}
\end{equation}
with $n$ is the dimension of the rotation matrix, while the plane
of rotation $\boldsymbol{J}$ can be obtained from the differential
map:
\begin{eqnarray*}
\left.\frac{d}{d\delta\phi}\right|_{\delta\phi=0}:\mathcal{G} & \rightarrow & \mathfrak{g}\\
\boldsymbol{H} & \mapsto & \boldsymbol{J}=\left.\frac{d\boldsymbol{H}}{d\delta\phi}\right|_{\delta\phi=0}.
\end{eqnarray*}
The map from the holonomy to the plane-angle representation is 1-to-1
and onto.

\subsubsection{Addition of Two Rotations.}

Another important property which is useful is the product of two rotations.
In the holonomy representation, the product of two holonomies is simply
the matrix multiplication between two holonomies as follows:
\begin{equation}
\boldsymbol{H}_{12}\left(\boldsymbol{A},\gamma_{12}\right)=\boldsymbol{H}_{1}\left(\boldsymbol{A},\gamma_{1}\right)\boldsymbol{H}_{2}\left(\boldsymbol{A},\gamma_{2}\right),\label{eq:8}
\end{equation}
which in general is not commutative: $\boldsymbol{H}_{1}\boldsymbol{H}_{2}\neq\boldsymbol{H}_{2}\boldsymbol{H}_{1}.$
This product defines the piecewise-linear aspect of discrete manifold.

In the plane-angle representation, the product formula is more complicated.
The total angle of rotation formula can be obtained by taking trace
of (\ref{eq:8}); this, in particular, depends on the dimension of
the space. As an example, for a special case $\mathcal{G}\sim SU\left(2\right)$,
the element of the group can be written as follows:
\begin{equation}
\boldsymbol{H}_{i}=\boldsymbol{I}\cos\theta_{i}+\boldsymbol{J}_{i}\sin\theta_{i},\label{eq:u}
\end{equation}
so that it gives the following total angle of rotation formula:
\begin{equation}
\cos\theta_{12}=\cos\theta_{1}\cos\theta_{2}-\cos\bar{\phi}_{12}\sin\theta_{1}\sin\theta_{2},\label{eq:22}
\end{equation}
with $\bar{\phi}_{12}$ is the angle between plane $\boldsymbol{J}_{1}$
and $\boldsymbol{J}_{2}$. The total plane of rotation $\boldsymbol{J}_{12}$
for $\theta_{12}$ can be obtained from the Baker-Campbell-Hausdorff
formula. For \textit{$\mathcal{G}=SU\left(2\right)$, }the total plane
formula is the following:
\begin{equation}
\boldsymbol{J}_{12}=\frac{\left(\boldsymbol{I}\cos\theta_{1}+\boldsymbol{J}_{1}\sin\theta_{1}\right)\left(\boldsymbol{I}\cos\theta_{2}+\boldsymbol{J}_{2}\sin\theta_{2}\right)-\boldsymbol{I}_{2\times2}\cos\theta_{12}}{\sin\theta_{12}}.\label{eq:2.13-1}
\end{equation}

\subsubsection{Conjugation and Adjoint Representation.}

Let $\boldsymbol{g},\boldsymbol{h}\in\mathcal{G}$, then suppose one
has the following conjugation induced by $\boldsymbol{g}$
as follows:

\begin{equation}
\boldsymbol{h}^{'}=\boldsymbol{g}^{-1}\boldsymbol{h}\boldsymbol{g}.\label{eq:gla}
\end{equation}
Using the exponential map on $\boldsymbol{h}$ and $\boldsymbol{h}'$,
one has:
\[
\exp\left(\boldsymbol{J}_{h}^{'}\phi\right)=\boldsymbol{g}^{-1}\exp\left(\boldsymbol{J}_{h}\phi\right)\boldsymbol{g}
\]
where $\phi\sim$$\textrm{tr}\boldsymbol{h}=\textrm{tr}\boldsymbol{h}'$
is invariant under conjugation. By Taylor expansion up to
the first order:
\[
\boldsymbol{J}_{h}^{'}\phi+O\left(\phi^{2}\right)=\boldsymbol{g}^{-1}\boldsymbol{J}_{h}\boldsymbol{g}\phi+\boldsymbol{g}^{-1}O\left(\phi^{2}\right)\boldsymbol{g},
\]
for each order $n^{th}$, one has $\boldsymbol{J}_{h}^{'n}=\boldsymbol{g}^{-1}\boldsymbol{J}_{h}^{n}\boldsymbol{g},$
which is equal to $\boldsymbol{J}_{h}^{'n}=\left(\boldsymbol{g}^{-1}\boldsymbol{J}_{h}\boldsymbol{g}\right)^{n}.$
Therefore, one obtains:
\begin{equation}
\boldsymbol{J}_{h}^{'}=\boldsymbol{g}^{-1}\boldsymbol{J}_{h}\boldsymbol{g},\label{eq:alg}
\end{equation}
which is the adjoint representation of the Lie group. Conjugation
on the group (\ref{eq:gla}) induces a transformation of the Lie algebra
by (\ref{eq:alg}). This will be useful when one consider a transformation
of planes with different origin.

\subsection{Loops, Hinges, and Contractibility}

To understand clearly the concept of curvatures in discrete geometry,
one needs to include the concept of \textit{contractible space}. As
a simple explanation,\textit{ a topological space is contractible
if it can be continuously shrunk to a point} \cite{key-65}. Let us
consider the following examples: All loops embedded in $\mathbb{R}^{2}$
or $\mathbb{S}^{2}$ are contractible. Some loops living in a torus
$\mathbb{T}^{2}$ are non-contractible. Some loops living in $\mathbb{R}^{2}-\left\{ 0\right\} $
are non-contractible. In higher dimension, all loops living in $\mathbb{R}^{3}-\left\{ 0\right\} $
are contractible, but some complete closed surface (2-dimensional
'loop', topologically equivalent to $\mathbb{S}^{2}$) living in $\mathbb{R}^{3}-\left\{ 0\right\} $
are non-contractible. This can be generalized to any dimension. See
FIG. 5. 
\begin{figure}[H]
\begin{centering}
\includegraphics[scale=0.55]{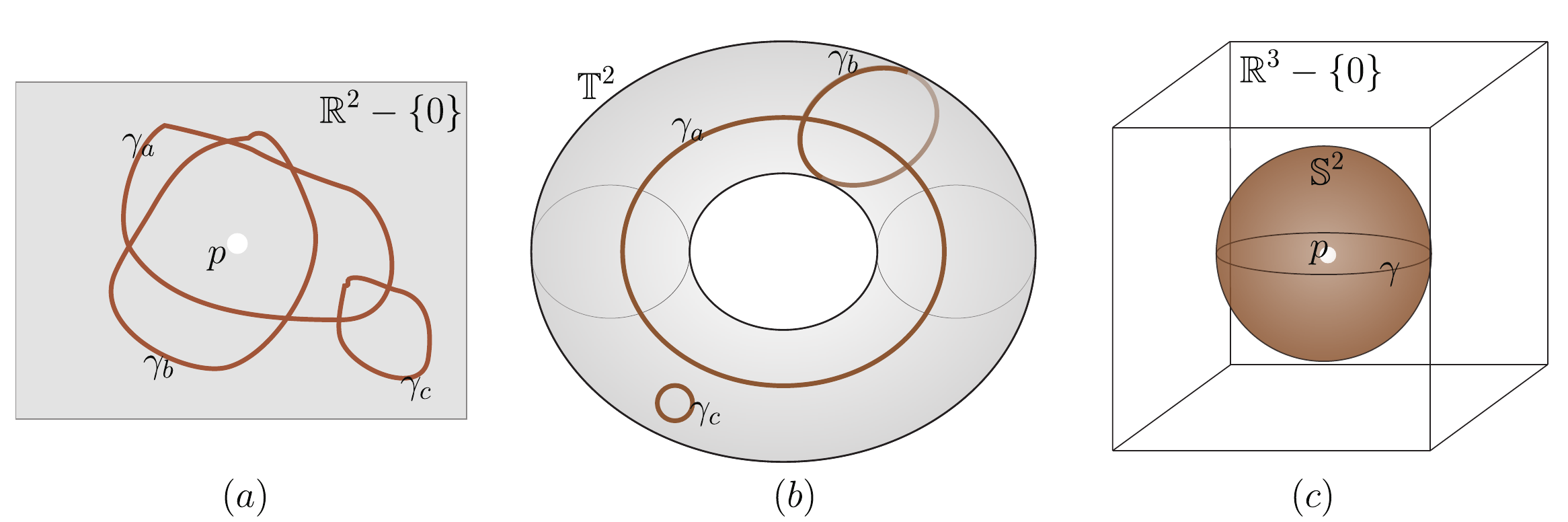}
\par\end{centering}

\caption{In $\mathbb{R}^{2}-\left\{ 0\right\} $, with $p$ as the origin (a),
and in a 2-dimensional torus $\mathbb{T}^{2}$ (b), $\gamma_{a}$
and $\gamma_{b}$ are non-contractible, $\gamma_{c}$ is contractible.
(c) In $\mathbb{R}^{3}-\left\{ 0\right\} $, a 2-sphere $\mathbb{S}^{2}$
centered at the origin is non-contractible, while any loop $\gamma$
is contractible.}
\end{figure}

Intuitively, the existence of a 'hole' contributes to the non-simply
connectedness of the manifold. In the context of Regge calculus, the
hinge, a $p$-form where the curvature (in the form of deficit angle)
is concentrated, acts as a $p$-dimensional 'hole'. To be precise,
in 2-dimensional dicrete geometry, the 'hole' is a point, in 3-dimension,
the 'hole' is an edge, in 4-dimension, is a triangle, in $n$-dimension,
the $\left(n-2\right)$ 'hole', is an $\left(n-2\right)$-simplex.
The existence of hinges in discrete manifold defines non-contractible
loops. These non-contractible loops are endowed with non-trivial holonomies
related to the deficit angles on the hinges, describing the curvature
of the discrete manifold. Two different non-contractible loops encircling
the same hinges are equivalent through an equivalence class defined
earlier in the previous section. Any contractible loop is endowed
with trivial holonomy. See FIG. 6. 
\begin{figure}[H]
\begin{centering}
\includegraphics[scale=0.55]{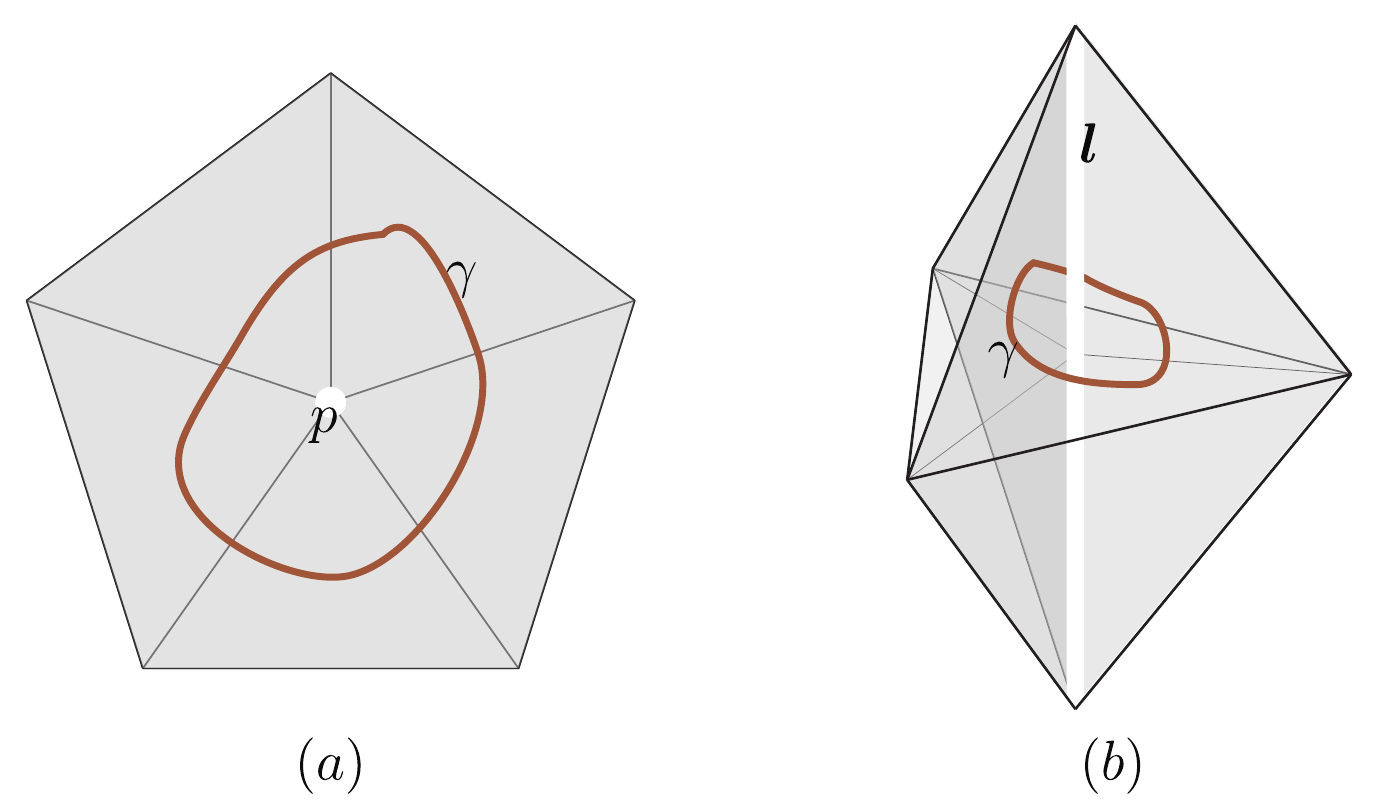}
\par\end{centering}

\caption{Loop $\gamma$ is non-contractible due to the existence of a hinge,
(a) in 2-dimensional triangulation example, and (b) in 3-dimensional
triangulation example.}
\end{figure}

\section{2+1 Regge Calculus}

Now we are ready to perform the ADM slicing on a 3-dimensional discrete
manifold. The procedure is important, in particular, as a lower dimensional
model for the (3+1) ADM slicing of 4-dimensional spacetime, which
is the first step to obtain the canonical quantization of gravity
\cite{key-40}. Works on this field are already developed, for instance,
in \cite{key-37,key-38,key-39,key-40,key-41,key-42,key-43,key-44,key-44a}.
We use the powerful tools of Regge calculus, where the simplices are
describe by coordinates-free variables, rather than vectorial elements.

\subsection{The Construction of (2+1) Lattice in First Order Formulation}

As a first step, we need to clarify and to gain insight of the geometrical
picture of the continuous first order formulation of gravity. Let
us use a local coordinate with orthonormal basis $\left(\partial_{x},\partial_{y},\partial_{z}\right)$
to characterize the 3-dimensional base manifold $M$. Let use take
point $\mathcal{O}$ as the origin. One could define the planes $dx\wedge dy$,
$dy\wedge dz$, $dz\wedge dx$ $\in\Lambda^{2}\left(T_{\mathcal{O}}M\right).$
These are the loop orientation planes, where the three infinitesimal
loops $\delta\gamma_{\mu\nu}$ are defined as the (square) boundary
of the plane $dx^{\mu}\wedge dx^{\nu},$ See FIG. 7(a).
\begin{figure}[H]
\begin{centering}
\includegraphics[scale=0.55]{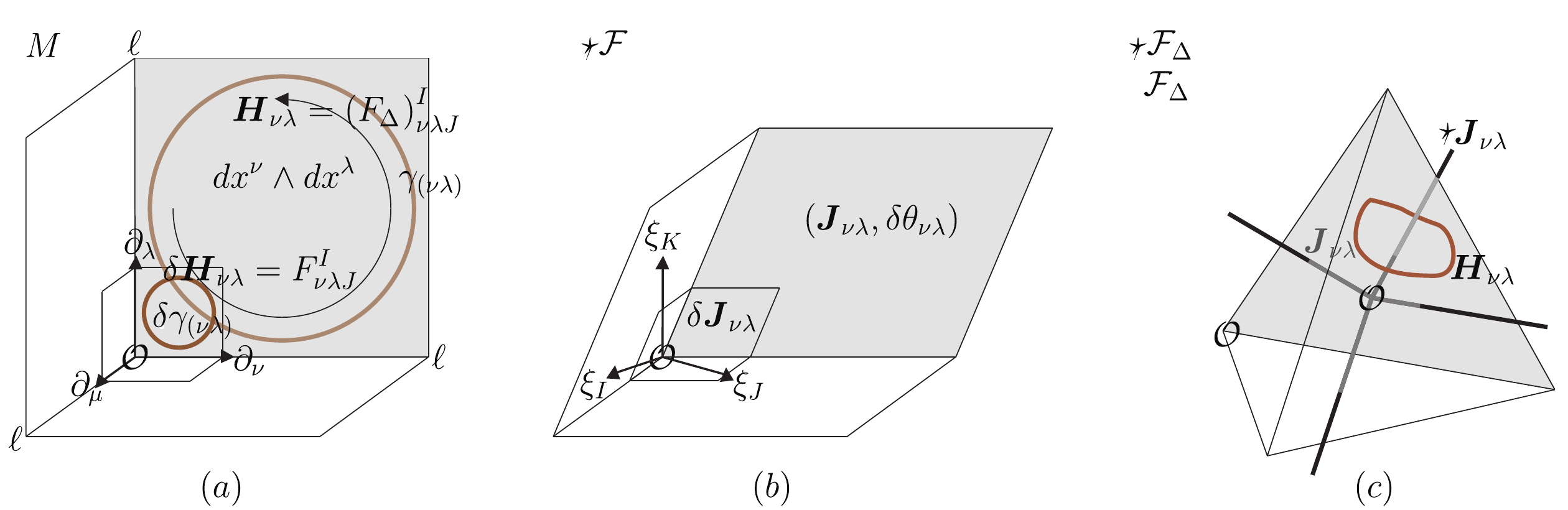}
\par\end{centering}

\caption{(a) The infinitesimal and finite loop orientation in base space $M$.
(b) The infinitesimal and finite plane of rotation in dual fibre $\star\mathcal{F}$.
(c) (b) is topologically equivalent with the tetrahedral lattice in
(c). Each triangle $\hat{\boldsymbol{J}}_{\mu\nu}$ of the dual lattice
is perpendicular to the hinge $\star\hat{\boldsymbol{J}}_{\mu\nu}$.}
\end{figure}
On these loops, the curvature 2-form components are attached: $F_{xyJ}^{I},$
$F_{yzJ}^{I},$ $F_{zxJ}^{I},$ which describe infinitesimal rotations,
namely the rotation bivectors (or plane of rotations). Let use relabel
these components as $\delta\hat{\boldsymbol{J}}_{xy}$, $\delta\hat{\boldsymbol{J}}_{yz}$,
$\delta\hat{\boldsymbol{J}}_{zx}$ $\in\Lambda^{2}\left(\star\mathcal{F}\sim\mathbb{R}^{3}\right)$.
These three rotation bivector planes, in general, are not orthogonal
to each other, see FIG. 7(b).

Now, let us clarify the geometrical picture of the first order formulation
of Regge calculus. This had been done partially in \cite{key-36}.
Let us define the \textit{finite} loop orientation planes $\ell^{2}dx^{\mu}\wedge dx^{\nu}$
$\in\Lambda^{2}\left(T_{\mathcal{O}}M\right),$ with loops $\gamma_{\mu\nu}$
as their boundaries. On these \textit{finite} loops, the \textit{finite}
curvature 2-form components are attached: $\left(F_{\Delta}\right)_{\mu\nu J}^{I},$
describing \textit{finite} rotation, which are indeed the holonomy.
Relabeling these finite rotation as holonomies $\boldsymbol{H}_{xy}$,
$\boldsymbol{H}_{yz}$, $\boldsymbol{H}_{zx}$, it is clear that they
satisfy (\ref{eq:7}), i.e., $\boldsymbol{H}_{\mu\nu}$ are exponential
map of $\hat{\boldsymbol{J}}_{\mu\nu}.$ The geometrical interpretation
of the finite version is similar with the infinitesimal ones, as compared
in FIG. 7(a)-(b).

For each loop defined in $M,$ there exist a corresponding plane of
rotation in $\Lambda^{2}\left(\star\mathcal{F}\sim\mathbb{R}^{3}\right)$.
The collection of planes of rotation $\hat{\boldsymbol{J}}_{\mu\nu}$
defined an (abstract) dual-lattice $\star\mathcal{F},$ see FIG. 7(c).
The corresponding primal lattice of $\star\mathcal{F},$ is (a simplicial
complex) $\mathcal{F}$, where dual of the planes of rotations $\hat{\boldsymbol{J}}_{\mu\nu}$
are the hinges $\star\hat{\boldsymbol{J}}_{\mu\nu},$ see FIG. 7(c).
For discrete geometry, it is convenient to drop the base manifold
picture and focus only on the fibre $\mathcal{F}_{\triangle}$. This
includes the 'moving' of holonomy $\boldsymbol{H}_{\mu\nu}$ (which
are located on $M)$ to $\star\mathcal{F}_{\triangle}$, circling
hinge $\star\hat{\boldsymbol{J}}_{\mu\nu},$ as in FIG. 7(c). For
the next section, we will only focus on the fibre lattice $\mathcal{F}_{\triangle}$
and its dual $\star\mathcal{F}_{\triangle}.$

\subsection{Terminologies of a 4-1 Pachner moves}

As already been explained in the previous sections, to describe completely
a curvature of a 3-dimensional space, one needs three hinges. On each
hinge, which in 3-dimension is a segment, a standard loop is defined
as the boundary of the faces in the $\star\mathcal{F}_{\triangle}$,
and the holonomy related to the curvature on the hinge is attached
on the loop. These three distinct holonomies are the finite version
of the three matrices elements of curvature 2-form in 3-dimension.
If in the previous section we label the curvature tensor components
by the infinitesimal loop orientation planes: $\boldsymbol{F}\left(\partial_{\mu},\partial_{\nu}\right),$
$\boldsymbol{F}\left(\partial_{\nu},\partial_{\lambda}\right),$ $\boldsymbol{F}\left(\partial_{\lambda},\partial_{\nu}\right),$
now we use the hinges to label the finite versions, say $\boldsymbol{H}_{i},$
with $\boldsymbol{l}_{i}$ describing  the finite hinge $i$. Therefore,
the simplest dicretization in three dimension which yield a complete
curvature is the discretization by the 4-1 Pachner move, see FIG.
8(a).
\begin{figure}[H]
\begin{centering}
\includegraphics[scale=0.55]{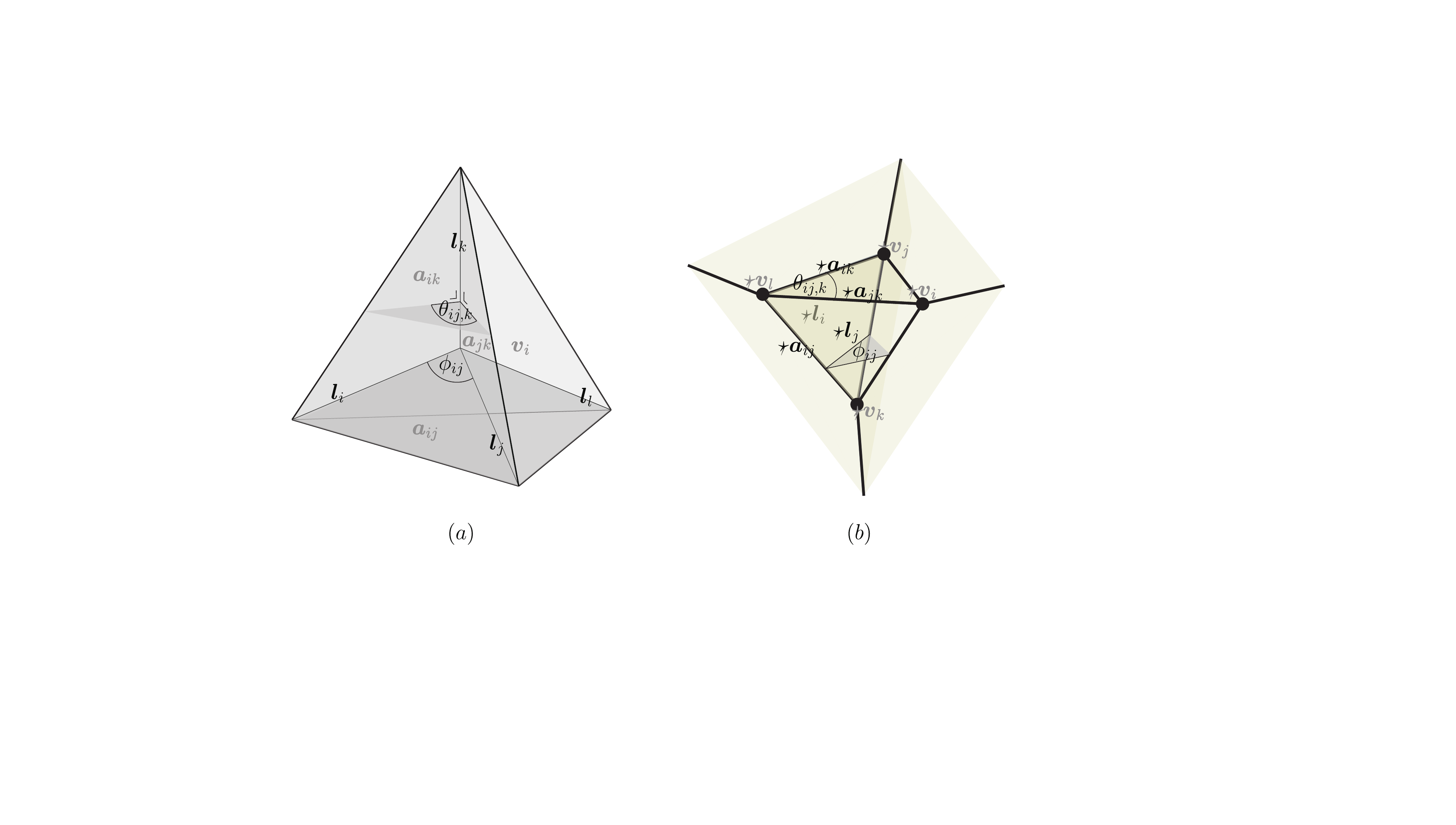}
\par\end{centering}

\caption{4-1 Pachner move (a) in primal lattice, and (b) in abstract-dual lattice.}
\end{figure}
The holonomies in 3-dimension are elements of rotation group $SO\left(3\right),$
but for our work, we use its complex counterpart, which is also its
double-cover, the group $SU\left(2\right)$. The reason for this,
is because the formulations can be written more compactly using the
$SU\left(2\right)$ group.

The 4-1 Pachner move is the boundary of a 4-simplex, where four tetrahedra
meet each other on their triangles. A 4-simplex, and similarly, its
boundary, can completely and uniquely be described by the length of
its ten segments \cite{key-59,key-66}. These variables are coordinate
free, i.e., they are not vectorial. Another different set of a complete
coordinate-free variables containing equivalent informations of the
move are the length of four internal segments and six internal 2D
angles \cite{key-66}; this will be our starting point. We define
the terminologies of 4-1 Pachner moves as follows. Each one of the
four internal segments of the move are 3-dimensional hinge. We label
them with $\boldsymbol{l}_{i}$, with $i=1,..,4$. The six remaining
variables are described by the six angles $\phi_{ij}$ between segment
$\boldsymbol{l}_{i}$ and $\boldsymbol{l}_{j}$, located at the center
point, see FIG. 8(a). These are 2-dimensional angles. Furthermore,
a triangle $\boldsymbol{a}_{ij}$ is the plane between segment $\boldsymbol{l}_{i}$
and $\boldsymbol{l}_{j}$. On each segment $\boldsymbol{l}_{k}$,
three 3-dimensional (dihedral) angles $\theta_{ij,k}$ are located,
which are the angles between plane $\boldsymbol{a}_{ik}$ and $\boldsymbol{a}_{jk}$.
The last geometric figures are tetrahedra $\boldsymbol{v}_{i},$ constructed
from the three segments $\boldsymbol{l}_{i}$, $\boldsymbol{l}_{j}$,
and $\boldsymbol{l}_{k}$. See FIG. 8(a). The abstract dual lattice
is described in FIG. 8(b). Vertices $\star\boldsymbol{v}_{i}$ are
dual to primal tetrahedra, edges $\star\boldsymbol{a}_{ij}$ are dual
to primal triangle, and faces $\star\boldsymbol{l}_{i}$ are dual
to primal segments.

The measure of the geometric quantities, such as length, area, and
volume, for the moment, is not included in our work, since we are
only interested in the cuvatures, which only needs the information
of the angles. But for further works including the dynamics of the
theory, it is possible to provide our construction with a geometric
measure, i.e, attaching 'norms' on the lattices by a well-defined
procedure; in particular, the hybrid cells introduced in \cite{key-46,key-49,key-54}.

\subsection{Curvatures, Closure Constraint, and Bianchi Identity}

A 3-dimensional holonomy of connection $\boldsymbol{A}\in\mathfrak{su\left(2\right)}$
along curve $\gamma$ with origin $\mathcal{O}$ is written as:
\[
\boldsymbol{H}_{\gamma}=\boldsymbol{H}_{\gamma}\left(\boldsymbol{A},\gamma_{\mathcal{O}}\right).
\]
We will simplify the notation as long as the meaning it describe is
clear and non-ambiguous.

As a first step, let us define the 3-dimensional holonomy on edge
$\star\boldsymbol{a}_{ij}$ between vertex $\star\boldsymbol{v}_{i}$
and $\star\boldsymbol{v}_{j}$ as $\boldsymbol{H}_{ij}$, see FIG.
9(a). 
\begin{figure}[H]
\begin{centering}
\includegraphics[scale=0.55]{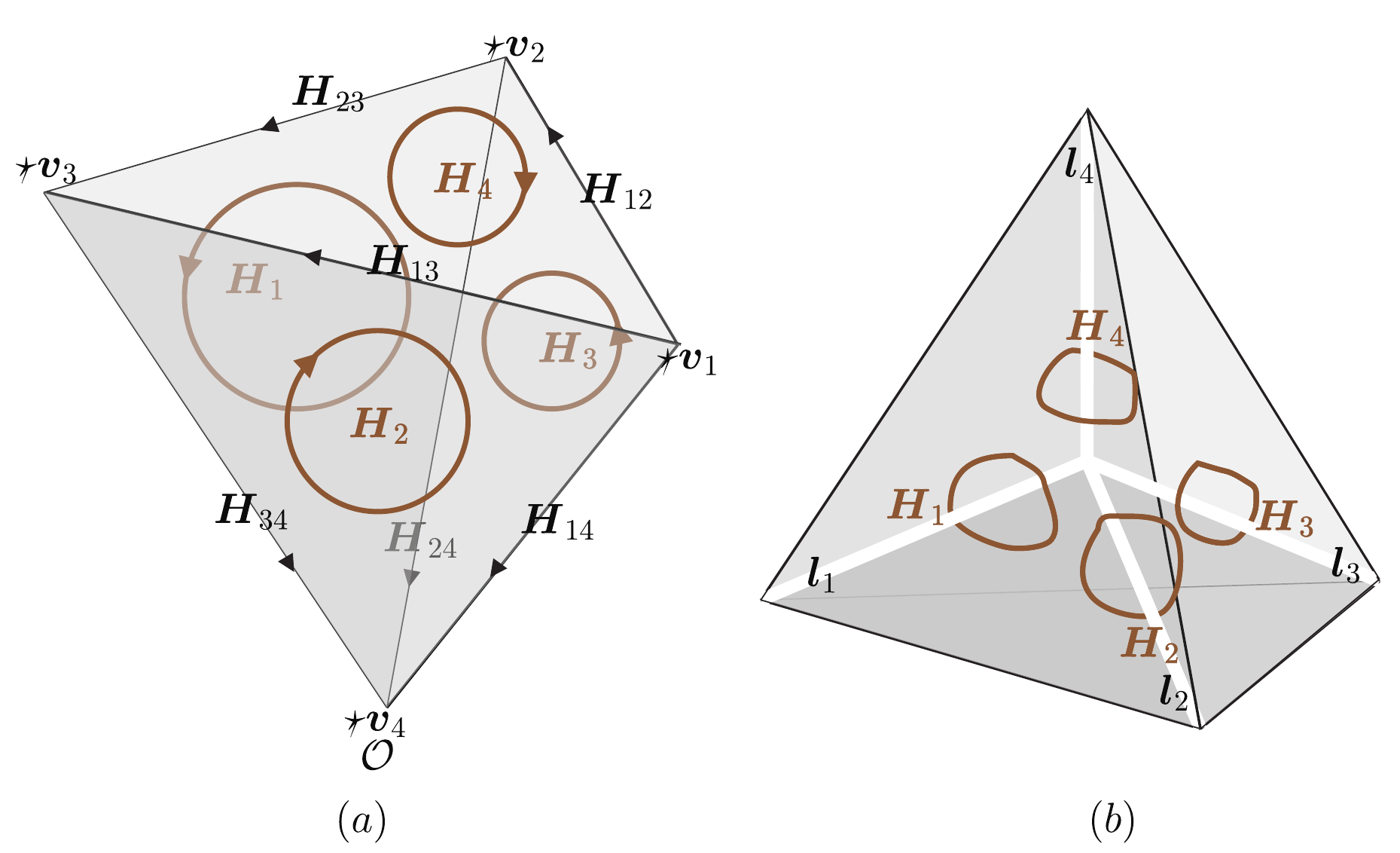}
\par\end{centering}

\caption{The generalized closure constraint in (a) the dual lattice, topologically
equivalent to a tetrahedron, and in (b) the primal lattice. }
\end{figure}
Notice that $\boldsymbol{H}_{ij}$ is attached on an \textit{open}
curve with the origin $\star\boldsymbol{v}_{i}$ towards $\star\boldsymbol{v}_{j}$,
so that it does not satisfies (\ref{eq:9}). The inverse is:
\[
\boldsymbol{H}_{ji}=\boldsymbol{H}_{ij}^{-1},
\]
with origin $\star\boldsymbol{v}_{j}$ towards $\star\boldsymbol{v}_{i}$.
The next step is to define the 3-dimensional holonomy on a \textit{closed}
loop, where the loop is the boundary of the faces $\star\boldsymbol{l}_{i}$:
a standard loop, as follows:
\begin{equation}
\boldsymbol{H}_{i}\left(\boldsymbol{A},\gamma_{i,\star\boldsymbol{v}_{j}}\right)=\boldsymbol{H}_{jk}\boldsymbol{H}_{kl}\boldsymbol{H}_{lj}.\label{eq:13}
\end{equation}
$\boldsymbol{H}_{i}$ is the holonomy around the loop $\gamma_{i}$
circling hinge $\boldsymbol{l}_{i}$ with origin $\star\boldsymbol{v}_{j}$.

\subsubsection{Generalized Closure Constraint.}

As already studied in \cite{key-67}, a curved tetrahedron satisfies
the generalized closure constraint governed by its holonomies. We
will use the result in this subsection. The 4-1 Pachner move contains
four internal hinges and therefore four planes of rotation dual to
the hinges, but only three of them are independent, such that the
following 'closure constraint' is satisfied:
\begin{equation}
\boldsymbol{H}_{1}\boldsymbol{H}_{2}\boldsymbol{H}_{3}\overline{\boldsymbol{H}}_{4}=\boldsymbol{1},\label{eq:12}
\end{equation}
This relation is taken with the vertex $\star\boldsymbol{v}_{4}$
as the origin. More precisely: 
\begin{eqnarray}
\boldsymbol{H}_{1}=\boldsymbol{H}_{1}\left(\boldsymbol{A},\gamma_{1,\star\boldsymbol{v}_{4}}\right) & = & \boldsymbol{H}_{42}\boldsymbol{H}_{23}\boldsymbol{H}_{34},\nonumber \\
\boldsymbol{H}_{2}=\boldsymbol{H}_{2}\left(\boldsymbol{A},\gamma_{2,\star\boldsymbol{v}_{4}}\right) & = & \boldsymbol{H}_{43}\boldsymbol{H}_{31}\boldsymbol{H}_{14},\label{eq:10}\\
\boldsymbol{H}_{3}=\boldsymbol{H}_{3}\left(\boldsymbol{A},\gamma_{3,\star\boldsymbol{v}_{4}}\right) & = & \boldsymbol{H}_{41}\boldsymbol{H}_{12}\boldsymbol{H}_{24},\nonumber \\
\overline{\boldsymbol{H}}_{4}=\overline{\boldsymbol{H}}_{4}\left(\boldsymbol{A},\gamma_{4,\star\boldsymbol{v}_{4}}\right) & = & \boldsymbol{H}_{42}\underset{\boldsymbol{H}_{4}\left(\boldsymbol{A},\gamma_{4,\star\boldsymbol{v}_{2}}\right)}{\underbrace{\boldsymbol{H}_{21}\boldsymbol{H}_{13}\boldsymbol{H}_{32}}}\boldsymbol{H}_{24}=\boldsymbol{H}_{123}^{-1}.\nonumber 
\end{eqnarray}
See FIG. 9(a). There is a gauge freedom in choosing path $\overline{\boldsymbol{H}}_{4}$,
which in this case, is gauge-fixed by taking the path through $\boldsymbol{H}_{42}$
from the origin. Other paths are possible,  see the explanation in
\cite{key-67}. Any tetrahedral lattice as in FIG. 9(a) will satisfy
(\ref{eq:12}). In the primal lattice point of view, the closure constraint
guarantees that a sets of four tetrahedra, connected to each other
on their internal faces, construct a closed, (in general) curved tetrahedron.
This will be explored in more detail in Subsection V A.

\subsubsection{3D Discrete Intrinsic Curvature.}

As explained earlier, one can choose three combinations of distinct
holonomies $\boldsymbol{H}_{i},\boldsymbol{H}_{j},\boldsymbol{H}_{k}$
from the four in (\ref{eq:10}) as the finite version of curvature
2-form. They contain the information of the 3-dimensional discrete
curvature as well as the curvature 2-form $\,^{3}\boldsymbol{F}_{IJ}$
contains for the continuous space. We label the components of discrete
intrinsic 3D curvature as the following three tuples of holonomies:
\begin{equation}
\left(\,^{3}\boldsymbol{F}_{\triangle}\right)_{\mu\nu}=\boldsymbol{H}_{\mu\nu}\left(\boldsymbol{A},\gamma_{\mu\nu,\mathcal{O}}\right)=\left(\boldsymbol{H}_{1}\left(\boldsymbol{A},\gamma_{1,\mathcal{O}}\right),\boldsymbol{H}_{2}\left(\boldsymbol{A},\gamma_{2,\mathcal{O}}\right),\boldsymbol{H}_{3}\left(\boldsymbol{A},\gamma_{3,\mathcal{O}}\right)\right).\label{eq:11}
\end{equation}
Furthermore, we will drop the indices $\mu,\nu$ and write the components
(\ref{eq:11}) as $\,^{3}\boldsymbol{F}_{\triangle}$ for simplicity.
The corresponding plane-angle representation can be obtained from
the trace and differential map of (\ref{eq:11}):
\[
\,^{3}\boldsymbol{F}_{\triangle}=\left\{ \left(\delta\theta_{1},\left.\hat{\boldsymbol{J}}_{1}\right|_{\mathcal{O}}\right),\left(\delta\theta_{2},\left.\hat{\boldsymbol{J}}_{2}\right|_{\mathcal{O}}\right),\left(\delta\theta_{3},\left.\hat{\boldsymbol{J}}_{3}\right|_{\mathcal{O}}\right)\right\} ,
\]
with the deficit angle on hinge $\boldsymbol{l}_{i}$ satisying (\ref{eq:trace})
and:
\begin{equation}
\delta\theta_{i}=2\pi-\left(\sum_{j,k}\theta_{jk,i}\right),\qquad j,k\neq i,\, j<k.\label{eq:16}
\end{equation}
and the rotation bivector with origin $\mathcal{O}$ satisfying:
\begin{equation}
\left.\hat{\boldsymbol{J}}_{i}\right|_{\mathcal{O}}=\left.\frac{d\boldsymbol{H}_{i}}{d\delta\theta_{i}}\right|_{\delta\theta_{i}=0}.\label{eq:penting}
\end{equation}
See FIG. 10. 
\begin{figure}[H]
\begin{centering}
\includegraphics[scale=0.55]{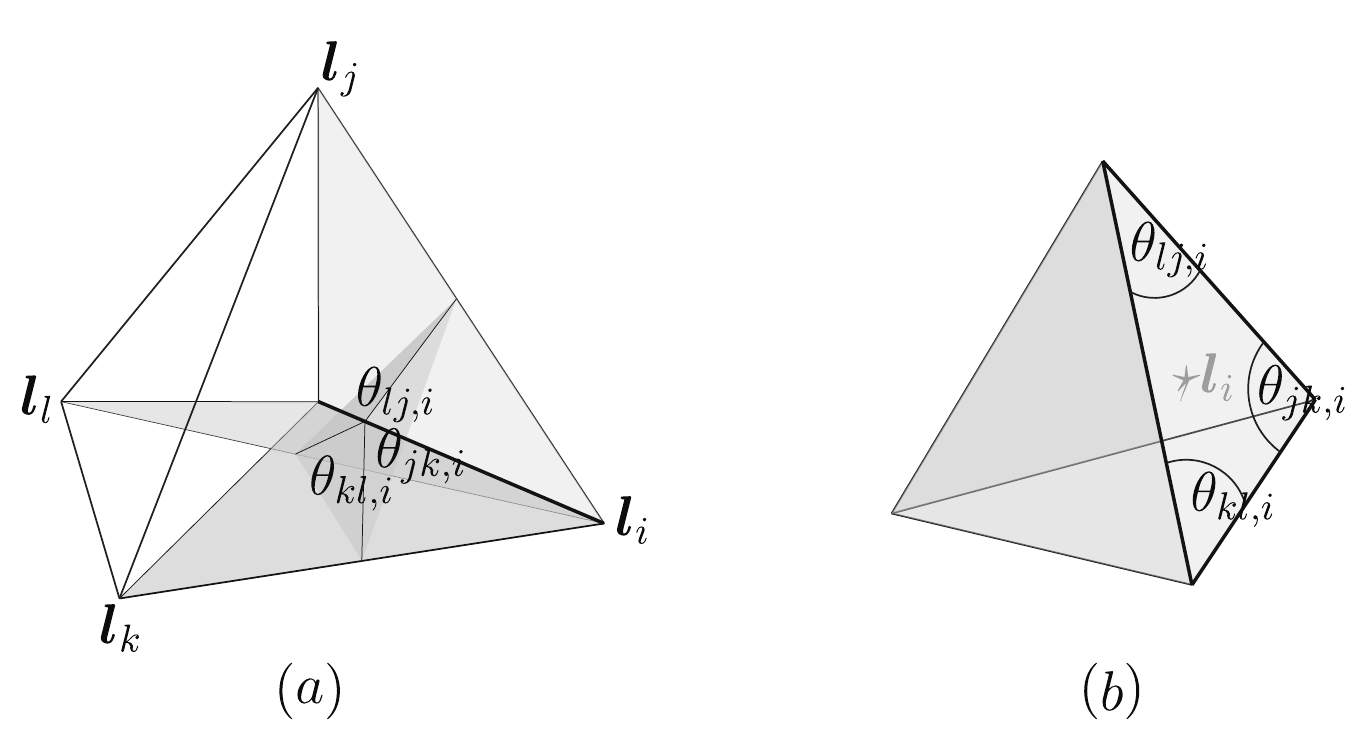}
\par\end{centering}

\caption{The 3-dimensional discrete intrinsic curvature on hinge in (a) primal
lattice, and (b) dual lattice.}
\end{figure}

\subsubsection{Bianchi Identity.}

Before arriving at the discrete version of Bianchi identity, we need
to proof an important relation. As explained earlier, any tetrahedral
lattice as in FIG. 9(a) always satisfy the generalized closure constraint
(\ref{eq:12}). By a straightforward calculation, (\ref{eq:12}) can
be written as:
\[
\boldsymbol{H}_{1}\boldsymbol{H}_{2}\boldsymbol{H}_{3}\overline{\boldsymbol{H}}_{4}=\underset{\boldsymbol{h}_{1}}{\underbrace{\boldsymbol{H}_{42}\boldsymbol{H}_{23}}}\underset{\boldsymbol{h}_{2}}{\underbrace{\boldsymbol{H}_{31}\boldsymbol{H}_{14}}}\underset{\boldsymbol{h}_{3}}{\underbrace{\boldsymbol{H}_{41}\boldsymbol{H}_{12}}}\underset{\boldsymbol{h}_{4}}{\underbrace{\boldsymbol{H}_{21}\boldsymbol{H}_{13}}}\underset{\boldsymbol{h}_{1}^{-1}}{\underbrace{\boldsymbol{H}_{32}\boldsymbol{H}_{24}}.}
\]
This immediately gives:
\[
\boldsymbol{h}_{1}\boldsymbol{h}_{2}\boldsymbol{h}_{3}\boldsymbol{h}_{4}\boldsymbol{h}_{1}^{-1}=\boldsymbol{1},
\]
where the two adjacents holonomies $\boldsymbol{H}_{jk}$ and $\boldsymbol{H}_{kl}$
are collected together as $\boldsymbol{h}_{i}$. In general, for every
point $\mathcal{O}$ in the lattice as the origin, the following relation,
which we called as \textit{'trivalent condition}', is valid:
\begin{equation}
\boldsymbol{h}_{i}\boldsymbol{h}_{j}\boldsymbol{h}_{k}=\boldsymbol{1},\quad i=1,..4,\: i\neq j\neq k.\label{eq:tri}
\end{equation}
Relation (\ref{eq:tri}) could be illustrated by the combinatorics
graph in FIG. 11(a).
\begin{figure}[H]
\centering{}\includegraphics[scale=0.55]{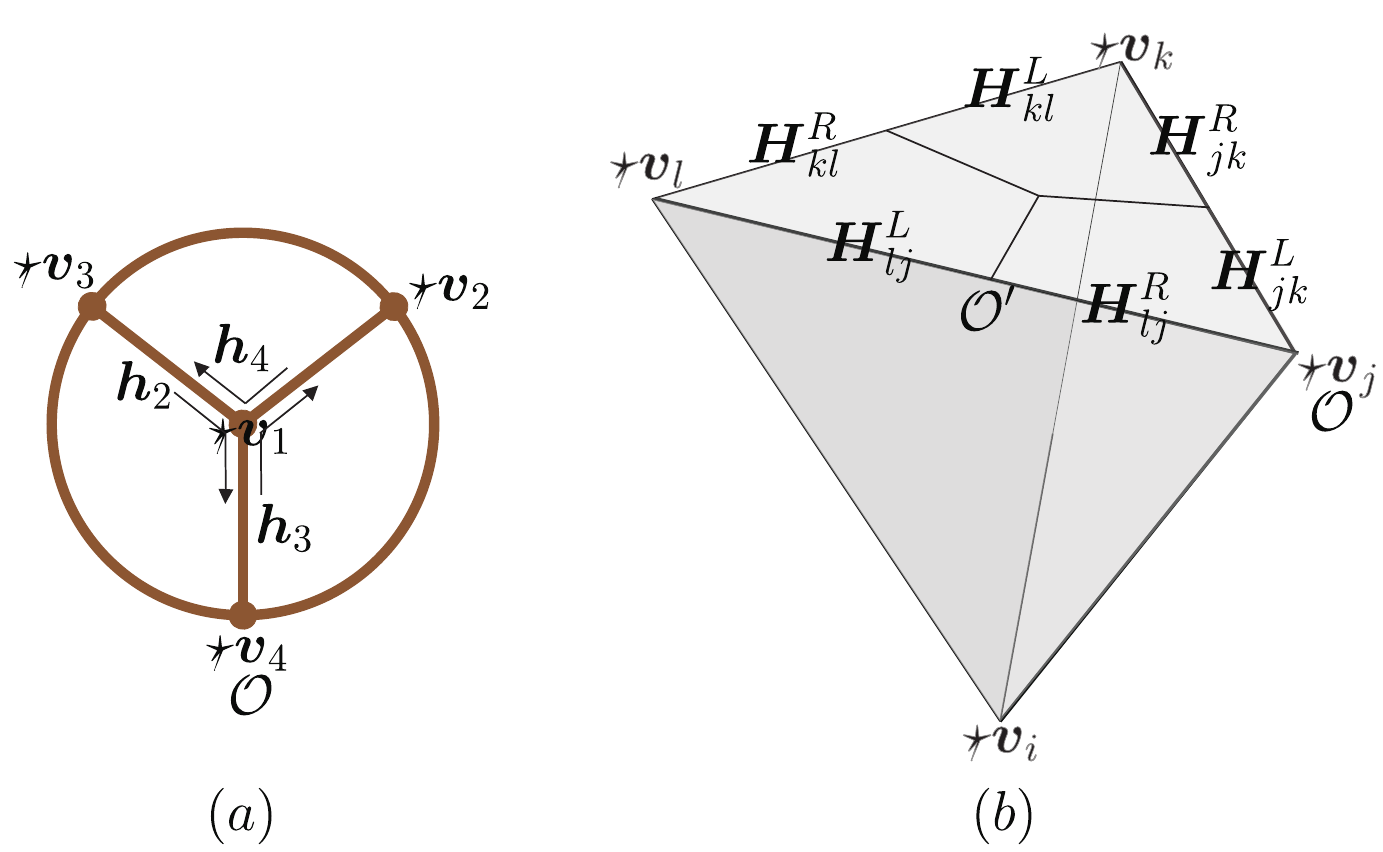}\caption{(a) An abstract tetrahedral lattice can be illustrated as above. Holonomies
on a trivalent vertex satisfies relation (\ref{eq:tri}), where in
this case, is $\boldsymbol{h}_{2}\boldsymbol{h}_{3}\boldsymbol{h}_{4}=\boldsymbol{1}$.
(b) Moving the origin from $\mathcal{O}$ to $\mathcal{O}$'.}
\end{figure}
Holonomies of any trivalent vertex satisfy relation (\ref{eq:tri}).
This relation will be important for the derivation of the Bianchi
identity in the following paragraph.

Let us split the holonomy $\boldsymbol{H}_{ij}$ as follows:
\begin{equation}
\boldsymbol{H}_{ij}=\boldsymbol{H}_{ij}^{\left(L\right)}\boldsymbol{H}_{ij}^{\left(R\right)},\label{eq:decompos}
\end{equation}
so that (\ref{eq:13}) can be rewritten as:
\[
\boldsymbol{H}_{i}=\boldsymbol{H}_{jk}^{\left(L\right)}\boldsymbol{H}_{jk}^{\left(R\right)}\boldsymbol{H}_{kl}^{\left(L\right)}\boldsymbol{H}_{kl}^{\left(R\right)}\boldsymbol{H}_{lj}^{\left(L\right)}\boldsymbol{H}_{lj}^{\left(R\right)}.
\]
Noted that these holonomies are originated at $\mathcal{O}=\star\boldsymbol{v}_{j}$.
Now we move the origin to point $\mathcal{O}'$, which is a point
on the edge between $\star\boldsymbol{v}_{j}$ and $\star\boldsymbol{v}_{l}$,
see FIG. 11(b). $\boldsymbol{H}_{i}$ is transformed into:
\begin{equation}
\boldsymbol{H}_{i}^{'}\left(\boldsymbol{A},\gamma_{i,\mathcal{O}'}\right)=\boldsymbol{H}_{lj}^{\left(R\right)}\boldsymbol{H}_{i}\left(\boldsymbol{A},\gamma_{i,\mathcal{O}}\right)\boldsymbol{H}_{lj}^{\left(R\right)-1}.\label{eq:s}
\end{equation}
Therefore, from point $\mathcal{O}'$, the holonomy circling hinge
$\boldsymbol{l}_{i}$ is:
\begin{equation}
\boldsymbol{H}_{i}^{'}=\boldsymbol{H}_{i}^{'}\left(\boldsymbol{A},\gamma_{i,\mathcal{O}'}\right)=\underset{\boldsymbol{h}_{kl,i}}{\underbrace{\boldsymbol{H}_{lj}^{\left(R\right)}\boldsymbol{H}_{jk}^{\left(L\right)}}}\underset{\boldsymbol{h}_{lj,i}}{\underbrace{\boldsymbol{H}_{jk}^{\left(R\right)}\boldsymbol{H}_{kl}^{\left(L\right)}}}\underset{\boldsymbol{h}_{jk,i}}{\underbrace{\boldsymbol{H}_{kl}^{\left(R\right)}\boldsymbol{H}_{lj}^{\left(L\right)}},}\label{eq:xbesar}
\end{equation}
splitted into three holonomies $\boldsymbol{h}_{jk,i}$ on \textit{open
curves}, see FIG. 12(a). 
\begin{figure}[H]
\begin{centering}
\includegraphics[scale=0.55]{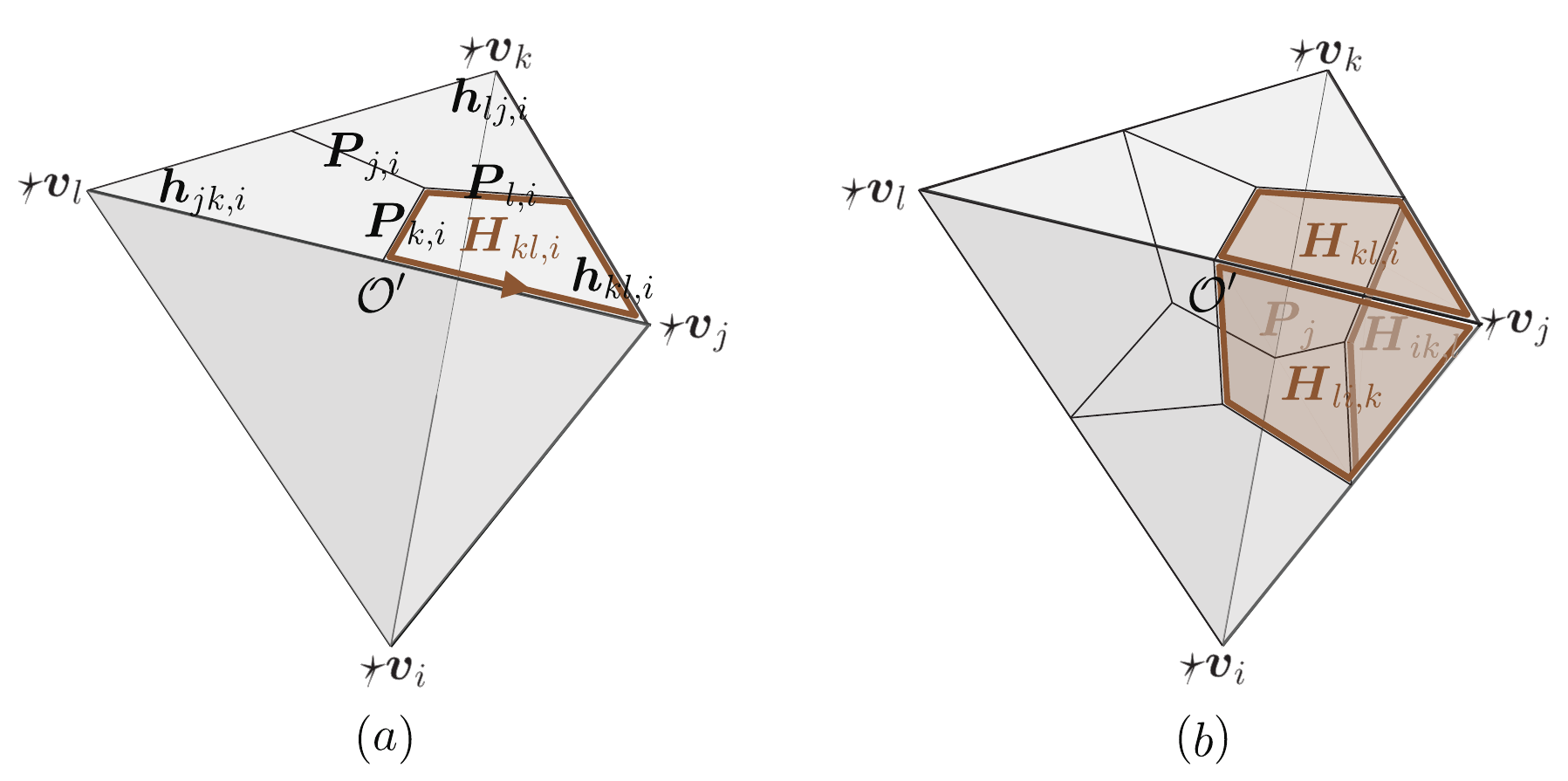}
\par\end{centering}

\caption{(a). Spliting holonomy into three holonomy on loops. (b) A tetrahedral
lattice on a vertex, satisfying the closure and trivalent condition.}
\end{figure}
The decomposition in (\ref{eq:decompos}) is chosen such that $\boldsymbol{h}_{kl,i}$,
$\boldsymbol{h}_{jl,i}$, and $\boldsymbol{h}_{jk,i}$, using the
trace (\ref{eq:trace}) and differential map (\ref{eq:penting}),
satisfy:
\begin{eqnarray}
\boldsymbol{h}_{kl,i} & = & \exp\left(\boldsymbol{H}_{lj}^{\left(R\right)}\hat{\boldsymbol{J}}_{i}\boldsymbol{H}_{lj}^{\left(R\right)-1}\theta_{kl,i}\right),\nonumber \\
\boldsymbol{h}_{lj,i} & = & \exp\left(\boldsymbol{H}_{lj}^{\left(R\right)}\hat{\boldsymbol{J}}_{i}\boldsymbol{H}_{lj}^{\left(R\right)-1}\theta_{lj,i}\right),\label{eq:nice}\\
\boldsymbol{h}_{jk,i} & = & \exp\left(\boldsymbol{H}_{lj}^{\left(R\right)}\hat{\boldsymbol{J}}_{i}\boldsymbol{H}_{lj}^{\left(R\right)-1}\theta_{jk,i}\right).\nonumber 
\end{eqnarray}
The origin of the rotation bivector plane $\hat{\boldsymbol{J}}_{i}$
is moved from $\mathcal{O}$ to $\mathcal{O}'$ using the adjoint
representation induced by (\ref{eq:s}).

The next step is to split $\boldsymbol{H}_{i}^{'}$ into three holonomies
\textit{on a loop}, with origin $\mathcal{O}'$, as follows:
\[
\boldsymbol{H}_{i}^{'}=\boldsymbol{H}_{kl,i}\boldsymbol{H}_{lj,i}\boldsymbol{H}_{jk,i},
\]
such that: 
\begin{eqnarray*}
\boldsymbol{H}_{kl,i} & = & \boldsymbol{h}_{kl,i}\boldsymbol{P}_{l,i}\boldsymbol{P}_{k,i},\\
\boldsymbol{H}_{lj,i} & = & \boldsymbol{P}_{k,i}^{-1}\boldsymbol{P}_{l,i}^{-1}\boldsymbol{h}_{lj,i}\boldsymbol{P}_{j,i}\boldsymbol{P}_{k,i},\\
\boldsymbol{H}_{jk,i} & = & \boldsymbol{P}_{k,i}^{-1}\boldsymbol{P}_{j,i}^{-1}\boldsymbol{h}_{jk,i},
\end{eqnarray*}
see FIG. 12(a). $\boldsymbol{P}_{k,i}$ are the gauge freedom which
can be fixed arbitrarily. Notice that on each vertex $\star\boldsymbol{v}$,
there exist a tetrahedral lattice defined by three holonomies on different
faces, see FIG. 12(b). The existence of the tetrahedral lattice on
each vertex is guaranteed as long as the decomposition (\ref{eq:decompos})
satisfies (\ref{eq:nice}). The tetrahedral lattice in vertex $\star\boldsymbol{v}_{j}$,
needs to satisfy the closure condition:
\[
\boldsymbol{H}_{kl,i}\overline{\boldsymbol{H}}_{ik,l}\boldsymbol{P}_{j}\boldsymbol{H}_{li,k}=1,
\]
with:
\begin{eqnarray*}
\overline{\boldsymbol{H}}_{ik,l} & = & \boldsymbol{P}_{k,i}^{-1}\boldsymbol{P}_{l,i}^{-1}\underset{\boldsymbol{h}_{ik,l}\boldsymbol{P}_{k,l}\boldsymbol{P}_{i,l}}{\underbrace{\boldsymbol{H}_{ik,l}}}\boldsymbol{P}_{l,i}\boldsymbol{P}_{k,i},\\
\boldsymbol{P}_{j} & = & \boldsymbol{P}_{k,i}^{-1}\boldsymbol{P}_{l,i}^{-1}\boldsymbol{P}_{i,l}^{-1}\boldsymbol{P}_{k,l}^{-1}\boldsymbol{P}_{l,k}\boldsymbol{P}_{i,k},\\
\boldsymbol{H}_{li,k} & = & \boldsymbol{P}_{l,k}^{-1}\boldsymbol{P}_{i,k}^{-1}\boldsymbol{h}_{li,k}.
\end{eqnarray*}
Since each tetrahedron on the 4-1 Pachner move is a flat 3-simplex,
we have:
\begin{equation}
\boldsymbol{P}_{j}=1,\label{eq:triv}
\end{equation}
this will be clear in Subsection V A. Moreover, the holonomies meeting
on vertex $\star\boldsymbol{v}_{j}$ also needs to satisfies the trivalent
condition:
\begin{equation}
\boldsymbol{h}_{kl,i}\boldsymbol{h}_{ik,l}\boldsymbol{h}_{li,k}=\boldsymbol{h}_{kl,i}\left(\boldsymbol{A},\gamma_{i,\mathcal{O}^{'}}\right)\boldsymbol{h}_{ik,l}\left(\boldsymbol{A},\gamma_{l,\mathcal{O}^{'}}\right)\boldsymbol{h}_{li,k}\left(\boldsymbol{A},\gamma_{k,\mathcal{O}^{'}}\right)=1,\label{eq:Bi}
\end{equation}
which is valid for every point on the lattice. We will show in Subsection
V A that (\ref{eq:Bi}) is indeed the discrete version of Bianchi
identity. This is consistent with a more general version of discrete
Bianchi identity defined by the product of $n$ holonomies in \cite{key-51}.
The Bianchi identity is satisfied universally in any dimension, and
in Subsection V A, we will show that in the discrete picture, it is
related to the spherical law of cosine and the dihedral angle relation
on a simplex.

\subsubsection{2D Intrinsic Curvature.}

Inside a 4-1 Pachner move, there exists four natural slicings of the
2-dimensional submanifold, see FIG. 13(a). These 2-dimensional surfaces
consist three triangles. 
\begin{figure}[H]
\centering{}\includegraphics[scale=0.55]{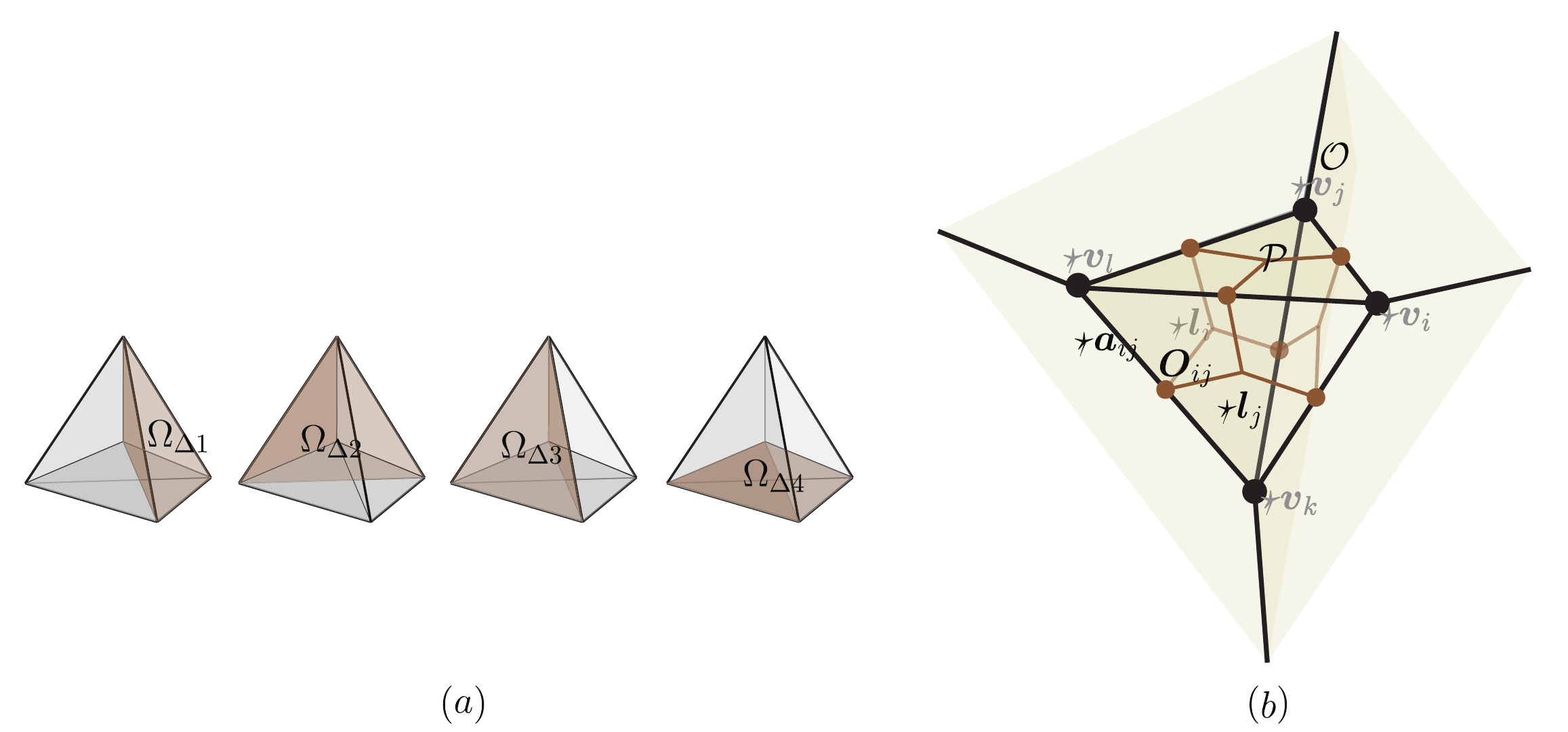} \caption{(a). 2-dimensional possible slicings of the 4-1 Pachner moves. (b).
The holonomy on each loops of the possible 2-dimensional slices.}
\end{figure}
 The 2-dimensional intrinsic curvature on each of these possible surfaces
contains a single rotation matrix and a plane, and therefore, contains
a single loop. To label the holonomy, we use similar terminologies
with the 3D version, but in one dimension lower: primal triangles
are dual to vertices, primal segments are dual to edges, both these
edges and vertices are called, respectively, as nodes and links, to
distinguish them from the edges and links of the 3D dual lattice. 

A 2-dimensional holonomy of connection $\boldsymbol{a}\in\mathfrak{so\left(2\right)}$
along curve $\lambda$ with origin $\mathcal{P}$ is written as:
\[
\boldsymbol{O}_{\lambda}=\boldsymbol{O}_{\lambda}\left(\boldsymbol{a},\lambda_{\mathcal{P}}\right).
\]
Let us define the 2-dimensional holonomy on a link crossing edge $\star\boldsymbol{a}_{ij}$
(embedded on the half of face $\star\boldsymbol{l}_{i}$ and $\star\boldsymbol{l}_{j}$,
in the direction from $\star\boldsymbol{l}_{i}$ to $\star\boldsymbol{l}_{j}$)
as $\boldsymbol{O}_{ij}\in\rho_{3}\left(SO(2)\right),$ an element
of representation of $SO\left(2\right)$ in three dimension, see FIG.
13(b). The holonomy around a loop, circling the 2D hinge which is
the center point $p$, is defined as: 
\[
\boldsymbol{O}_{i}=\boldsymbol{O}_{jk}\boldsymbol{O}_{kl}\boldsymbol{O}_{lj}.
\]
Each $\boldsymbol{O}_{i}$ represents loop on different slice $\Omega_{i}.$
The four loops are connected to each other, and similar with (\ref{eq:12}),
they also satisfy the closure constraint, where $\star\boldsymbol{l}_{4}$
is chosen to be the origin $\mathcal{P}$:
\begin{equation}
\boldsymbol{O}_{1}\boldsymbol{O}_{2}\boldsymbol{O}_{3}\overline{\boldsymbol{O}}_{4}=1,\label{eq:xx}
\end{equation}
with: 
\begin{eqnarray}
\boldsymbol{O}_{1}=\boldsymbol{O}_{1}\left(\boldsymbol{a},\lambda_{1,\star\boldsymbol{l}_{4}}\right) & = & \boldsymbol{O}_{42}\boldsymbol{O}_{23}\boldsymbol{O}_{34},\nonumber \\
\boldsymbol{O}_{2}=\boldsymbol{O}_{2}\left(\boldsymbol{a},\lambda_{2,\star\boldsymbol{l}_{4}}\right) & = & \boldsymbol{O}_{43}\boldsymbol{O}_{31}\boldsymbol{O}_{14},\nonumber \\
\boldsymbol{O}_{3}=\boldsymbol{O}_{3}\left(\boldsymbol{a},\lambda_{3,\star\boldsymbol{l}_{4}}\right) & = & \boldsymbol{O}_{41}\boldsymbol{O}_{12}\boldsymbol{O}_{24},\label{eq:14}\\
\overline{\boldsymbol{O}}_{4}=\overline{\boldsymbol{O}}_{4}\left(\boldsymbol{a},\lambda_{4,\star\boldsymbol{l}_{4}}\right) & = & \boldsymbol{O}_{42}\underset{\boldsymbol{O}_{4}\left(\boldsymbol{a},\lambda_{4,\star\boldsymbol{l}_{2}}\right)}{\underbrace{\boldsymbol{O}_{21}\boldsymbol{O}_{13}\boldsymbol{O}_{32}}}\boldsymbol{O}_{24}=\boldsymbol{O}_{123}^{-1}.\nonumber 
\end{eqnarray}
See FIG. 13(b). 

Let us choose a specific slicing, orthogonal to the hinge $\boldsymbol{l}_{4}$,
which is labeled as $\Omega_{4}$. The 2-dimensional intrinsic curvature
of $\Omega_{4}$ has a single components, written as follows: 
\begin{equation}
\,^{2}\boldsymbol{F}_{\triangle}=\overline{\boldsymbol{O}}_{4}=\overline{\boldsymbol{O}}_{4}\left(\boldsymbol{a},\lambda_{4,\mathcal{P}}\right).\label{eq:15}
\end{equation}
The corresponding plane-angle representation is:
\begin{equation}
\,^{2}\boldsymbol{F}_{\Delta}=\left(\delta\phi_{4},\left.\boldsymbol{j}_{4}\right|_{\mathcal{P}}\right)=\left(\delta\phi_{4},\boldsymbol{O}_{42}\left.\frac{d\boldsymbol{O}_{4}}{d\delta\phi_{4}}\right|_{\delta\phi_{4}=0}\boldsymbol{O}_{24}\right),\label{eq:18}
\end{equation}
with the deficit angle on hinge $p$ satisying (\ref{eq:trace}) and:
\[
\delta\phi_{4}=2\pi-\left(\sum_{j,k}\phi_{jk}\right),\qquad j,k\neq4,\, j<k.
\]
and the rotation bivector with origin $\mathcal{P}$ as:
\[
\left.\boldsymbol{j}_{4}\right|_{\mathcal{P}}=\left.\frac{d\overline{\boldsymbol{O}}_{4}}{d\delta\phi_{4}}\right|_{\delta\phi_{4}=0}.
\]

As for the trivalent condition, we split holonomy $\boldsymbol{O}_{ij}$
in a similar way with the 3D holonomy as follows:
\[
\boldsymbol{O}_{ij}=\boldsymbol{O}_{ij}^{\left(L\right)}\boldsymbol{O}_{ij}^{\left(R\right)},
\]
so that (\ref{eq:xx}) can be rewritten as:
\[
\boldsymbol{O}_{i}=\boldsymbol{O}_{jk}^{\left(L\right)}\boldsymbol{O}_{jk}^{\left(R\right)}\boldsymbol{O}_{kl}^{\left(L\right)}\boldsymbol{O}_{kl}^{\left(R\right)}\boldsymbol{O}_{lj}^{\left(L\right)}\boldsymbol{O}_{lj}^{\left(R\right)}.
\]
The next step is to move the origin from point $\mathcal{P}$ to point
$\mathcal{O}'$, which is, as explained earlier, a point between $\star\boldsymbol{v}_{j}$
and $\star\boldsymbol{v}_{l}$, see FIG. 14(a). $\boldsymbol{O}_{i}$
is transformed into:
\[
\boldsymbol{O}_{i}^{'}=\boldsymbol{O}_{lj}^{\left(R\right)}\boldsymbol{O}_{i}\boldsymbol{O}_{lj}^{\left(R\right)-1}
\]
Therefore, in a similar way with the 3D holonomy, the holonomy circling
hinge $p$ according to $\mathcal{O}'$ is:
\begin{equation}
\boldsymbol{O}_{i}^{'}=\underset{\boldsymbol{o}_{kl,i}}{\underbrace{\boldsymbol{O}_{lj}^{\left(R\right)}\boldsymbol{O}_{jk}^{\left(L\right)}}}\underset{\boldsymbol{o}_{jl,i}}{\underbrace{\boldsymbol{O}_{jk}^{\left(R\right)}\boldsymbol{O}_{kl}^{\left(L\right)}}}\underset{\boldsymbol{o}_{jk,i}}{\underbrace{\boldsymbol{O}_{kl}^{\left(R\right)}\boldsymbol{O}_{lj}^{\left(L\right)}},}\label{eq:yay}
\end{equation}
splitted into three holonomies $\boldsymbol{o}_{jk,i}$ on open curves,
see FIG. 14(b).
\begin{figure}[H]
\begin{centering}
\includegraphics[scale=0.55]{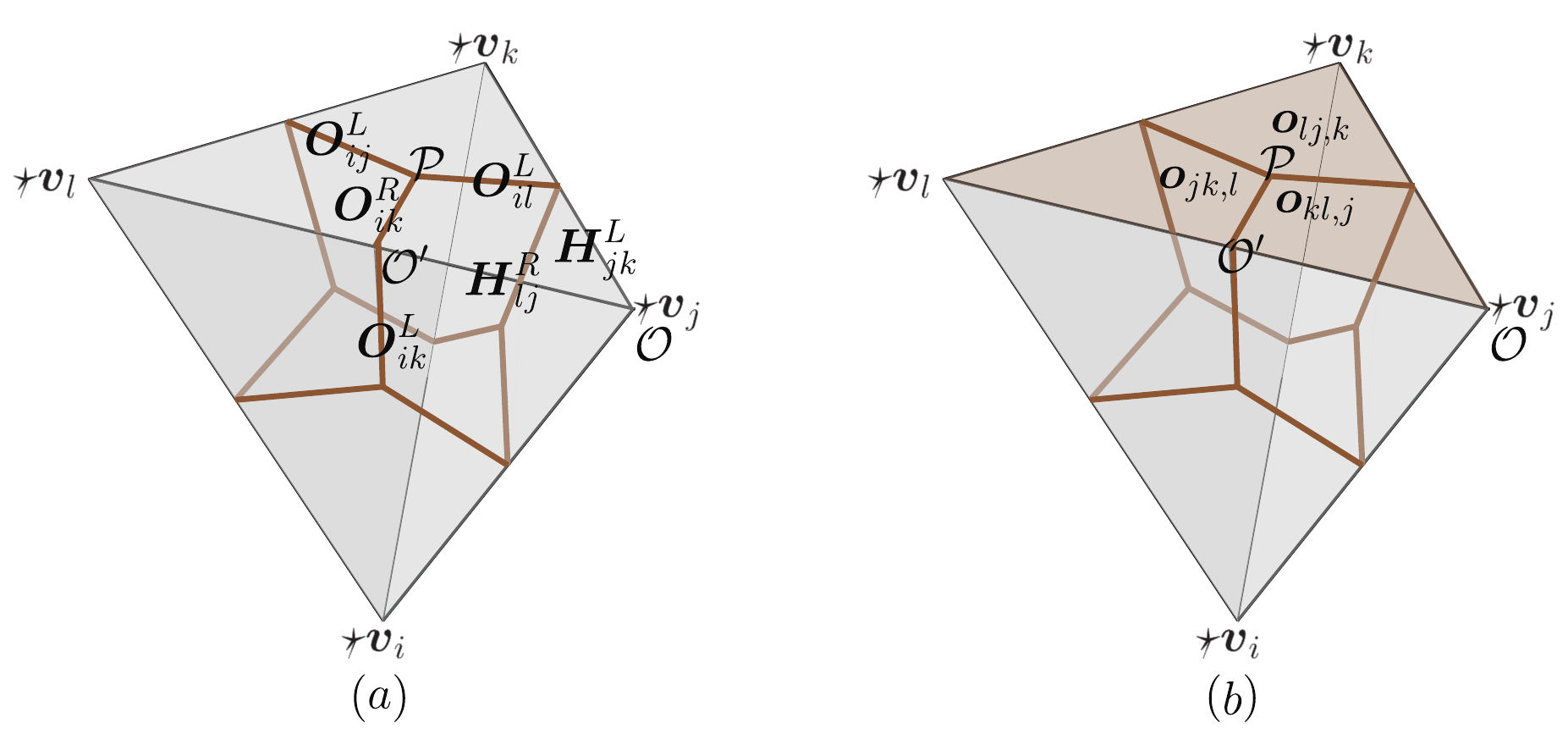}
\par\end{centering}

\caption{(a). Transformation from $\mathcal{O}$ to $\mathcal{O}'$ to $\mathcal{P}$
. (b) Trivalent condition for 2D holonomies.}
\end{figure}
As a representation of $SU\left(2\right)$ in 3-dimension, the 2D
holonomies satisfy the trivalent condition:
\begin{equation}
\boldsymbol{o}_{kl,i}\boldsymbol{o}_{ik,l}\boldsymbol{o}_{li,k}=\boldsymbol{o}_{kl,i}\left(\boldsymbol{a},\gamma_{i,\mathcal{O}^{'}}\right)\boldsymbol{o}_{ik,l}\left(\boldsymbol{a},\gamma_{l,\mathcal{O}^{'}}\right)\boldsymbol{o}_{li,k}\left(\boldsymbol{a},\gamma_{k,\mathcal{O}^{'}}\right)=1,\label{eq:r}
\end{equation}
see FIG. 14(b).

\subsubsection{Extrinsic Curvature.}

The definition of extrinsic curvature in discrete geometry is not
entirely clear \cite{key-29,key-44a}. Attempts had been done to give
it a well-defined definition, in particular, \cite{key-48}, and more
recently, \cite{key-49,key-50}. Nevertheless, we choose a different
approach for the definition of extrinsic curvature as follows. 

The extrinsic curvature (\ref{eq:zoz}) of a given slice $\Omega$,
in a general gauge condition, is defined as follows:
\begin{equation}
\boldsymbol{K}\left(\boldsymbol{s},\boldsymbol{V}\right)=g\left(D_{\boldsymbol{s}}\tilde{\boldsymbol{n}},\boldsymbol{V}\right),\label{eq:exgen}
\end{equation}
with $\boldsymbol{s}\in T_{p}M$, $\tilde{\boldsymbol{n}},\boldsymbol{V}\in E_{p}$.
$\tilde{\boldsymbol{n}}$, which is a section of a bundle, is a vector
normal to the fibre hypersurface $\Omega\subset\mathcal{F}$, moreover,
$\boldsymbol{K}\left(\boldsymbol{s},\boldsymbol{V}\right)$ can be
geometrically interpreted as the change of normal $\tilde{\boldsymbol{n}}$
in the direction of $\boldsymbol{s}$. One could construct the Lie
derivative of the extrinsic curvature, which is an element of $\mathfrak{so\left(n\right)}$
by Theorem I:
\begin{eqnarray*}
\boldsymbol{k} & = & \left[\boldsymbol{K},\boldsymbol{K}\right]=\left(K_{\mu}^{I}K_{\nu}^{J}-K_{\nu}^{I}K_{\mu}^{J}\right)\:\xi_{I}\wedge\xi_{J}\otimes dx^{\mu}\wedge dx^{\nu},\\
K_{\mu}^{I} & = & \boldsymbol{K}\left(\partial_{\mu},\xi^{I}\right)=g\left(D_{\mu}\tilde{\boldsymbol{n}},\xi^{I}\right).
\end{eqnarray*}
$\boldsymbol{k}$ is an infinitesimal rotation on the boundary of
plane $a=\ell^{2}dx^{\mu}\wedge dx^{\nu}$. Therefore, we could define
the holonomy of extrinsic curvature $\boldsymbol{k}$ along the loop,
by (\ref{eq:5}) as follows:
\begin{equation}
\mathcal{\boldsymbol{K}}=\hat{P}\exp\int_{a}\boldsymbol{k},\label{eq:huumm}
\end{equation}
but it is not clear if $\boldsymbol{k}$ comes from a connection,
i.e., if $\boldsymbol{k}$ is a differential of a 1-form $\boldsymbol{a}_{k}$
such that $d_{D}\boldsymbol{a}_{k}=\boldsymbol{k}$. (\ref{eq:huumm})
can be expanded into:
\begin{equation}
\mathcal{\boldsymbol{K}}\left(\gamma_{\mathcal{O}}\right)=\boldsymbol{1}+\ell^{2}\boldsymbol{k}+O\left(\ell^{3}\right).\label{eq:ban}
\end{equation}

We could obtain a discrete version of $\mathcal{\boldsymbol{K}}$
as follows. For the first step, we will obtain the corresponding angle
of rotation. Given a prefered slicing $\Omega_{4}$, relation (\ref{eq:16}),
which describe the 3-dimensional deficit angle $\delta\theta_{i}$
on each internal segment of the move, can be rewritten as follows:
\begin{equation}
\delta\theta_{i}=2\pi-\left(\underset{\theta_{i}}{\underbrace{\theta_{jk,i}}}+\underset{\bar{\theta}_{i}}{\underbrace{\theta_{j4,i}+\theta_{k4,i}}}\right),\qquad i,j,k=1,2,3,\: i\neq j\neq k,\label{eq:19}
\end{equation}
where $\theta_{i}$, with respect to tetrahedron $\boldsymbol{v}_{4}$,
is the \textit{internal} dihedral angle, and $\bar{\theta}_{i}$ is
the \textit{external} dihedral angle coming from the dihedral angles
of \textit{other} tetrahedra \cite{key-44a}, see FIG. 15. 
\begin{figure}[H]
\begin{centering}
\includegraphics[scale=0.55]{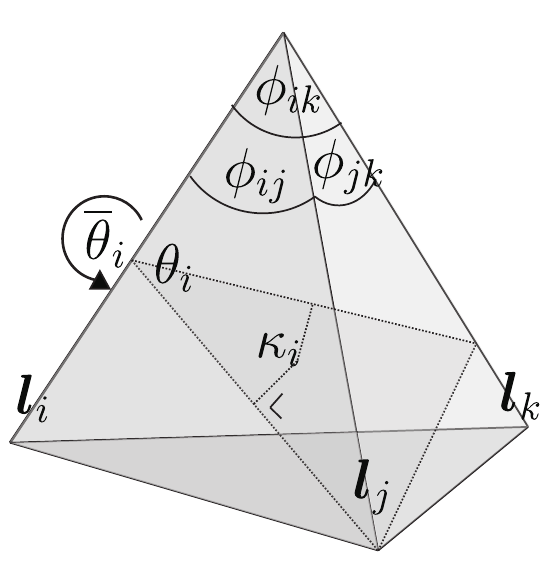}
\par\end{centering}

\caption{Internal and external dihedral angles. From the dihedral angle relation,
we could obtain the \textit{internal} dihedral angle $\theta_{i}$
on a segment. $\bar{\theta}_{i}$ is the \textit{external} dihedral
angle, while $\kappa_{i}$ is the angle between \textit{normals} of
the two triangles.}
\end{figure}

Following the definition in our previous work \cite{key-44a}, let
us introduce the quantity: 
\begin{equation}
\kappa_{i}=\bar{\theta}_{i}-\pi.\label{eq:key}
\end{equation}
For the case where $\mathcal{F}_{\Delta}$ is flat, $\delta\theta_{i}=0$.
This causes $\theta_{i}+\bar{\theta}_{i}=2\pi,$ and using definition
(\ref{eq:key}), we obtain: 
\[
\theta_{i}+\kappa_{i}=\pi.
\]
In this flat case, it is clear that $\kappa_{i}$ is the angle between
the normals of two triangles, see FIG. 15.

We can write $\kappa_{i}$ as: 
\begin{equation}
\kappa_{i}=\pi-\left(\delta\theta_{i}+\theta_{i}\right),\label{eq:hex}
\end{equation}
with $\theta_{i}$ is the internal dihedral angle. Therefore, we define
$\kappa_{i}$ as the 2-dimensional deficit angle of the \textit{extrinsic
curvature}, because it is in accordance with the definition of extrinsic
curvature (\ref{eq:exgen}), where $\boldsymbol{\mathcal{K}}$ is
defined as the covariant derivative of the normal $\tilde{\boldsymbol{n}}$
to the hypersurface $\Omega$. It will inherit the curvature of the
3-dimensional manifold. 

The discrete holonomy of extrinsic curvature on hinge $\boldsymbol{l}_{i}$
can be written as:
\begin{equation}
\mathcal{\boldsymbol{K}}_{\Delta i}\left(\gamma_{\mathcal{O}}\right)=\exp\left(\left.\boldsymbol{J}_{i}\right|_{\mathcal{O}}\kappa_{i}\right),\label{eq:exdis}
\end{equation}
with $\left.\boldsymbol{J}_{i}\right|_{\mathcal{O}}$ satisfying (\ref{eq:penting})
and $\kappa_{i}$ satisfying (\ref{eq:hex}).

Another way to obtain the discrete extrinsic curvature exist, which
yields the following relation:

\begin{equation}
\boldsymbol{\mathcal{K}}_{ij,k}\left(\gamma_{\mathcal{O}'}\right)=\boldsymbol{h}_{kj,i}^{'}\boldsymbol{O}_{ij}^{'-1}\boldsymbol{h}_{ik,j}.\label{eq:exdis2}
\end{equation}
(\ref{eq:exdis2}) is a consequence of the Gauss-Codazzi equation.
In the next sections, we will show that (\ref{eq:exdis}) and (\ref{eq:exdis2})
yields similar deficit angles, but located on different hinges; and
both of these definitions will coincide in the continuum limit.

\subsection{The Discrete Gauss-Codazzi Equation}

\subsubsection{Geometrical Settings}

The continuous Gauss-Codazzi equation is defined on each point on
an arbitrary manifold $\mathcal{F}$. By introducing a regulator which
'blows' a points into $n$-dimensional 'bubbles' (that is, an $n$-dimensional
simplex in the dual lattice $\star\mathcal{F}_{\triangle}$, see FIG.
3(d)), and using the fact that bubbles is constructed from several
loops meeting together, we could define the discrete Gauss-Codazzi
equation on each loop of triangulation $\mathcal{F}_{\Delta}$. To
do this, we need to choose a specific loop lying on the submanifold
$\Omega_{\Delta}$, see FIG. 16.

\begin{figure}[H]
\centering{}\includegraphics[scale=0.55]{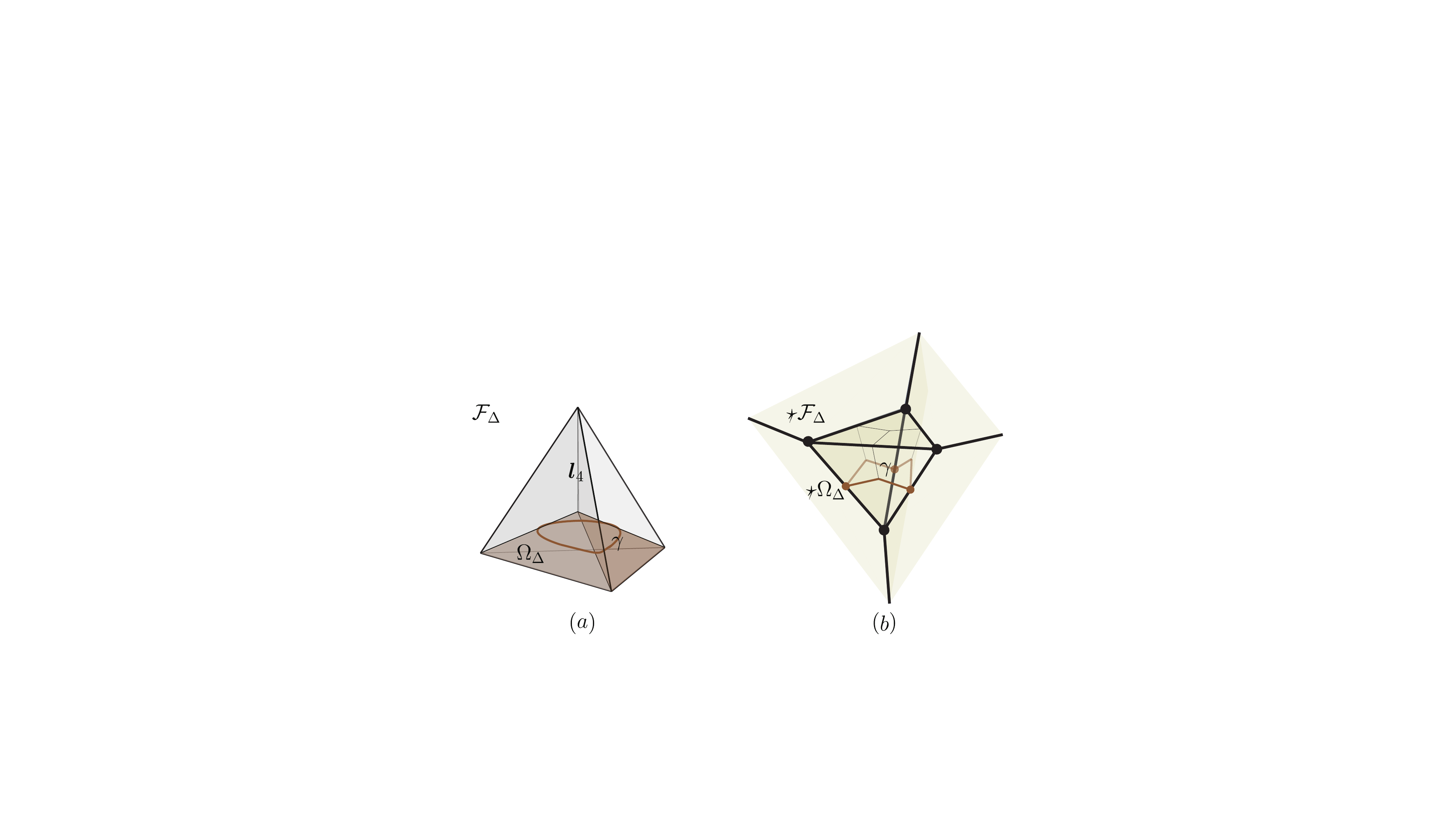} \caption{Suppose we have a 3-dimensional curved manifold $\mathcal{F}_{\triangle}$
discretized by four tetrahedron in the figure (a) (in flat case it
is known as \textit{1-4 Pachner move}). Then we take an embedded slice
$\Omega_{\triangle}$ as the surface of one tetrahedron (the dark
blue surface discretized by three triangle). Embedded on $\Omega_{\triangle},$
we take a loop $\gamma$ circling a point of a tetrahedron. Attached
to $\gamma$, are the $SU(2)$ and $SO(2)$ holonomy, which are related
to the 3D and 2D intrinsic curvature, respectively. (b) is the dual
lattice of (a). The use of abstract combinatorial dual guarantees
$\star\Omega_{\Delta}$ to be embedded in $\star\mathcal{F}.$}
\end{figure}
The first step is to define the holonomy of the projected 3-dimensional
and 2-dimensional intrinsic curvature along the loop. The following
are several quantities we had on the simplicial complex, each of them
will be illustrated geometrically on the primal and dual lattices.

\subsubsection{The Curvatures}

\paragraph{Projected 3D intrinsic curvature: deficit angle on segment $\boldsymbol{l}_{4}$. }

Loop $\gamma_{4}$ circles hinges $\boldsymbol{l}_{1}$, $\boldsymbol{l}_{2}$,
and $\boldsymbol{l}_{3}$, and therefore, the total 3-dimensional
holonomy on $\gamma_{4}$ is the product of holonomies on each hinge
it contains. By (\ref{eq:10}), it is clear that the 3-dimensional
holonomy around loop $\gamma_{4}$ on $\Omega_{4}$ with origin $\mathcal{O}=\star\boldsymbol{v}_{4}$,
is:
\[
\boldsymbol{H}_{123}\left(\boldsymbol{A},\gamma_{4,\mathcal{O}}\right)=\boldsymbol{H}_{42}\boldsymbol{H}_{4}^{-1}\boldsymbol{H}_{24}.
\]
Therefore, the projected 3D intrinsic curvature on $\Omega_{4}$ in
the holonomy representation, is:
\[
\left.\,^{3}\boldsymbol{F}_{\Delta}\right|_{\Sigma}=\boldsymbol{H}_{123}\left(\boldsymbol{A},\gamma_{4,\mathcal{O}}\right).
\]
Written in the plane-angle representation, we have the deficit angle
and plane which are functions of $\delta\theta_{i},\boldsymbol{l}_{i}$,
$i=1,2,3$ by the closure constraint (\ref{eq:12}):

\paragraph*{\textup{
\begin{eqnarray*}
\left.\,^{3}\boldsymbol{F}_{\Delta}\right|_{\Sigma} & = & \left(\delta\theta_{4},\left.\hat{\boldsymbol{J}}_{4}\right|_{\mathcal{O}}\right)=\left(\delta\theta_{4},\boldsymbol{H}_{42}\left.\frac{d\boldsymbol{H}_{4}}{d\delta\theta_{4}}\right|_{\delta\theta_{4}=0}\boldsymbol{H}_{24}\right),\protect\\
\delta\theta_{4} & = & 2\pi-\left(\theta_{ij,4}+\theta_{ik,4}+\theta_{jk,4}\right),\, i,j,k\neq4,i<j<k,\protect\\
\left.\hat{\boldsymbol{J}}_{4}\right|_{\mathcal{O}} & = & \left.\frac{d\boldsymbol{\overline{H}}_{4}}{d\delta\theta_{4}}\right|_{\delta\theta_{4}=0}.
\end{eqnarray*}
}}

see FIG. 17. 
\begin{figure}[H]
\centering{}\includegraphics[scale=0.55]{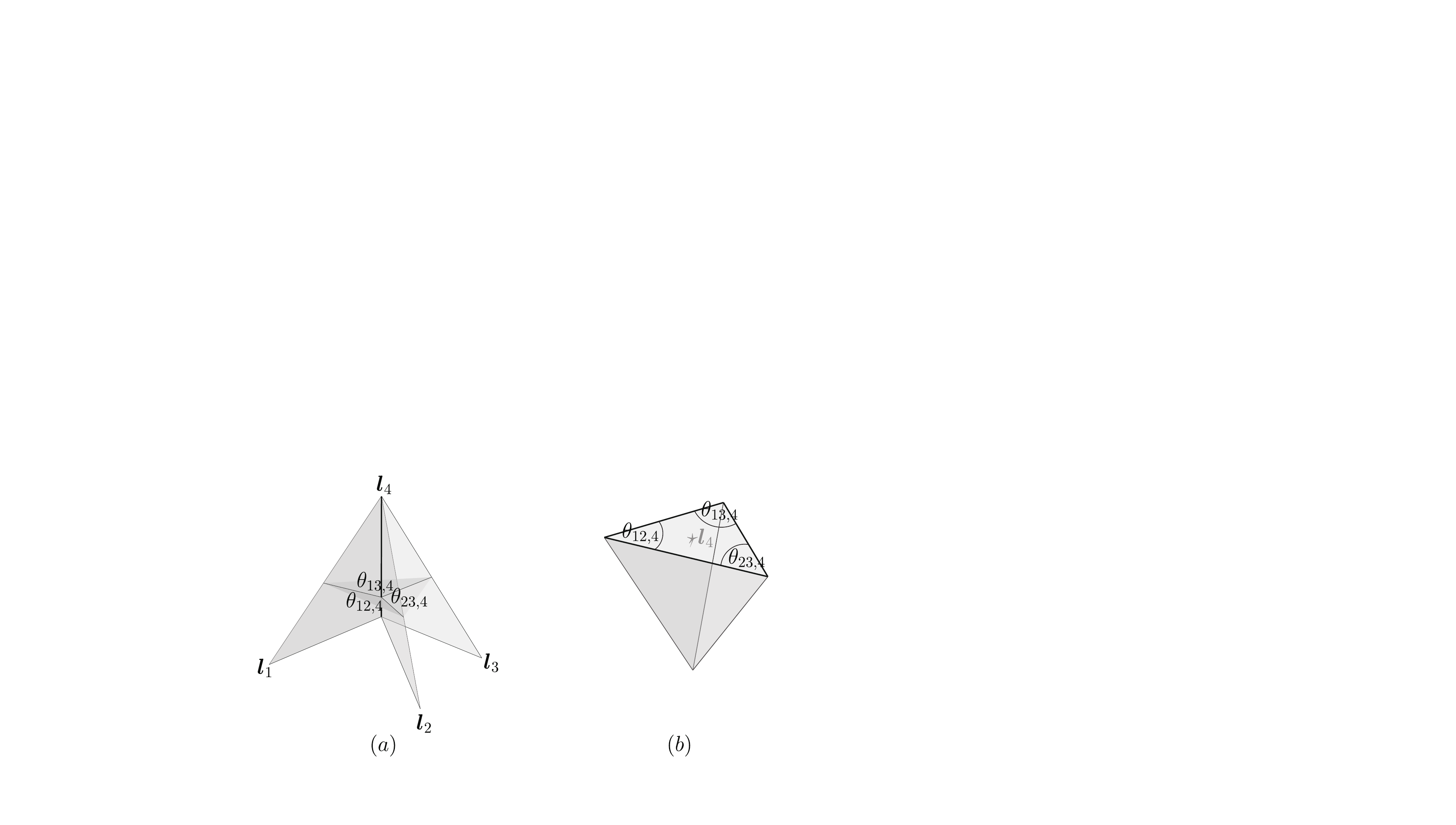}\caption{a) the projected 3D intrinsic curvature in primal lattice, (b) the
projected 3D intrinsic curvature in dual lattice.}
\end{figure}

\paragraph{2D intrinsic curvature: deficit angle on point $p$.}

It is clear from (\ref{eq:18}) that the 2-dimensional holonomy around
loop $\gamma_{4}$ on $\Omega_{4}$ with origin $\mathcal{P}$, is:
\[
\,^{2}\boldsymbol{F}_{\triangle}=\boldsymbol{O}_{4}\left(\boldsymbol{a},\gamma_{4,\mathcal{P}}\right),
\]
with the plane-angle representation as:

\paragraph*{\textup{
\begin{eqnarray*}
\,^{2}\boldsymbol{F}_{\Delta} & = & \left(\delta\phi_{4},\left.\boldsymbol{\hat{j}}_{4}\right|_{\mathcal{P}}\right),\protect\\
\delta\phi_{4} & = & 2\pi-\left(\phi_{ij}+\phi_{ik}+\phi_{jk}\right),\, i,j,k\neq4,i<j<k,
\end{eqnarray*}
}}

see FIG. 18. 
\begin{figure}[H]
\centering{}\includegraphics[scale=0.55]{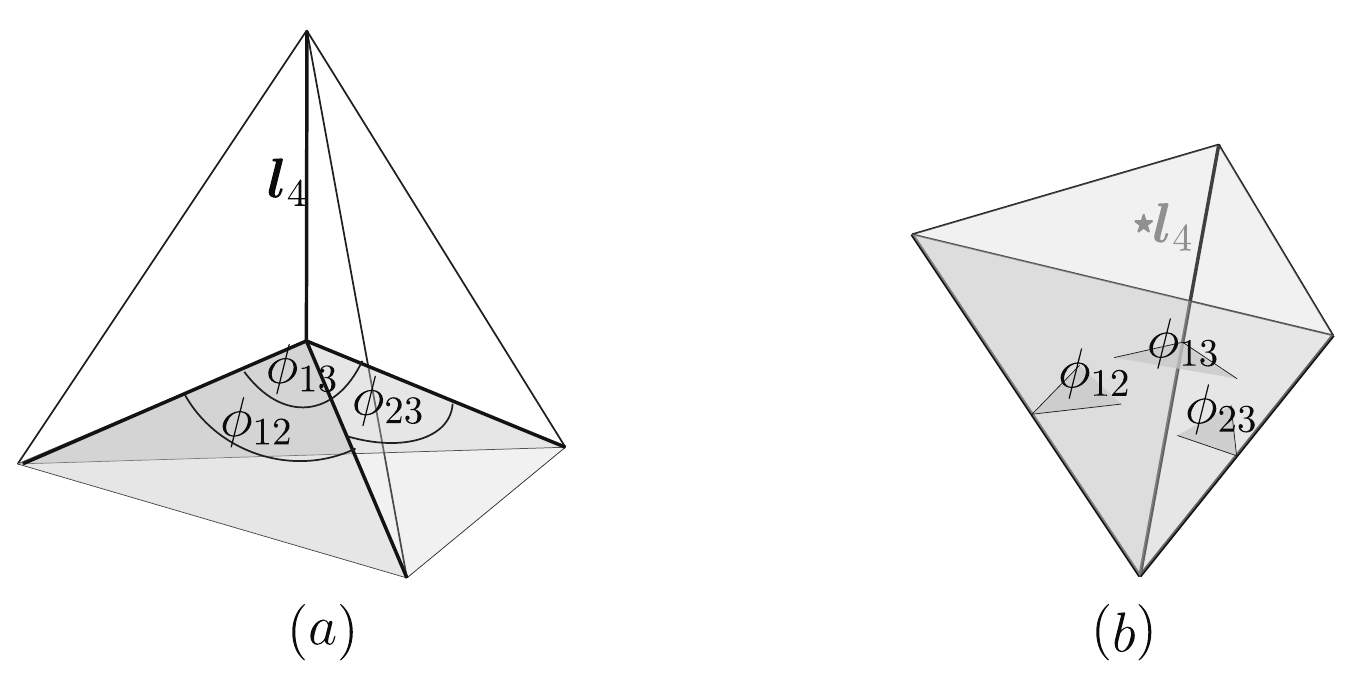}\caption{(a) the 2D intrinsic curvature in primal lattice, (b) the 2D intrinsic
curvature in dual lattice.}
\end{figure}

\paragraph{2D extrinsic curvature.}

The total extrinsic curvature circling loop $\gamma_{4}$ on slice
$\Omega_{4}$ with origin $\mathcal{O}$, is the product of $\boldsymbol{\mathcal{K}}_{i}$
on the three hinge it crosses:
\[
\boldsymbol{\mathcal{K}}_{123}=\boldsymbol{\mathcal{K}}_{1}\boldsymbol{\mathcal{K}}_{2}\boldsymbol{\mathcal{K}}_{3},
\]
Each $\boldsymbol{\mathcal{K}}_{i}$ satisfies (\ref{eq:key}) and
(\ref{eq:exdis}), where each external dihedral angles, from (\ref{eq:19}),
are:
\begin{equation}
\bar{\theta}_{i}=\theta_{j4,i}+\theta_{k4,i},\qquad i,j,k=1,2,3,\: i\neq j\neq k,\label{eq:huhuk}
\end{equation}
see FIG. 19. 
\begin{figure}[H]
\centering{}\includegraphics[scale=0.55]{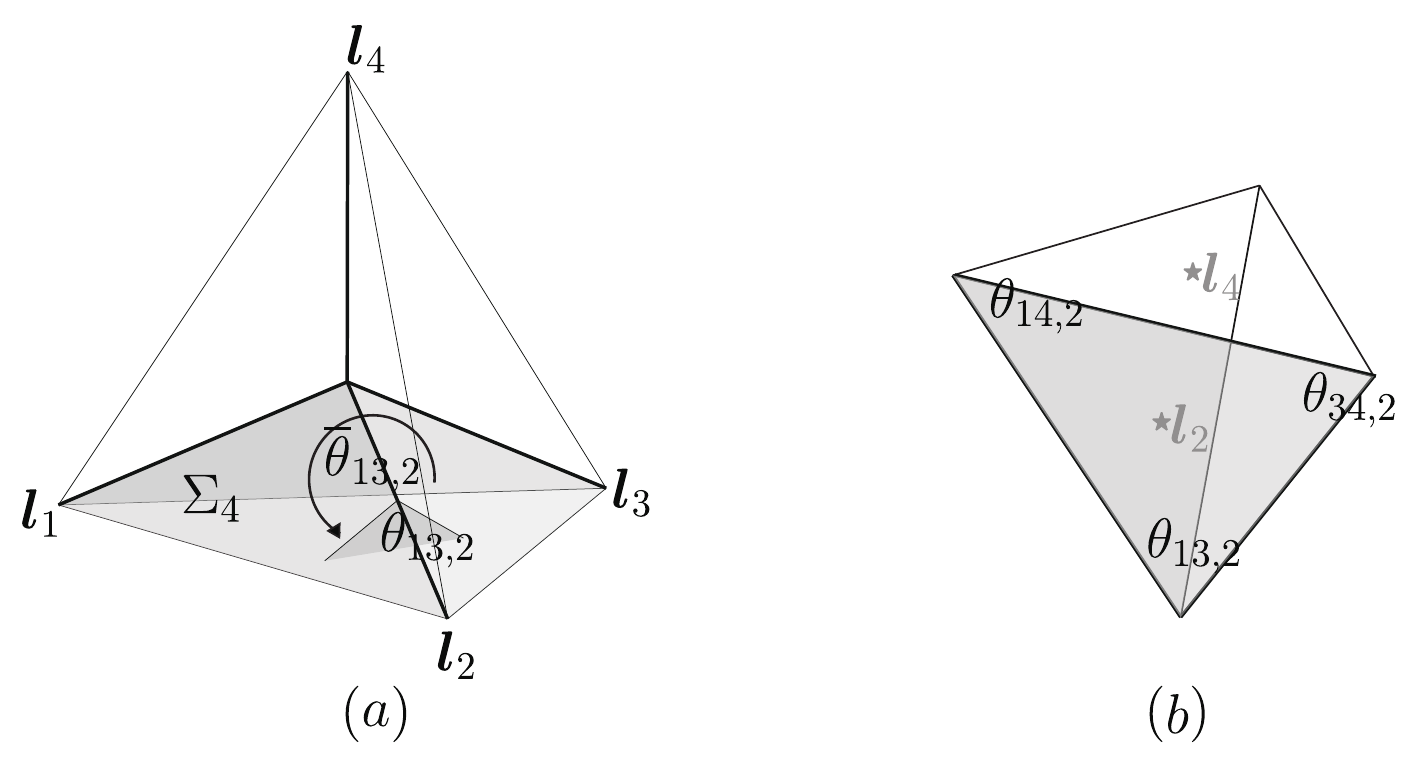}\caption{(a) the 2D extrinsic curvature in primal lattice, (b) the 2D extrinsic
curvature in dual lattice.}
\end{figure}

Therefore, the 2D extrinsic curvature, written in holonomy and plane-angle
representation, are:
\begin{eqnarray*}
\,^{2}\boldsymbol{K}_{\triangle} & = & \boldsymbol{\mathcal{K}}_{123}\left(\gamma_{4,\mathcal{O}}\right),\\
\,^{2}\boldsymbol{K}_{\Delta} & = & \left(\delta\kappa_{123}\sim\textrm{tr \ensuremath{\boldsymbol{\mathcal{K}}_{123}}},\left.\frac{d\boldsymbol{\mathcal{K}}_{123}}{d\delta\kappa_{123}}\right|_{\delta\kappa_{123}=0}\right)
\end{eqnarray*}
These definitions of discrete curvatures are natural, in the sense
that we did not use any assumption to derive them, besides the assumption
of small loop approximation. An important fact that arise from these
definition is that the extrinsic and intrinsic curvature can not be
obtained simultaneously; which will be clear in the next subsection.

\subsubsection{Dihedral Angle Relation as the Discrete Gauss-Codazzi Equation}

To derive the Gauss-Codazzi equation, we need these following quantities:
3D curvature, 2D curvature, and the Bianchi identity. The holonomies
of these three quantities have different 'natural' points of origin;
the origin $\mathcal{O}$ of 3D holonomy is naturally located at the
vertex, the origin $\mathcal{P}$ of 2D holonomy at (the middle of)
the face, while the origin $\mathcal{O}'$ of the trivalent loops
on (the middle of) the edge. To obtain the correct relation, all of
them need to have a same origin. The following transformation of an
arbitrary holonomy $\boldsymbol{H}_{\gamma}$ on loop $\gamma$ will
be useful:
\begin{equation}
\boldsymbol{H}_{\gamma}\left(\gamma_{\mathcal{P}}\right)=\boldsymbol{O}_{ik}^{\left(R\right)-1}\boldsymbol{H}_{\gamma}\left(\gamma_{\mathcal{O}'}\right)\boldsymbol{O}_{ik}^{\left(R\right)}=\boldsymbol{O}_{ik}^{\left(R\right)-1}\boldsymbol{H}_{lj}^{\left(R\right)}\boldsymbol{H}_{\gamma}\left(\gamma_{\mathcal{O}}\right)\boldsymbol{H}_{lj}^{\left(R\right)-1}\boldsymbol{O}_{ik}^{\left(R\right)}.\label{eq:tr}
\end{equation}
 See FIG. 14(a).

From the discrete Bianchi identity, we could write:
\begin{equation}
\boldsymbol{h}_{ij,k}\left(\boldsymbol{A},\gamma_{k,\mathcal{O}^{'}}\right)=\boldsymbol{h}_{kj,i}\left(\boldsymbol{A},\gamma_{i,\mathcal{O}^{'}}\right)\boldsymbol{h}_{ik,j}\left(\boldsymbol{A},\gamma_{j,\mathcal{O}^{'}}\right),\label{eq:bagus}
\end{equation}
with point $\mathcal{O}'$ as the origin. There exist a beautiful
geometrical interpretation of relation (\ref{eq:bagus}) as follows.
Since they are 3D holonomies, $\boldsymbol{h}_{ij,k}$'s are elements
of $SU\left(2\right)$, and therefore, could be written explicitly
by (\ref{eq:u}). Taking the trace of (\ref{eq:bagus}), gives exactly
the following relation: 
\begin{equation}
\cos\theta_{ij,k}=\cos\theta_{ik,j}\cos\theta_{kj,i}-\cos\bar{\phi}_{ij}\sin\theta_{ik,j}\sin\theta_{kj,i},\label{eq:32}
\end{equation}
which is indeed the total angle relation formula (\ref{eq:22}). The
$\theta_{ij,k}$'s are the angle of rotation of holonomy $\boldsymbol{h}_{ij,k}$'s,
while $\bar{\phi}_{ij}$ is the angle between plane of rotation $\hat{\boldsymbol{J}}_{i}$
and $\hat{\boldsymbol{J}}_{j}$:
\begin{eqnarray*}
\left.\hat{\boldsymbol{J}}_{i}\right|_{\mathcal{O}^{'}} & = & \left.\frac{d\boldsymbol{h}_{kj,i}}{d\theta_{kj,i}}\right|_{d\theta_{kj,i}=0},\\
\left.\hat{\boldsymbol{J}}_{j}\right|_{\mathcal{O}^{'}} & = & \left.\frac{d\boldsymbol{h}_{ik,j}}{d\theta_{ik,j}}\right|_{d\theta_{ik,j}=0},
\end{eqnarray*}
which are (\ref{eq:penting}) parallel-transported to $\mathcal{O}^{'}$
(remember that $\boldsymbol{H}_{k}$ and $\boldsymbol{h}_{ij,k}$
share the same hinge $\boldsymbol{l}_{k},$and therefore share the
same plane of rotation $\boldsymbol{J}_{k}$, but different angle
of rotation). In the vectorial picture viewed from $\mathcal{O}'$,
plane of rotation $\boldsymbol{J}_{i}$ and $\boldsymbol{J}_{j}$
are indeed dual to segment $\boldsymbol{l}_{i}$ and $\boldsymbol{l}_{j}$.
With $\phi_{ij}=\pi-\bar{\phi}_{ij}$ as the angle between hinge $\boldsymbol{l}_{i}$
and $\boldsymbol{l}_{j}$, we could write (\ref{eq:32}) as the \textit{dihedral
angle formula}, which is a relation between $\phi$, the 2D angles
between segments of a Euclidean tetrahedron, and $\theta,$ the 3D
angles between planes of the same tetrahedron:
\begin{equation}
\cos\phi_{ij}=\frac{\cos\theta_{ij,k}-\cos\theta_{ik,j}\cos\theta_{kj,i}}{\sin\theta_{ik,j}\sin\theta_{kj,i}}.\label{eq:inverse}
\end{equation}
see FIG. 20. 
\begin{figure}[H]
\centering{}\includegraphics[scale=0.55]{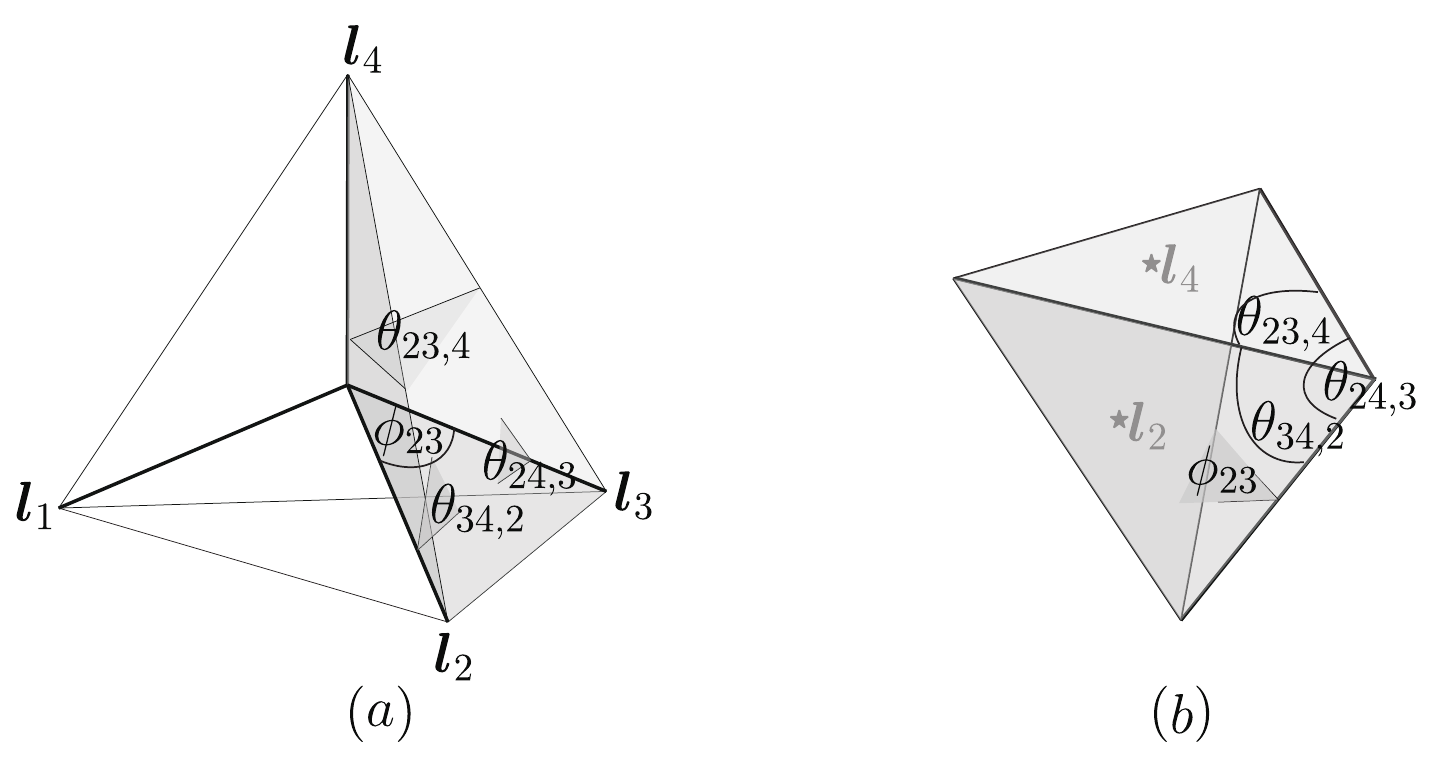} \caption{(a) Given angles $\phi_{ij}$, $\phi_{ik}$, $\phi_{jk}$ at point
$p$ of a tetrahedron, we could obtain the dihedral angle $\theta{}_{ij,k}$.
In fact, $\theta{}_{ij,k}$ is only $\phi_{ij}$ projected on the
plane normal to segment $\boldsymbol{l}_{k}$. (b) In the dual lattice.}
\end{figure}

Remarkably, as shown in \cite{key-59}, the dihedral angle formula
is valid for any dimension, relating $p$-dimensional angle (the angle
between $\left(p-1\right)$-simplices) with $\left(p-1\right)$-dimensional
angle (the angle between $\left(p-2\right)$-simplices). The formula
can also be written in the inverse form:
\begin{equation}
\cos\theta{}_{ij,k}=\frac{\cos\phi{}_{ij}-\cos\phi_{ik}\cos\phi_{kj}}{\sin\phi_{ik}\sin\phi_{kj}},\label{eq:8-1}
\end{equation}
We will show that this formula is the discrete Gauss-Codazzi relation
for angles, which will give the continuous Gauss-Codazzi relation
(\ref{eq:20}) in the continuum limit. 

Since $\theta_{ij,k}$ and $\phi_{ij}$ are, respectively, the parts
of 3D and 2D intrinsic curvature, the dihedral angle formula relates
these intrinsic curvatures together. Therefore, it is reasonable to
expect the remaining term to be the extrinsic curvature; we will check
if it coincides with our definition in (\ref{eq:exdis}).

The next step, is to write the elements of 3D and 2D curvature in
the point of view of $\mathcal{O}'$ as the origin. For the 3D curvature,
it is done by equation (\ref{eq:xbesar}), while for 2D, it is done
by (\ref{eq:yay}). Returning to relation (\ref{eq:bagus}), an important
remarks we need to emphasize is: the $\boldsymbol{h}$'s are the holonomy
circling segments, which is a 3-dimensional properties. The 2-dimensional
property $\bar{\phi}_{ij}$ comes \textit{implicitly} from the relation
between two 3D holonomies $\boldsymbol{h}_{kj,i}$ and $\boldsymbol{h}_{ik,j}$.
We could explicitly insert the 2-dimensional property by gauge fixing:
sending one of the holonomy, say $\boldsymbol{h}_{kj,i}$ at hinge
$\boldsymbol{l}_{i}$ to $\boldsymbol{l}_{j}$. The corresponding
2D holonomy connecting these two hinges is:
\[
\boldsymbol{O}_{ij}=\boldsymbol{O}_{ij}\left(\boldsymbol{a},\gamma_{\mathcal{P}}\right)=\exp\left(\left.\frac{d\boldsymbol{O}_{ij}}{d\phi_{ij}}\right|_{\phi_{ij}=0}\phi_{ij}\right),
\]
which can be written in $\mathcal{O}^{'}$ point of view, using transformation
(\ref{eq:tr}):
\begin{eqnarray}
\boldsymbol{O}_{ij}^{'} & = & \boldsymbol{O}_{ij}^{'}\left(\boldsymbol{a},\gamma_{\mathcal{O}'}\right)=\exp\left(\boldsymbol{O}^{\left(R\right)}\left.\frac{d\boldsymbol{O}_{ij}}{d\phi_{ij}}\right|_{\phi_{ij}=0}\boldsymbol{O}^{\left(R\right)-1}\phi_{ij}\right).\label{eq:cc}
\end{eqnarray}
Relation (\ref{eq:bagus}) can be rewritten as:

\begin{equation}
\boldsymbol{h}_{ij,k}\left(\boldsymbol{A},\gamma_{\mathcal{O}'}\right)=\boldsymbol{O}_{ij}^{'}\left(\boldsymbol{a},\gamma_{\mathcal{O}'}\right)\underset{\boldsymbol{h}_{kj,i}^{'}\left(\boldsymbol{A},\gamma_{\mathcal{O}'}\right)}{\underbrace{\boldsymbol{O}_{ij}^{'-1}\left(\boldsymbol{a},\gamma_{\mathcal{O}'}\right)\boldsymbol{h}_{kj,i}\left(\boldsymbol{A},\gamma_{\mathcal{O}'}\right)\boldsymbol{O}_{ij}^{'}\left(\boldsymbol{a},\gamma_{\mathcal{O}'}\right)}}\boldsymbol{O}_{ij}^{'-1}\left(\boldsymbol{a},\gamma_{\mathcal{O}'}\right)\boldsymbol{h}_{ik,j}\left(\boldsymbol{A},\gamma_{\mathcal{O}'}\right).\label{eq:hue}
\end{equation}
with:
\begin{eqnarray}
\boldsymbol{h}_{kj,i}^{'} & = & \boldsymbol{O}_{ij}^{'-1}\boldsymbol{h}_{kj,i}\boldsymbol{O}_{ij}^{'}=\boldsymbol{O}_{ij}^{'-1}\exp\left(\left.\boldsymbol{J}_{i}\right|_{\mathcal{O}^{'}}\theta_{kj,i}\right)\boldsymbol{O}_{ij}^{'},\nonumber \\
 & = & \exp\left(\boldsymbol{O}_{ij}^{'-1}\left.\boldsymbol{J}_{i}\right|_{\mathcal{O}^{'}}\boldsymbol{O}_{ij}^{'}\theta_{kj,i}\right),\nonumber \\
 & = & \exp\left(\left.\boldsymbol{J}_{j}\right|_{\mathcal{O}^{'}}\theta_{kj,i}\right).\label{eq:bb}
\end{eqnarray}
$\boldsymbol{h}_{kj,i}^{'}$ is the holonomy $\boldsymbol{h}_{kj,i}$
at hinge $\boldsymbol{l}_{i}$, sent to $\boldsymbol{l}_{j}$. Therefore,
(\ref{eq:hue}) can be written as:
\begin{equation}
\boldsymbol{h}_{ij,k}\left(\boldsymbol{A},\gamma_{\mathcal{O}'}\right)=\boldsymbol{O}_{ij}^{'}\left(\boldsymbol{a},\gamma_{\mathcal{O}'}\right)\boldsymbol{h}_{kj,i}^{'}\left(\boldsymbol{A},\gamma_{\mathcal{O}'}\right)\boldsymbol{O}_{ij}^{'-1}\left(\boldsymbol{A},\gamma_{\mathcal{O}'}\right)\boldsymbol{h}_{ik,j}\left(\boldsymbol{A},\gamma_{\mathcal{O}'}\right),\label{eq:hgh}
\end{equation}
with $\boldsymbol{O}{}_{ij}^{'}$ satisfying (\ref{eq:cc}) and $\boldsymbol{h}_{kj,i}^{'}$
satisfying (\ref{eq:bb}).

If $\boldsymbol{h}_{ij,k}$ and $\boldsymbol{O}_{ij}^{'}$ describe
the 3D and 2D intrinsic curvature, then the extrinsic curvature term
should be:
\begin{equation}
\boldsymbol{\mathcal{K}}_{ij,k}\left(\gamma_{\mathcal{O}'}\right)=\boldsymbol{h}_{kj,i}^{'}\boldsymbol{O}_{ij}^{'-1}\boldsymbol{h}_{ik,j},\label{eq:26}
\end{equation}
which define the extrinsic curvature of triangle $\boldsymbol{a}_{ij}$,
located at $\boldsymbol{l}_{i}$. (\ref{eq:hgh}) can be written as:
\begin{equation}
\boldsymbol{h}_{ij,k}=\boldsymbol{O}_{ij}^{'}\boldsymbol{\mathcal{K}}_{ij,k}.\label{eq:GCholonomy}
\end{equation}
Notice that $\boldsymbol{\mathcal{K}}_{ij,k}$ and $\boldsymbol{O}_{ij}^{'}$
do not commute in general. Another way of writting (\ref{eq:GCholonomy})
exist, which is a consequence of the freedom in choosing the order
of $\boldsymbol{\mathcal{K}}_{ij,k}$ and $\boldsymbol{O}_{ij}^{'}$,
and the freedom in choosing a fixed hinge $\boldsymbol{l}_{i}$ or
$\boldsymbol{l}_{j}$. 

If (\ref{eq:32}) is the discrete Gauss-Codazzi in terms of angle,
then (\ref{eq:GCholonomy}) is the discrete Gauss-Codazzi in terms
of holonomy. It must be kept in mind that (\ref{eq:GCholonomy}) is
\textit{not} the Gauss-Codazzi equation on a \textit{full} loop $\gamma$,
but on the \textit{third half} of the loop (or the red tetrahedral
lattice in FIG. 12(b)).

Now let us check if $\boldsymbol{\mathcal{K}}_{ij,k}$ coincides with
our definition of extrinsic curvature $\boldsymbol{\mathcal{K}}_{i\Delta}$
in (\ref{eq:exdis}), or at least, with the external dihedral angle
$\bar{\theta}_{i}$ in (\ref{eq:huhuk}). Taking trace of (\ref{eq:26})
(and using the fact that $\textrm{tr}ABC=\textrm{tr}CAB,$ together
with equation (\ref{eq:u}), (\ref{eq:22}), and some trigonometric
identities) gives:
\begin{equation}
\cos\kappa_{ij,k}=\cos\left(\underset{\theta_{j}^{'}}{\underbrace{\theta_{ik,j}+\theta_{kj,i}}}\right)\cos\phi_{ij}.\label{eq:alfa}
\end{equation}
Comparing (\ref{eq:huhuk}) and (\ref{eq:alfa}), it is clear that
they are not equivalent, with the geometrical picture illustrated
in FIG. 21 as follows.
\begin{figure}[H]
\begin{centering}
\includegraphics[scale=0.55]{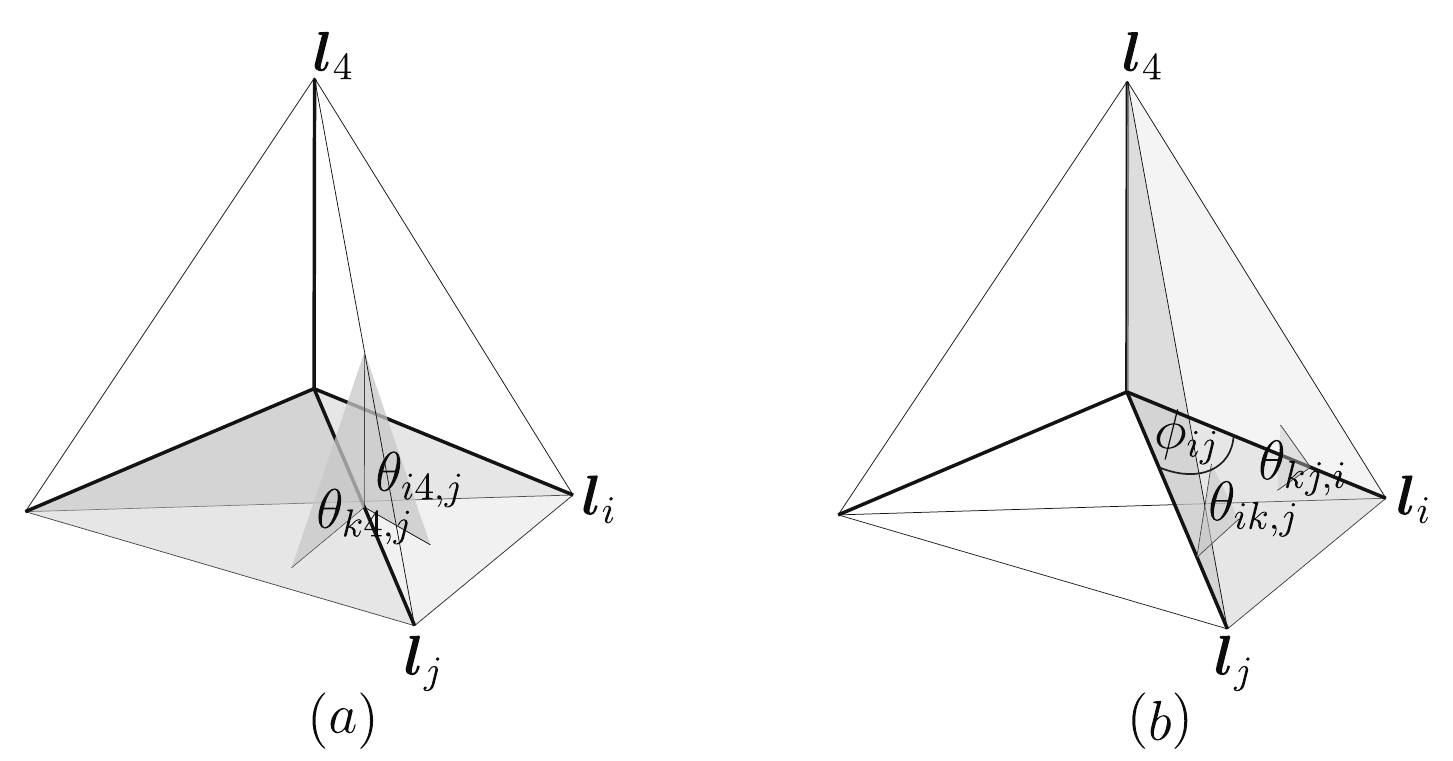}
\par\end{centering}

\caption{(a) The discrete extrinsic curvature from its continuous counterpart,
(\ref{eq:huhuk}), (b) The extrinsic curvature from discrete Gauss-Codazzi
equation, (\ref{eq:alfa}).}
\end{figure}
The discrepancy is caused by the different natural location of the
extrinsic and intrinsic curvature. The extrinsic curvature is located
naturally on segment $\boldsymbol{l}$ (FIG. 21(a), where the extrinsic
curvature is defined by the external angle $\bar{\theta}$ of (\ref{eq:huhuk})),
while the 2D intrinsic curvature lies naturally on point $p$. Since
the Gauss-Codazzi relation needs to be defined on a same (part) of
the loop, the extrinsic curvature is forced to be located at a same
place where the intrinsic curvature lies. This is illustrated in FIG.
21(b), where the extrinsic curvature is defined by angle $\theta^{'}$
of (\ref{eq:alfa}), on different hinge $i,j$, which defines the
tetrahedron. The factor $\cos\phi_{ij}$ describe the relation between
different parts of extrinsic curvature. It is impossible to obtain
the intrinsic and extrinsic curvature simultaneously, in the sense,
the sharpness of one of them will cause the spread in other, since
they live in different hinges. This 'non-commutativity' occurs because
of the discreteness. Nevertheless, it is clear that if we refine the
discretization, in the continuum limit where $\phi\rightarrow0$,
the tetrahedron will shrink to a single point, thus the intrinsic
and extrinsic curvature will be located on the same place and $\bar{\theta}\approx\theta^{'}$.
The 'non-commutativity' between the quantities will dissapear.

\section{The Continuum Limit and Discussions}

\subsection{Recovering the Continuum Limit}

Let us collect all together the results obtained from the previous
sections in the following table: 
\begin{table}[H]
\centering{}%
\begin{tabular}{|c|c|}
\hline 
2D discrete curvature 2-form & $\,^{2}\boldsymbol{F}_{\triangle}=\boldsymbol{O}_{4}\left(\boldsymbol{a},\lambda_{4,\mathcal{P}}\right)$\tabularnewline
\hline 
3D discrete curvature 2-form & $\,^{3}\boldsymbol{F}_{\triangle}=\left(\boldsymbol{H}_{1}\left(\boldsymbol{A},\gamma_{1,\mathcal{O}}\right),\boldsymbol{H}_{2}\left(\boldsymbol{A},\gamma_{2,\mathcal{O}}\right),\boldsymbol{H}_{3}\left(\boldsymbol{A},\gamma_{3,\mathcal{O}}\right)\right).$\tabularnewline
\hline 
Closure constraint & $\boldsymbol{H}_{1}\left(\boldsymbol{A},\gamma_{1,\mathcal{O}}\right)\boldsymbol{H}_{2}\left(\boldsymbol{A},\gamma_{2,\mathcal{O}}\right)\boldsymbol{H}_{3}\left(\boldsymbol{A},\gamma_{3,\mathcal{O}}\right)\overline{\boldsymbol{H}}_{4}\left(\boldsymbol{A},\gamma_{4,\mathcal{O}}\right)=\boldsymbol{1},$\tabularnewline
\hline 
Bianchi identity & $\boldsymbol{h}_{kl,i}\left(\boldsymbol{A},\gamma_{i,\mathcal{O}^{'}}\right)\boldsymbol{h}_{ik,l}\left(\boldsymbol{A},\gamma_{l,\mathcal{O}^{'}}\right)\boldsymbol{h}_{li,k}\left(\boldsymbol{A},\gamma_{k,\mathcal{O}^{'}}\right)=1,$\tabularnewline
\hline 
Gauss-Codazzi equation & $\boldsymbol{h}_{ij,k}\left(\boldsymbol{A},\gamma_{i,\mathcal{O}^{'}}\right)=\boldsymbol{O}_{ij}^{'}\left(\boldsymbol{A},\gamma_{i,\mathcal{O}^{'}}\right)\boldsymbol{\mathcal{K}}_{ij,k}\left(\boldsymbol{A},\gamma_{i,\mathcal{O}^{'}}\right).$\tabularnewline
\hline 
\end{tabular}\caption{The discrete geometrical quantities and relations.}
\end{table}

We will show that these discrete geometrical variables and relations
will yield their standard infinitesimal and continuous counterparts.

\subsubsection{Recovering the Infinitesimal Curvature 2-Form}

The 3D discrete curvature 2-form is written as follows:

\begin{equation}
\left(\,^{3}\boldsymbol{F}_{\triangle}\right)_{\mu\nu}=\boldsymbol{H}_{\mu\nu}\left(\boldsymbol{A},\gamma_{\mu\nu,\mathcal{O}}\right)=\left(\boldsymbol{H}_{1}\left(\boldsymbol{A},\gamma_{1,\mathcal{O}}\right),\boldsymbol{H}_{2}\left(\boldsymbol{A},\gamma_{2,\mathcal{O}}\right),\boldsymbol{H}_{3}\left(\boldsymbol{A},\gamma_{3,\mathcal{O}}\right)\right),\label{eq:x}
\end{equation}
where we recover the indices $\mu\nu$ which had been dropped for
simplicity in the previous section. The first step is to expand (\ref{eq:x})
near the origin $\mathcal{O}$ in the direction $\ell dx^{\mu}\times\ell dx^{\nu}$.
By relation (\ref{eq:6}), the holonomies can be written as: 
\[
\boldsymbol{H}_{\mu\nu}\left(\boldsymbol{A},\gamma_{a,\mathcal{O},\partial_{\mu}\partial_{\nu}}\right)=1+\frac{\ell^{2}}{2}\boldsymbol{F}\left(\partial_{\mu},\partial_{\nu}\right)+O\left(\ell^{3}\right).
\]
Therefore, (\ref{eq:x}) can be written as:
\begin{eqnarray*}
\left(\,^{3}\boldsymbol{F}_{\triangle}\right)_{\mu\nu} & = & 1+\frac{\ell^{2}}{2}\boldsymbol{F}\left(\partial_{\mu},\partial_{\nu}\right)+O\left(\ell^{3}\right),\\
 & = & \left(1+\frac{\ell^{2}}{2}\boldsymbol{F}\left(\partial_{2},\partial_{3}\right)+O\left(\ell^{3}\right),1+\frac{\ell^{2}}{2}\boldsymbol{F}\left(\partial_{3},\partial_{1}\right)+O\left(\ell^{3}\right),1+\frac{\ell^{2}}{2}\boldsymbol{F}\left(\partial_{1},\partial_{2}\right)+O\left(\ell^{3}\right)\right).
\end{eqnarray*}
Taking only the first order terms, which is equivalent with taking
a small loop by setting $\ell\ll1$, we have:
\[
\left(\,^{3}\boldsymbol{F}_{\triangle}\right)_{\mu\nu}\approx1+\frac{\ell^{2}}{2}\boldsymbol{F}\left(\partial_{\mu},\partial_{\nu}\right)=\left(1+\frac{\ell^{2}}{2}\boldsymbol{F}\left(\partial_{2},\partial_{3}\right),1+\frac{\ell^{2}}{2}\boldsymbol{F}\left(\partial_{3},\partial_{1}\right),1+\frac{\ell^{2}}{2}\boldsymbol{F}\left(\partial_{1},\partial_{2}\right)\right).
\]
Now, to take the continuum limit, we differentiate $\,^{3}\boldsymbol{F}_{\triangle}$
with respect to a parameter, which we choose to be the norm of the
vector, $\ell$. This is analog to a differentiation of a curve by
a differential operator to obtain a vector:
\begin{equation}
\frac{d\left(\,^{3}\boldsymbol{F}_{\triangle}\right)_{\mu\nu}}{d\ell}\approx\ell\boldsymbol{F}\left(\partial_{\mu},\partial_{\nu}\right)=\ell\left(\boldsymbol{F}\left(\partial_{2},\partial_{3}\right),\left(\partial_{3},\partial_{1}\right),\left(\partial_{1},\partial_{2}\right)\right).\label{eq:aa}
\end{equation}
Moreover, (\ref{eq:aa}) can be written as: 
\begin{equation}
\frac{1}{\ell}\frac{d\,^{3}\boldsymbol{F}_{\triangle}}{d\ell}\approx\boldsymbol{F}.\label{eq:bbb}
\end{equation}
with $\,^{3}\boldsymbol{F}_{\triangle}$ is the discrete curvature
2-form and $\boldsymbol{F}$ is its continuous counterpart. The same
procedure could be applied to $\,^{2}\boldsymbol{F}_{\triangle}$
and $\,^{2}\boldsymbol{K}_{\triangle}$.

\subsubsection{Recovering the Closure Constraint of a Flat Tetrahedron}

The generalized closure constraint: 
\begin{equation}
\boldsymbol{H}_{1}\boldsymbol{H}_{2}\boldsymbol{H}_{3}\overline{\boldsymbol{H}}_{4}=1,\label{eq:4clos}
\end{equation}
guarantees the closure of, in general, a curved tetrahedron \cite{key-67}.
For a special case where the gauge group $\mathcal{G}=SU\left(2\right)$,
(\ref{eq:4clos}) can be written in the plane-angle representation
using (\ref{eq:u}), which in general, does not gives zero for the
total summations of the planes. But according to \cite{key-29,key-67},
in a small loop approximation, the Taylor expansion of (\ref{eq:4}),
will give:
\begin{equation}
\boldsymbol{H}\left(\boldsymbol{A},\gamma_{\mathcal{O},\ell}\right)=\boldsymbol{1}+\ell\boldsymbol{A}+O\left(\ell^{2}\right),\label{eq:27}
\end{equation}
such that (\ref{eq:4clos}) gives the closure of a \textit{flat} tetrahedron:
\begin{equation}
\sum_{i=1}^{4}\boldsymbol{A}_{i}=0,\label{eq:closflat}
\end{equation}
with $\boldsymbol{A}_{i}$ define the hinges of a flat 4-1 Pachner
move.

Let us take one of the planes in (\ref{eq:closflat}), say $\boldsymbol{A}_{4}$,
to be zero. This will give: 
\begin{equation}
\sum_{i=1}^{3}\boldsymbol{A}_{i}=0,\label{eq:yah}
\end{equation}
which is geometrically interpreted as a closure of a flat triangle.
Thus we could conclude that a special case of (\ref{eq:closflat}),
where one of the plane is trivial, is a condition for a \textit{flat}
tetrahedron with a \textit{zero} volume, practically, a \textit{flat}
triangle. Now let us take a special case of (\ref{eq:4clos}), where
one of the holonomy, say $\overline{\boldsymbol{H}}_{4}$, is trivial.
This gives: $\boldsymbol{H}_{1}\boldsymbol{H}_{2}\boldsymbol{H}_{3}=1,$
which is the Bianchi identity. Following the analogy with the small
loop approximation case, a special case of (\ref{eq:4clos}), where
one of the holonomy is trivial, is a condition for a \textit{curved}
tetrahedron with a \textit{zero} volume: a \textit{curved} triangle.
But a curved triangle can always be constructed from three flat triangles
meeting one another on their segments, which is an open portion of
a surface of a flat tetrahedron.

Another fact which strengthen our claim is, (\ref{eq:yah}), which
can be written as $-\boldsymbol{A}_{3}=\boldsymbol{A}_{1}+\boldsymbol{A}_{2},$
gives:
\begin{equation}
\left|\boldsymbol{A}_{3}\right|^{2}=\left|\boldsymbol{A}_{1}+\boldsymbol{A}_{2}\right|^{2}=\left|\boldsymbol{A}_{1}\right|^{2}+\left|\boldsymbol{A}_{2}\right|^{2}+2\left\langle \boldsymbol{A}_{1},\boldsymbol{A}_{2}\right\rangle ,\label{eq:yeah}
\end{equation}
as their norms relation (by taking traces), which is clearly the flat
law of cosine. The curved version of this, is remarkably the trace
of Bianchi identity, namely relation (\ref{eq:32}), which is the
spherical law of cosine. Taking small angle (which is equivalent with
taking small loop) approximation, (\ref{eq:32}) becomes:
\[
\theta_{ij,k}^{2}\approx\theta_{ik,j}^{2}+\theta_{kj,i}^{2}+2\theta_{ik,j}\theta_{kj,i}\cos\bar{\phi}_{ij},
\]
which is clearly the flat law of cosine in the form of (\ref{eq:yeah}).

\subsubsection{Recovering the Bianchi Identity}

Let us consider the tetrahedral lattice in FIG. 11(a) and 12(b). Since
we assume the tetrahedra are flat in the interior, relation (\ref{eq:triv})
is satisfied, and the generalized closure constraint reduces to Bianchi
identity, which we rewrite as follows:
\begin{equation}
\boldsymbol{h}_{\mu\nu,\lambda}\boldsymbol{h}_{\lambda\mu,\nu}\boldsymbol{h}_{\nu\lambda,\mu}=1.\label{eq:haix}
\end{equation}
The total holonomy $P_{j}$ is trivial so that the loop can be shrunk
into a point, such that it gives lattice in FIG. 22(b). 
\begin{figure}[H]
\begin{centering}
\includegraphics[scale=0.55]{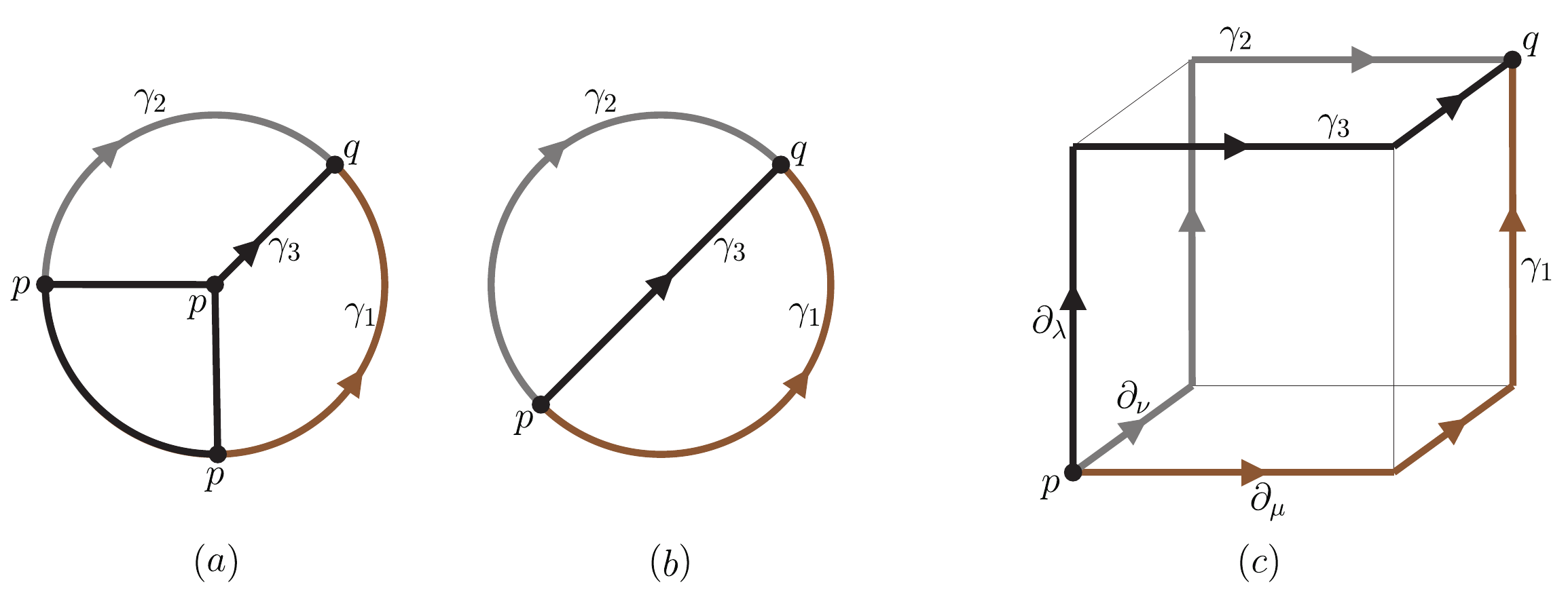}\caption{(a). A tetrahedral lattice where one of the four loop has trivial
holonomy such that it can be shrunk to a point $p$. (b) Figure (a)
is topologically equivalent with (b). (c) Figure (b) is topologically
equivalent with (c).}

\par\end{centering}

\end{figure}
 FIG. 22(a), usually called as the theta-graph \cite{key-22}, is
topologically equivalent to holonomies on the segments of the cube,
see FIG. 22(c). Let us define the holonomies on the segments of the
cube as follows, see FIG. 23. 
\begin{figure}[H]
\begin{centering}
\includegraphics[scale=0.55]{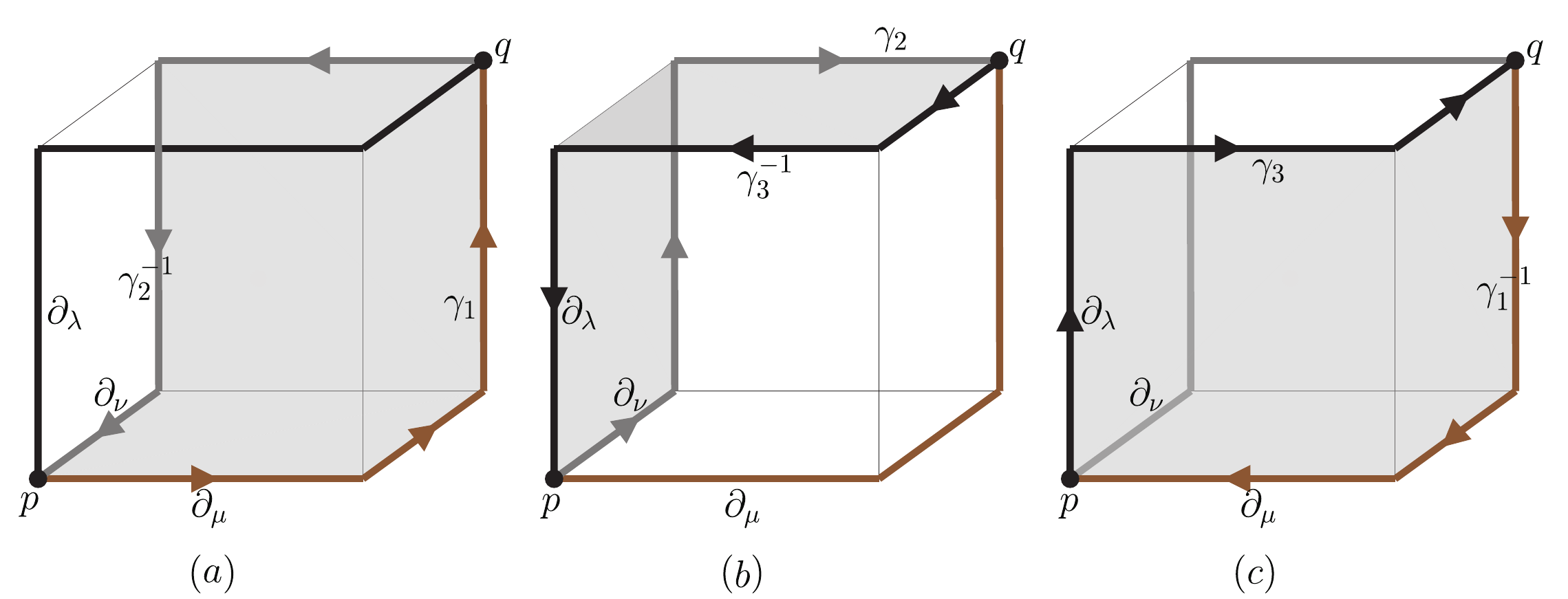}
\par\end{centering}

\caption{Three path of the holonomies on the bones of the cube.}
\end{figure}

The holonomy on path $\gamma_{\mu}\gamma_{\nu}^{-1}$ is written as:
\[
\boldsymbol{h}_{\mu\nu,\lambda}=\boldsymbol{h}_{\mu\nu,\lambda}\left(\boldsymbol{A},\gamma_{\mu}\gamma_{\nu,\mathcal{O}^{'}}^{-1}\right).
\]
For path $\gamma_{\mu}\gamma_{\nu}^{-1}$ , the holonomy could be
Taylor expand near point $p$, in the direction of plane $\ell dx^{\mu}\times\ell dx^{\nu}$,
up to the third order, as:
\[
\boldsymbol{h}_{\mu\nu,\lambda}\left(\boldsymbol{A},\gamma_{\mu}\gamma_{\nu,\mathcal{O}^{'},\partial_{\mu}\partial_{\nu}}^{-1}\right)=1-\frac{\ell^{2}}{2}F_{\mu\nu}+\frac{\ell^{2}}{2}F_{\nu\lambda}+\frac{\ell^{3}}{3!}D_{\nu}F_{\lambda\mu}+O\left(\ell^{4}\right).
\]
It is clear that the holonomies on the three paths in FIG. 21 satisfy
the Bianchi identity (\ref{eq:haix}). Inserting the expansion to
the discrete Bianchi identity yields:
\[
\boldsymbol{h}_{\mu\nu,\lambda}\boldsymbol{h}_{\lambda\mu,\nu}\boldsymbol{h}_{\nu\lambda,\mu}=1+\frac{\ell^{3}}{3!}\left(D_{\mu}F_{\nu\lambda}+D_{\nu}F_{\lambda\mu}+D_{\lambda}F_{\mu\nu}\right)+O\left(\ell^{4}\right)=1.
\]
Taking the small loop limit, which is equal with neglecting the terms
up to the fourth order, gives:
\[
1+\frac{\ell^{3}}{3!}\left(D_{\mu}F_{\nu\lambda}+D_{\nu}F_{\lambda\mu}+D_{\lambda}F_{\mu\nu}\right)\approx1,
\]
or:
\[
\left(D_{\mu}F_{\nu\lambda}+D_{\nu}F_{\lambda\mu}+D_{\lambda}F_{\mu\nu}\right)\approx0,
\]
which is exactly relation (\ref{eq:prev}), or geometrically, (\ref{eq:0}).
(\ref{eq:prev}) could also be written as the Jacobi identity:
\[
\left[D_{\mu},\left[D_{\nu},D_{\lambda}\right]\right]+\left[D_{\nu},\left[D_{\lambda},D_{\mu}\right]\right]+\left[D_{\lambda},\left[D_{\mu},D_{\nu}\right]\right]=0.
\]
The geometrical interpretation of Jacobi identity is the altitude
of a trihedron have three planes meeting in a line \cite{key-1.14}, which guarantees
the flat law of cosine to be satisfied for a triangle. This is in
accordance with the fact that the Bianchi identity for angles (\ref{eq:32})
is indeed the dihedral angle relation, or the spherical law of cosine,
which is satisfied by a curved triangle.

\subsubsection{Recovering the Gauss-Codazzi equation}

Let us take the discrete Gauss-Codazzi relation (\ref{eq:GCholonomy})
as follows:
\[
\boldsymbol{h}_{ij,k}\left(\boldsymbol{A},\gamma_{i,\mathcal{O}^{'}}\right)=\boldsymbol{O}_{ij}^{'}\left(\boldsymbol{A},\gamma_{i,\mathcal{O}^{'}}\right)\boldsymbol{\mathcal{K}}_{ij,k}\left(\boldsymbol{A},\gamma_{i,\mathcal{O}^{'}}\right).
\]
This relation is defined only on a third half part of loop $\gamma_{4}$.
To obtain the full relation on loop $\gamma_{4}$, one need to take
the piecewise linear product of all parts of $\gamma_{4}$. Following
relation (\ref{eq:xbesar}), the total 3D holonomy on $\gamma_{4}$
is:
\[
\left.\,^{3}\boldsymbol{H}\right|_{\Sigma_{4}}=\boldsymbol{h}_{ij,4}\boldsymbol{h}_{jl,4}\boldsymbol{h}_{li,4}.
\]
Inserting the Gauss-Codazzi relation (\ref{eq:GCholonomy}) to the
previous equation, we have:
\begin{equation}
\left.\,^{3}\boldsymbol{H}\right|_{\Sigma_{4}}=\boldsymbol{O}_{ij}^{'}\boldsymbol{\mathcal{K}}_{ij,4}\boldsymbol{O}_{jl}^{'}\boldsymbol{\mathcal{K}}_{jl,4}\boldsymbol{O}_{li}^{'}\boldsymbol{\mathcal{K}}_{li,4}.\label{eq:xoxox}
\end{equation}
Notice that it is impossible to explicitly obtain $\left.\,^{3}\boldsymbol{H}\right|_{\Sigma}$
simultaneously as a function of $\boldsymbol{O}$ or/and $\boldsymbol{\mathcal{K}}$,
where: 
\begin{eqnarray*}
\boldsymbol{O}^{'} & = & \boldsymbol{O}'_{ij}\boldsymbol{O}'_{jl}\boldsymbol{O}'_{li},\\
\boldsymbol{\mathcal{K}} & = & \boldsymbol{\mathcal{K}}_{ij,4}\boldsymbol{\mathcal{K}}_{jl,4}\boldsymbol{\mathcal{K}}_{li,4}.
\end{eqnarray*}
because of the non-commutativity between $\boldsymbol{O}_{ij}^{'}$
and $\boldsymbol{\mathcal{K}}_{ij,4}$. 

But as explained earlier, taking the continuum limit, i.e., taking
small loop approximation, will give simplification. Using (\ref{eq:9})
and (\ref{eq:ban}), we can write (\ref{eq:xoxox}) as:
\[
\boldsymbol{1}-\frac{\ell^{2}}{2}\left.\,^{3}\boldsymbol{F}\right|_{\Sigma}+O\left(\ell^{3}\right)=\prod_{l=1}^{3}\left(1-\frac{\ell^{2}}{2}\,^{2}\boldsymbol{F}_{l}+O\left(\ell^{3}\right)\right)\left(1-\frac{\ell^{2}}{2}\boldsymbol{k}_{l}+O\left(\ell^{3}\right)\right),
\]
where:
\begin{eqnarray*}
\,^{2}\boldsymbol{F}_{l} & = & \,^{2}\boldsymbol{F}\left(\hat{\boldsymbol{s}}_{i},\hat{\boldsymbol{s}}_{j}\right),\quad l\neq i,j,k.\\
\boldsymbol{k}_{l} & = & \left[\mathcal{\boldsymbol{K}}_{l},\mathcal{\boldsymbol{K}}_{l}\right].
\end{eqnarray*}
Neglecting up to the third order terms (which is equivalent with using
small loop) gives:
\[
\boldsymbol{1}-\ell^{2}\left.\,^{3}\boldsymbol{F}\right|_{\Sigma}\approx1-\ell^{2}\left(\left[\mathcal{\boldsymbol{K}}_{1},\mathcal{\boldsymbol{K}}_{1}\right]+\left[\mathcal{\boldsymbol{K}}_{2},\mathcal{\boldsymbol{K}}_{2}\right]+\left[\mathcal{\boldsymbol{K}}_{3},\mathcal{\boldsymbol{K}}_{3}\right]\right)-\ell^{2}\left(\,^{2}\boldsymbol{F}\left(\hat{\boldsymbol{s}}_{1},\hat{\boldsymbol{s}}_{2}\right)+\,^{2}\boldsymbol{F}\left(\hat{\boldsymbol{s}}_{2},\hat{\boldsymbol{s}}_{3}\right)+\,^{2}\boldsymbol{F}\left(\hat{\boldsymbol{s}}_{3},\hat{\boldsymbol{s}}_{1}\right)\right),
\]
and writing in terms of coordinates gives: 
\[
\boldsymbol{1}-\ell^{2}\left.\,^{3}\boldsymbol{F}\right|_{\Sigma}\approx1-\ell^{2}\left(\left[\mathcal{\boldsymbol{K}},\boldsymbol{\mathcal{K}}\right]+\underset{a_{123}^{\mu\nu}}{\underbrace{\left(s_{1}^{\mu}s_{2}^{\nu}+s_{2}^{\mu}s_{3}^{\nu}+s_{3}^{\mu}s_{1}^{\nu}\right)}}\,^{2}F_{\mu\nu}\right).
\]
Using the fact that $a_{123}^{\mu\nu}\,^{2}F_{\mu\nu}$ is the curvature
of $\Omega_{4}$, namely, $\,^{2}\boldsymbol{F}$, finally we have:
\[
\left.\,^{3}\boldsymbol{F}\right|_{\Sigma}=\,^{2}\boldsymbol{F}+\left[\mathcal{\boldsymbol{K}},\boldsymbol{\mathcal{K}}\right],
\]
which is exactly the Gauss-Codazzi equation (\ref{eq:20}) for $\left(2+1\right)$-dimension.

\subsection{The Fundamental Fuzziness in Discrete Geometry}

A careful reader will notice an ambiguity arise in the choice of the
loop used in construction IV C1. Loop $\gamma_{4}$ contains \textit{two}
holonomies, the trivial one, which is $\boldsymbol{P}_{j}$ from relation
(\ref{eq:triv}), and the non-trivial one, which is the projection
of the 3D holonomy $\left.\,^{3}\boldsymbol{F}\right|_{\Sigma}$.
In other words, the loop is simultaneously contractible and non-contractible.
How could this be possible? We try to remove this ambiguity by the
explanation as follows. 

First, the loop is embedded on a 2D slice, while the 2D slice is constructed
by three triangles meeting each other on their edges, which in turn,
are the 3D hinges, see Fig 24(a).
\begin{figure}[H]
\begin{centering}
\includegraphics[scale=0.55]{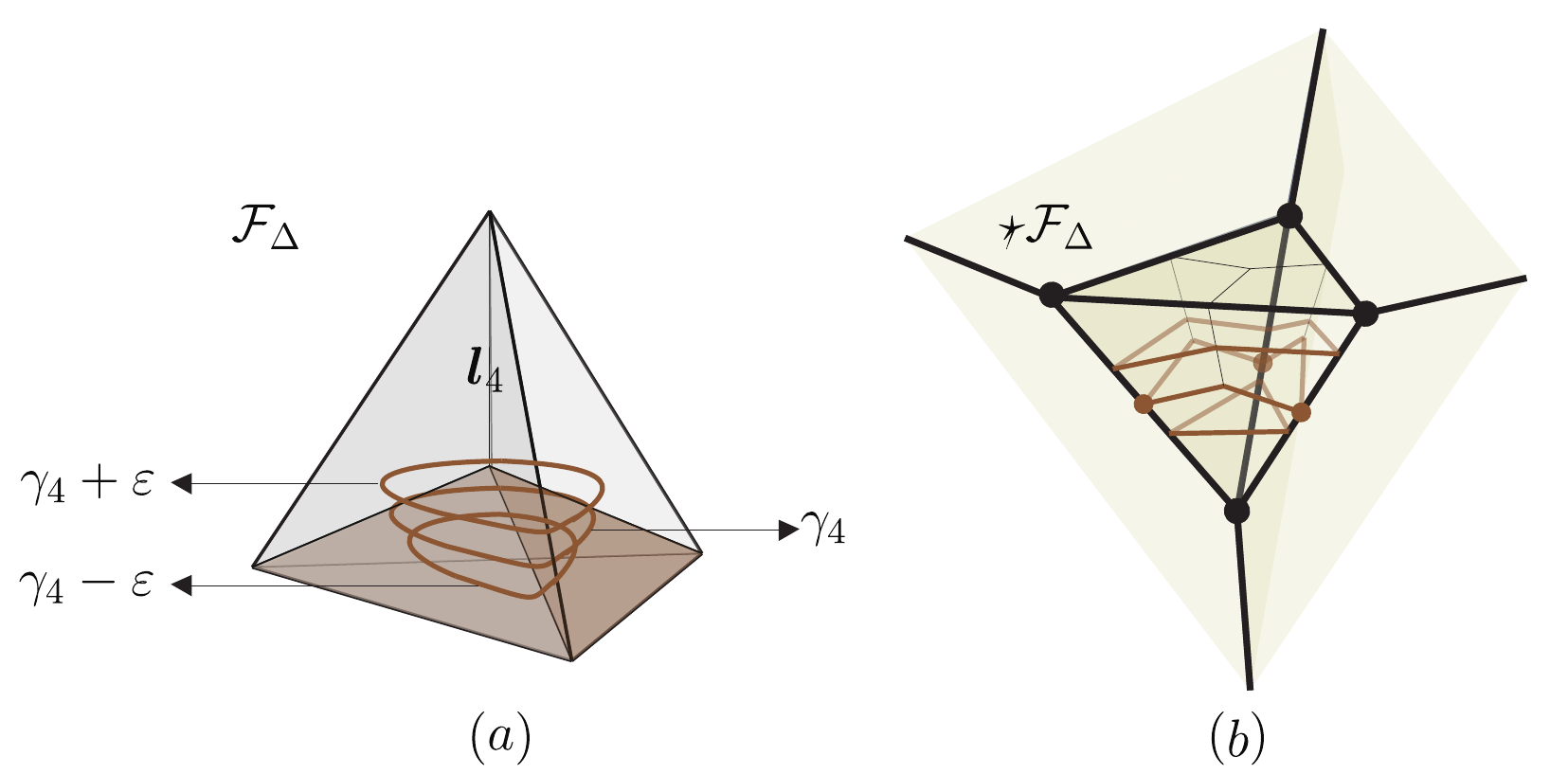}
\par\end{centering}

\caption{(a) Three loops $\gamma_{4}+\varepsilon$, $\gamma_{4}$, and $\gamma_{4}-\varepsilon$
in the primal lattice and (b) in the dual lattice. }
\end{figure}
 We choose the loop such that it circles a point where these 3D hinges
meet, say, point $p,$ which is the 2D hinge. In other words, our
choice of loop will \textit{always} cross these three hinges, so that
the 3D hinges in neither 'outside' nor 'inside' the loop (or both
outside and inside). This is the origin of the ambiguity arise in
our construction.

To solve this, let us define other loops, which is $\gamma_{4}+\varepsilon$
and $\gamma_{4}-\varepsilon$ through a homotopy map, with $\varepsilon$
is small. See FIG. 24. The loop $\gamma+\varepsilon$ is non-contractible,
since it circles the three 3D hinges, while the loop $\gamma_{4}-\varepsilon$
is contractible. It is clear that $\gamma_{4}+\varepsilon$ is the
loop where $\left.\,^{3}\boldsymbol{F}\right|_{\Sigma}$ is located,
while $\gamma_{4}-\varepsilon$ is the loop satisfying relation (\ref{eq:triv}).

Both of these argument are equally correct and well-defined, so it
force us to interpret that there exist a fundamental fuzziness, that
is, an impossibility to obtain sharps variables simultaneously, in
discrete geometry. In particular, it is impossible to obtain the 2D
and 3D holonomy simultaneously; to obtain the sharp 2D holonomy, one
need to place the loop on the 2D surface, that is, loop $\gamma_{4}$,
and this will lead to ambiguity in the holonomy of the 3D curvature.
Meanwhile, to obtain a sharp 3D holonomy, one need to move the loop
sligthly \textit{outside} the 2D surface, which is $\gamma_{4}+\varepsilon$.
Both of these holonomy can not be placed together on a same loop.

We interpret this as a fundamental fuzziness or 'non-commutativity',
which occurs due to the discrete nature of the geometries. In the
(asymptotical) continuum limit, where the hinges become infinitesimal,
$\varepsilon\sim0$, and the two loops will coincide, therefore, the
non-commutativity trait between the 3D and 2D curvature will dissapear,
which is reflected through the continuous Gauss-Codazzi equation.

Another fact which strengthen our argument about the existence of
the fundamental fuzziness in discrete geometry is already explained
in Subsection IV C, which is the impossibility in  obtaining the discrete
2D intrinsic and extrinsic curvature simultaneously.

\subsection{Recovering the Second Order Formulation}

To obtain the second order formulation of gravity, one needs the triads
$e$ coming from the local trivialization between the bundle and its
standard. With the triads satisfying torsionless condition: 
\[
d_{D}e=0,
\]
one could obtain the following relation: 
\[
e\left(\boldsymbol{F}\right)=\boldsymbol{R},
\]
 with $\boldsymbol{R}$ are the Riemann curvature tensor of the base
manifold $M$.

The torsionless condition guarantees the Bianchi identity (\ref{eq:Bi}).
With the torsionless triads, one could obtain the second order variables
of general relativity. It must be kept in mind that the triads maps
alter the coordinate of the plane or rotation, but not the angles
relation, since the trace of holonomy is invariant under diffeomorphism.
The torsionless condition also reduce the degrees of freedom in the
3-dimensional system, from nine components of $F_{\mu\nu J}^{I}$
to six components of $R_{\mu\nu\beta}^{\alpha}$. 

But even in the second order formulation point of view, we still have
a similar geometrical interpretation as explained in Subsection IV
A, but with all geometrical quantities embedded in the base space
$M.$ This is due to the fact that the loop orientation plane and
the rotation bivector in general do not coincide. In fact, using a
specific coordinate, one could write the Riemann tensor such that
all the components are zero, except $R_{\nu\mu\nu}^{\mu}.$ This means
there exist a coordinate where the two planes coincide. As a consequence,
in discrete geometry, it is convenient to treat the loop orientation
plane purely as a coordinate property, or a pure gauge. The use of
the base space is cumbersome in discrete geometries, which may indicate
that the base space is related to a dependent background structure.

\subsection{Conclusions}

We have clarified the definitions and the geometrical interpretation
of curvatures, Bianchi identity, and Gauss-Codazzi equation in the
first order Regge calculus setting. Our variables and relations converge
to their continuous counterparts in the continuum limit. The case
studied in this work is $\left(2+1\right)$-dimensional. A generalization
to higher dimension of these results is possible and highly encouraged.
In particular, it is interesting to see if it is possible to obtain
a compactly written formula for a $\left(3+1\right)$-dimensional
case. Furthermore, as a more ambitious goal, the trivalent condition
could be geometrically interpreted as a spherical triangle, which
could be use as a building blocks for higher dimensional \textit{spherical}
simplex.

\end{document}